%% file: manuscript.tex
\documentclass[manuscript,screen]{acmart}
\usepackage{enumitem}
\usepackage{graphics}
\usepackage{multirow}
\usepackage{ctable} 
\usepackage{ragged2e}
\usepackage{doi}
\usepackage{dirtytalk}
\usepackage[pagewise]{lineno}

\setcopyright{none}
\usepackage{wrapfig}
\newcolumntype{P}[1]{>{\centering\arraybackslash}p{#1}}
\renewcommand\footnotetextcopyrightpermission[1]{} 
\AtBeginDocument{%
  \providecommand\BibTeX{{%
    \normalfont 
    B\kern-0.5em{\scshape i\kern-0.25em b}\kern-0.8em\TeX}}}

\begin{document}

\title{Pre-deployment Analysis of Smart Contracts - A Survey}

\author{Sundas Munir}
\email{sundas.munir@hh.se}
\affiliation{
  \institution{Halmstad University}
  \streetaddress{Kristian IV:s väg 3}
  \city{Halmstad}
  \state{Halland}
  \country{Sweden}
  \postcode{30118}
}

\author{Walid Taha}
\affiliation{%
  \institution{Halmstad University}
  \streetaddress{Kristian IV:s väg 3}
  \city{Halmstad}
  \state{Halland}
  \country{Sweden}
  }
\email{walid.taha@hh.se}

\begin{abstract}
Smart contracts are programs that execute transactions involving independent parties and  cryptocurrencies. As programs, smart contracts are susceptible to a wide range of errors and vulnerabilities. Such vulnerabilities can result in significant losses. Furthermore, by design, smart contract transactions are irreversible. This creates a need for methods to ensure the correctness and security of contracts pre-deployment. Recently there has been substantial research into such methods. The sheer volume of this research makes articulating state-of-the-art a substantial undertaking. To address this challenge, we present a systematic review of the literature. A key feature of our presentation is to factor out the relationship between vulnerabilities and methods through properties. Specifically, we enumerate and classify smart contract vulnerabilities and methods by the properties they address. The methods considered include static analysis as well as dynamic analysis methods and machine learning algorithms that analyze smart contracts before deployment. Several patterns about the strengths of different methods emerge through this classification process.
\end{abstract}



\maketitle
\section{Introduction}
\label{intro}
Smart contracts are computer programs that execute transactions involving independent parties and cryptocurrencies \cite{blockchain_Smartcontracts, lamela2018marlowe}. Their applications are diverse and include the Internet of Things (IoT), insurance policies, lottery schemes, healthcare, games, financial transactions, and supply chain agreements. The most common applications are related to accounts on blockchain platforms such as wallets. Wallet contracts can execute significant financial transactions involving numerous cryptocurrencies \cite{cryptocurrencies}. Today, cryptocurrencies and smart contract applications are gaining increasing adoption.
At the time of writing this paper, one report estimates that there are 22,153 available cryptocurrencies with approx USD 870.5 billion market capitalization \cite{Coinranking}. The combination of immutability and economic significance means that programming errors and flaws in translating business logic into code can have adverse consequences. 
\begin{wrapfigure}{r}{0.45\textwidth}
\begin{center}
 \includegraphics[width=0.42\textwidth]{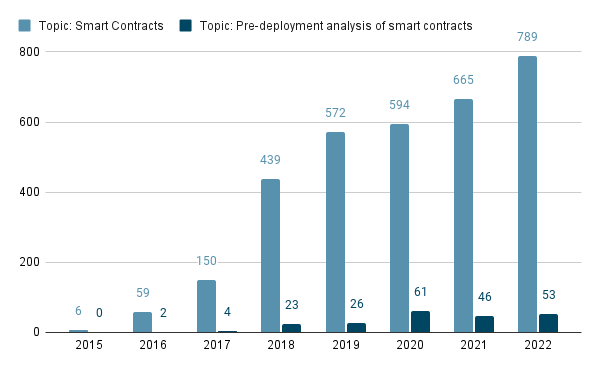}
 \end{center}
  \Description{Publications on smart contracts and on pre-deployment analysis from 2015 to 2022}
    \caption{Publications on smart contracts and on pre-deployment analysis from 2015 to 2022.
}
  \label{fig:graphOfGrowthInResearch}
\end{wrapfigure}
This is far from being a hypothetical concern, as has been demonstrated by multiple incidents, such as the DAO hack \cite{DAOAttack}, the Parity multi-sig wallet library hack \cite{ParityWalletBug}, and the attack on the King of Ether Throne contract \cite{KOTH}. This creates a need for methods to ensure the correctness and security of smart contracts before deployment. Indeed, the last few years have witnessed an increasing volume of research efforts that develop methods to analyze, verify, and test smart contracts before they execute in the real environment. The chart in Figure~\ref{fig:graphOfGrowthInResearch} reflects a surge in activity in this area over the past seven years.
The chart plots, for each year, both the total number of publications on the topic of smart contracts (papers with the term “smart contracts” as a part of the title) as well as the number of papers on pre-deployment analysis that are covered by this survey (details about the criteria for selection are presented in Sec. \ref{Scope and Methodology}).\\
The sheer volume of work in this area makes it challenging to grasp the overall state of the art.\\
\textbf{{Contributions and Organization of the paper. }}
 To address this problem, we present a classification of smart contract vulnerabilities and pre-deployment analysis methods. For transparency and reproducibility, we describe the scope and methodology for the reviewed literature in detail (Sec. \ref{Scope and Methodology}). Compared to existing surveys, a key feature of our presentation is the use of the notion of program properties in our classification (Sec. \ref{relatedWork}). Intuitively, a program property is any statement about the syntax or behavior of a given program. More formally, a property is a set of syntactic programs selected either on the basis of their syntax or their semantics. Examples of properties include depositing only positive values or always selecting the highest bid among a set of offers. Properties can be used to fully or partially  specify the correctness of a contract. It should also be noted that pursuing this approach is not without its own challenges. For example, a practical issue concerning properties is that different papers treat properties with varying degrees of detail and formality. For instance, some introduce a property only by a high-level description or reference, whereas others might introduce it using formal logic or natural language. 
\input{T1_PropertiesSpecifications.tex}
Table~\ref{tab:Specifications} illustrates this by comparing the informal specifications of one specific property (Sec. \ref{P_statepreservation}) in the studies \cite{tolmach2020survey, schneidewind2020good}. In developing this survey, we worked to strike a balance in dealing with such variations. For example, the first two rows of this table show how a property is described in other studies, and the last row shows how we leverage the fact that single-entrancy is a special case of preservation of the state.\\
To help ensure the accessibility of the classification, we provide a brief review of the basic concepts of smart contracts (Sec. \ref{background}). We then move to identify and classify relevant program properties (Sec. \ref{properties}), which allows us to factor out the relationship between vulnerabilities and analysis methods in a manner that brings out interesting patterns in the distinguishing features and relative strengths of analysis methods (Sec. \ref{methods}). We conclude by summarizing key trends that this work helped surface (Sec. \ref{discussion}).

\section{Scope and Methodology}
\label{Scope and Methodology}
For transparency and reproducibility, we detail here the scope and the methodology used.

\textbf{{Scope. }} This survey covers the \textit{pre-deployment analysis} of smart contracts. Concretely, we consider static program analysis, formal verification (excluding runtime verification applied to the deployed contracts), dynamic analysis (pre-deployment), and machine learning approaches that statically analyze smart contracts. Methods that analyze and verify the source code or bytecode of a contract without running it are called static verification methods. We include static analysis and formal verification methods in this terminology but, again, exclude runtime verification methods. A \textit{static analysis} is an always-terminating algorithm that examines a program to determine whether it satisfies a particular property \cite{cousot2010gentle}. A \textit{formal verification} (possibly a non-terminating \cite{cousot2010gentle}) process takes a program and a property and attempts to mechanically prove that the program satisfies that property \cite{praitheeshan2019security}. It is useful to note that pre-deployment methods are not limited to static analysis and formal verification and can include methods traditionally considered “dynamic,” such as testing, simulation, and dynamic analysis. Domain Specific Languages (DSLs) are also being actively investigated for writing secure contracts and machine learning algorithms that statically analyze contracts. We focus on investigating program properties addressed by such methods and the vulnerabilities these properties could capture. In particular, the objective is {\em not} to assess the approaches or make suggestions about tools but rather build a framework to facilitate identifying related efforts for future individual studies on specific methods. Finally, this work surveys only smart contracts that execute as on-chain code and not off-chain services.

\textbf{{Methodology. }} The first step in defining the literature we review in this survey is to search for and aggregate citations to publications recorded in \romannumeral 1) ACM Digital Library, \romannumeral 2) Dimensions, \romannumeral 3) Elsevier ScienceDirect, Elsevier Engineering Village, \romannumeral 4) Google Scholar, \romannumeral 5) IEEE Xplore, \romannumeral 6) Microsoft Academic, \romannumeral 7) Semantic Scholar, and \romannumeral 8) Springer. We select the publications most relevant to the survey's goal in the following five phases.
\begin{itemize}
\item Phase1: The synonyms of "smart contracts" we observed in the literature are “digital contracts” and “crypto contracts.” Also, we use pre-deployment analysis as an umbrella term that covers all approaches that can analyze, verify and test smart contracts before they are deployed on the blockchains. So, the keywords commonly used in publications related to such approaches are “program analysis”, “static analysis”, “dynamic analysis,” “formal methods,” “verification,” and “machine learning.” Using all these keywords, we build the following broad query:
    \begin{description}
    \item (Smart contracts OR digital contracts OR crypto contracts) 
    \item AND 
    \item (program analysis OR static analysis OR dynamic analysis OR formal methods OR verification OR machine learning)
    \end{description}
We searched the databases listed above using this query~\footnote{We have also evaluated relaxations of this query such as replacing the clause “(Smart contracts OR digital contracts OR crypto contracts)” with the strictly more inclusive clause “((Smart contracts OR digital contracts OR crypto contracts) OR (“contracts” AND solidity)”. Such relaxations do return additional papers, but none that pass the additional steps of the methodology.} and the timeline from January 2016 to December 2022. The initial search resulted in thousands of citations, not all closely related to the goal of the survey.
    \item {Phase 2:} The results of the Phase 1 search also include papers that are not related to the survey's goals, for example, secondary studies (such as surveys) and studies on post-deployment analysis. Thus, we used a second search string by adding NOT with keywords in the titles to exclude unrelated citations. The second search query, "Phase-1-results AND NOT [inTitle: (survey OR literature review OR empirical study OR analysis tool evaluation OR post-deployment OR runtime verification OR code repair OR forensics analysis)]," results in 669 publications, including duplicates.
    \item {Phase 3:} The results of Phase 2 contain redundancies both between and within databases. In this step, we manually removed redundancies and arrived at 237 unique publications. 
    \item {Phase 4:} At this point, we switched to a manual selection step, inspecting each publication's title, abstract, and contributions. A publication is included for the survey if it analyzes contracts before deployment using \romannumeral 1) static analysis or verification, or \romannumeral 2) dynamic analysis, or \romannumeral 3) machine learning algorithms. This curation step resulted in 232 citations.
    \item {Phase 5:} Finally, we manually excluded publications that focus primarily on other topics.\footnote{References have been dropped from this section to keep the length of the paper under the recommended 35-page limit.}
    Examples of excluded topics include post-deployment and runtime verification techniques, code repairs, clone detection, code metrics, and other publications that do not meet any of the previously mentioned inclusion criteria. This curation step resulted in the 215 papers that we have studied in detail to build this survey.
\end{itemize}
%
\section{Related Work}
\label{relatedWork}
Related work consists of secondary studies concerned with the pre-deployment analysis of smart contracts. Existing surveys primarily focus on vulnerabilities, as shown in column three of Table~\ref{tab:surveys}. In contrast, we propose that {\em program properties} are a useful tool to classify and connect vulnerabilities and methods --- they can be viewed as a declarative specification of the problem and can, in general, give an indication of the computability and/or computational complexity of the problem of checking a program for this property. This perspective was only taken by one other survey, but it had a slightly different focus, namely, the formal specification and verification of smart contracts \cite{tolmach2020survey}.
\input{T2_RelatedSurveys.tex}
Table~\ref{tab:surveys} gives an overview of related surveys.
We group studies based on the extent to which they address the notions of program properties, vulnerabilities, and analysis methods and whether they focus on the peer-reviewed research literature and consider multiple platforms. The last point is included because many studies focus exclusively on the Ethereum platform. The cells of the table are populated with one of three possible symbols, \romannumeral 1) a filled circle to indicate that the study treats this aspect, \romannumeral 2) a half-filled circle to indicate that the study either provides a brief treatment of this aspect (for example, without any classification) or addresses essentially different areas, for example, formal specifications, or formal modeling, and \romannumeral 3) an empty circle to indicate that the study does not address this aspect.
These criteria allow us to identify different groups of studies (shown in the table) as follows:\\
\textbf{{Surveys that address smart contract properties, vulnerabilities, and methods. }} Several surveys adopt a classification similar to the one proposed in this paper, with the main difference in the scope of properties and methods covered. Tolmach et al. \cite{tolmach2020survey} analyze and classify existing approaches to formal modeling, specification, and verification of smart contracts, outline common properties from different application domains and correlate them with the capabilities of existing verification techniques. That work is similar in spirit to ours, with the main differences in scope. Specifically, they review formal verification techniques and exclude static type checking and machine learning-based approaches. They also consider properties of smart contracts primarily for formal control or program specification. In contrast, we focus on properties for the classification of vulnerabilities and pre-deployment analysis methods.
In contrast, Schneidewind et al. \cite{schneidewind2020good} do review existing approaches to automated, sound, static analysis of contracts, and Grishchenko et al. \cite{grishchenko2018foundations} review  tools for formal verification of contracts. However, both works focus on a single property (Sec. \ref{P_statepreservation}) of Ethereum smart contracts.\\
\textbf{{Surveys that address vulnerabilities and methods. }}
Some surveys connect vulnerabilities in smart contracts with the methods that can be used to detect them. Praitheeshan et al. \cite{praitheeshan2019security} review security vulnerabilities in Ethereum smart contracts and categorize analysis tools according to their applied detection methods (static analysis, dynamic analysis, and formal verification) and investigate their limitations. That work focuses only on the Ethereum platform, whereas the present survey does not limit its focus to any particular blockchain technology. 
Also, it does not provide any specification or classification of the properties of smart contracts.
Tang et al. \cite{tang2021vulnerabilities} review and classify security vulnerabilities into Solidity, EVM, and blockchain levels, and Lopez Vivar et al. \cite{lopez2020smart} overview key vulnerabilities with two relevant attacks. Both works do review analysis methods and vulnerability detection methods, including static analysis, dynamic analysis, and formal verification. However, they focus only on the Ethereum blockchain and do not consider the properties of smart contracts.\\
\textbf{{Surveys that address vulnerabilities.}}
Whereas our focus is on pre-deployment analysis, several surveys focus primarily on vulnerabilities. That said, it is still informative to consider the extent to which they use vulnerabilities for classification. H. Chen et al.
\cite{chen2020survey} take a different approach to classify vulnerabilities and possible detection methods for Ethereum smart contracts. Their classification is based on \romannumeral 1) root causes, such as programming error, compiler bugs, improper language design, or EVM design issues, \romannumeral 2) occurrence layers, for instance, at the Ethereum infrastructure level, and \romannumeral 3) status, such as whether they are already eliminated or not. In that respect, their focus is on vulnerabilities originating from infrastructure rather than from the models themselves. Other surveys examine vulnerabilities and their causes in terms of actual attacks \cite{atzei2017survey, li2020research, staderini2020classification, zhu2018research}, their prevention, detection, and defense approaches \cite{demir2019security, He9143290, chen2020Defects}, and evaluate the performance of analysis tools \cite{dika2018security, menseVulnerabilities, gupta2020insecurity, khan2020survey, yamashitaPotentialRisks, rahimian2021tokenhook, ji2021evaluating, krupa2021security, groce2020actual}. All of these surveys focus on Ethereum smart contracts written in Solidity, except the study of Yamashita, Kazuhiro et al. \cite{yamashitaPotentialRisks}, which investigates the potential risks of Hyperledger Fabric blockchain.\\
\textbf{{Surveys that address analysis methods. }} Surveys exist that focus on pre-deployment analysis methods. Wang et al. \cite{wang2021security} review symbolic execution, abstract interpretation, fuzz testing, formal verification, deep learning, and privacy enhancement on smart contracts and compare various tools and methods that address security issues. That work does not survey properties verified using these approaches.
Other surveys differ from the present work in terms of the types of methods covered. For example, they classify formal verification methods \cite{liu2019survey, almakhour2020verification}, static and dynamic analysis methods \cite{kim9230065} or review existing approaches of formal verification \cite{garfatta2021survey, singh2020blockchain}.\\ 
\textbf{{Surveys that have limited coverage of properties, vulnerabilities, and methods. }} Many prior surveys address analysis aspects investigated in the present work with a different perspective. For instance, there are surveys on \romannumeral 1) vulnerable contracts to evaluate the number of already exploited contracts \cite{perez2021smart}, \romannumeral 2) attacks and vulnerability detection tools \cite{sayeed2020Attacks, saad2019exploring, dai2022superdetector, 9667515, rameder2022review, piantadosi2022detecting}, \romannumeral 3) security issues detected through analysis tools \cite{praitheeshan2020security, tantikul2020exploring, Onur2022},
and \romannumeral 4) classification of research publications into various topics, such as testing, security analysis, and smart contracts issues \cite{vacca2020systematic, alharby2018blockchain}.  Thus, numerous surveys may fit into this category but cannot be cited due to space limitations (in terms of page count).\\
Finally, some surveys address aspects of smart contract analysis that are complementary to the present survey and are therefore not addressed here. Examples include languages for smart contract developments, design patterns for smart contracts, problematic code patterns, and other more specialized aspects, such as privacy, concurrency improvements, potential challenges, and applications of smart contracts.
%
\section{Background}
\label{background}
This section reviews basic concepts relating to blockchains and smart contracts.\\
A blockchain is a distributed ledger replicated and shared between a network of peer-to-peer nodes. The popularity of blockchains began with the Bitcoin cryptocurrency in 2008 \cite{nakamoto2008bitcoin}. Later, Ethereum \cite{wood2014ethereum} expanded blockchain applications by supporting smart contracts \cite{tolmach2020survey}. Smart contracts are programs stored along with their local state on a blockchain and executed in response to events, such as when a transaction is submitted to execute a function defined on a smart contract \cite{introSC}. Nodes in a blockchain network, sometimes called miners, package such transactions in batches called blocks. Each block also contains a link to the previous block, thus creating a chain of blocks; hence the term “blockchain” \cite{blockchain_Smartcontracts}.
Each deployed contract on a blockchain is assigned a unique address used to perform various transactions, such as sending funds (for example, ether or tokens) or calling other on/off-chain contracts to complete a transaction. 
Smart contracts can hold and transfer cryptocurrencies and tokens and typically serve as open-source libraries and open API services. Smart contracts also provide the backend logic of decentralized applications (dApps) supported by blockchains. Once deployed, smart contracts cannot be changed due to the immutable nature of the blockchain \cite{andesta2020testing}. Smart contracts are also transparent; all nodes share the same copy of the transactions and are irreversible; all the interactions with smart contracts are time-stamped and permanently recorded on the ledger.
%
\input{T_Blockchains.tex}
\label{blockchainsTokens}
Table~\ref{tab:Blockchains} gives some examples of blockchain platforms, showing for each platform the languages for writing smart contracts, its native cryptocurrency, whether it supports non-fungible tokens (filled circle means yes and empty circle means no), and its implementation standards. Each node on these platforms runs a virtual machine, for example, Ethereum Virtual Machine (EVM), which runs smart contracts that have been compiled into low-level machine instructions \cite{blockchain_Smartcontracts}. Cryptocurrencies, such as bitcoin (BTC) or ether (ETH), are payment-related digital assets native to the blockchains on which they operate. Non-fungible tokens are implemented using technical standards \cite{BlockchainStandards} to support multiple applications such as participating in decentralized finance (DeFi) mechanisms, accessing platform-specific services, and even playing games \cite{cryptopedia}.\\
Some programming languages have certain properties that give them advantages over other languages supported by the same platform. For example, Ethereum supports two languages, Solidity and Vyper. Vyper performs arithmetic checks that protect against integer over/underflows (Sec. \ref{P_integeroverunderflow}) which are only specific to Solidity (lower than 0.8.0). State preservation (Sec. \ref{P_statepreservation}) is also more pertinent to Solidity due to its support for inheritance, which is not present in Vyper. As a result, Vyper is considered more secure than Solidity because it restricts the use of certain features that could potentially introduce security vulnerabilities, including reentrancy.
Hyperledger Fabric supports two languages, Go, a statically typed language, and JavaScript, a dynamically typed language. The strict typing and error-handling features of Go make it less susceptible to coding mistakes that could potentially result in security vulnerabilities.
Cardano blockchain supports multiple programming languages, including Marlowe, Plutus, and Haskell. Marlow is specifically designed to write financial contracts and offers unique properties like Transfer amount boundedness (Sec. \ref{P_amountBoundedness}) that are not found in other Cardano-supported languages. These properties make Marlowe contracts more robust against security vulnerabilities that may impact contracts written in other languages.
Tezos blockchain also supports multiple languages, including Michelson, SmartPy, Ligo, and Fi. Michelson is specifically designed for Tezos and features an integrated formal verification process that guarantees the correctness and security of smart contracts before deployment. Michelson is a strongly typed language that statically analyzes smart contract code, estimating the maximum amount of gas needed for execution to prevent potential out-of-gas errors.\\
Blockchains differ greatly in design and impose different programming paradigms for their smart contracts with different kinds of program errors \cite{chen2020survey}. For example, Ethereum and Bitcoin adopt different transaction models for transferring cryptocurrencies. In Bitcoin's transaction model, each transaction generates (multiple) outputs corresponding to the transferred (spent) and remaining (unspent) cryptocurrencies. This transaction model is called the unspent transaction output (UTxO) model, in which both spent and unspent transaction outputs represent the global state. In contrast, Ethereum adopts an account-based transactions model in which assets are represented as balances within accounts. In the account model, only the current accounts and their balances represent the global state. The UTxO model lacks expressiveness, and Ethereum's account-based ledger and the associated notion of contract accounts have been motivated by the desire to overcome those limitations \cite{chakravarty2020extended}.\\
Blockchain platforms charge a fee to execute transactions in smart contracts. Transactions fees are calculated using different metrics for each blockchain; for example, Ethereum charges transaction fees in terms of gas units and price per unit, and their product is the total cost that the user will pay \cite{atzei2017survey}, and Tezos charges two costs for a transaction, one for the amount of gas consumed and the second for the amount of data permanently stored on the blockchain \cite{reis2020tezla}. The gas consumption mechanism also ensures that the contracts will eventually terminate because gas is a limited resource. 
Smart contract users specify transaction fees or prices per gas unit they are willing to pay for their transactions. The amount of gas a transaction utilizes and the user-specified price per gas unit both play an important role in a transaction’s execution order and failures. For example, expensive transactions involve complex or often failing operations and may require more gas than the user offered to pay; consequently, these are more likely to fail to execute. Also, when transactions are ordered to execute, the transactions with higher user-specified gas prices get executed first. On the other hand, each block in the Ethereum blockchain can only utilize a limited amount of gas to execute transactions in smart contracts. The block gas limit bounds the gas requirement of the contract; that is, a contract cannot use more gas than allowed by the block gas limit.
%
\section{Properties and Vulnerabilities}
\label{properties}
This section relates vulnerabilities and properties, covering 52 vulnerabilities and 35 properties that studies have shown can be mitigated or established, respectively, using program analysis.
\subsection{Properties}
Properties are often classified as either safety or liveness \cite{Lamport77, lamport1983specifying, lamport1983good, alpern1987recognizing}. This classification is a useful indication of the type of analysis (or, more generally, proof technique) needed to establish that a given program has the property. Not all properties have to be safety or liveness properties, but in certain logics, such as linear temporal logic (LTL), one can factor any property into safety, and liveness components \cite{alpern1987recognizing}. \textit{Safety properties} assert that an event does not occur during execution. An example is that the balance of a contract does not go below zero \cite{sergey2018SCProperties}. \textit{Liveness properties} assert that progress is always made -- according to some notion of progress. Examples would be that a contract will eventually perform a transaction or that a contract will eventually terminate. Such properties can help establish the correctness of smart contracts, and failing to have them can indicate vulnerability. For an accessible discussion of liveness vs. safety property, we refer the reader to the Liveness Manifesto \cite{TheLivenessManifestor}. 
\subsection{From Vulnerabilities to Properties}
%
\input{T4_PropertiesDetails.tex}
In this section, we introduce specific smart contract properties and use them to group vulnerabilities. 
We also point out some cases where one vulnerability can be viewed in more than one way. Table~\ref{tab:propDetails} lists properties that have been used in studies that address specific vulnerabilities. We use the "$^\star$" symbol to highlight the properties that have been formalized by the studies listed in column 5 of Table~\ref{tab:propDetails}. We address formalization and rigorous treatment of properties in detail in Sec. \ref{discussion}.  
 The table uses "-" for properties generally applicable to any language the underlying platform supports. Language versions such as "< 0.8.0" show that the corresponding property is concerned with a language version lower than 0.8.0. Omitted language version means the property generally applies to any language version at the time of writing this paper.
\subsection{Liveness Properties}
First, we review four liveness properties introduced in the literature.
\subsubsection{Deposit Acceptance}
\label{P_deposit}
If a pre-compiled contract takes only a certain amount of hard-coded gas units as input, it should not take more or less than those units, even if the user provides more gas with the transaction. Thusly hard-coded gas unit amounts may be insufficient when a future hard fork increases the gas fee schedule \cite{vyperIssue}. The pre-compiled contract, in this case, will not be able to accept a new deposit, even if it is valid, and will always fail in such a future hard fork due to the out-of-gas exception \cite{park2020end}. \textit{Deposit acceptance} is established to recognize this issue and asserts that a contract is always eventually able to accept a new valid deposit as long as a sufficient amount of gas is provided. 
This property is specific to the Ethereum 2.0 deposit smart contract \cite{EthereumDepositContract}, 
which accepts deposits and includes a one-way function to move funds from Ethereum 1.0 to Ethereum 2.0.
\subsubsection{Eventual Contract Removal}
\label{P_wallet}
Three events are involved in a transaction of the \texttt{WalletLibrary} contract, 'begin transaction,' 'finish transaction,' and 'remove killed contracts.' The 'remove killed contracts' transition removes any contract that was self-destructed during the transaction. This event is performed through the \texttt{destroy} instruction in the \texttt{Wallet} library. Not invoking this instruction may result in a vulnerability involving keeping references to the destroyed contracts.
\textit{Eventual contract removal} asserts that the \texttt{WalletLibrary.destroy} will eventually be invoked after 'begin transaction' \cite{nelaturu2020verified}.
\subsubsection{Eventual Contract Termination}
\label{P_eventualTermination}
A non-terminating contract is a DoS risk. \textit{Eventual contract termination} asserts that the contract will eventually terminate its execution \cite{genet2020termination, seijas2020efficient, Seijas2020Marlowe}. 
\subsubsection{Withdrawal Acceptance}
\label{P_withdrawal}
A \textit{locking, locked, or freezing Ether} vulnerability is present when a smart contract can receive but not send ethers \cite{lu2021neucheck, nelaturu2020verified, ashouri2020etherolic, xing2020new, wei2020smart, Li2021Clue, stephens2021smartpulse, tikhomirov2018smartcheck,jiang2018contractfuzzer, DefectChecker2022}. This can happen because of non-existent or unreachable withdrawal functions or if the library containing the withdrawal function is destroyed. \textit{Withdrawal acceptance} asserts that the function(s) that perform ether withdrawal will eventually execute.
\subsection{Safety Properties}
Next, we review 27 safety properties introduced in the literature. 
\subsubsection{Argument Precondition Validation} 
\label{P_argumentsValidation}
Arguments of a method could come from untrusted, user-provided input and should be validated before any use, including storing in memory. \textit{Argument precondition validation} asserts that the contract checks whether all arguments of a transaction meet their desired preconditions \cite{tsankov2018securify}.
\subsubsection{Balance Comparison}
\label{P_balance}
Checking that a contract has enough balance with strict equality operator ‘==’ affirms that the contract will have exactly that balance to proceed with a transaction \cite{lu2021neucheck, arganaraz2020detection,tikhomirov2018smartcheck, DefectChecker2022}.
However, if the contract has more balance than the checking amount, the transaction will be halted due to a failed check. If an adversary observes a check for balance with strict equality, he could forcibly send ethers beforehand, using \texttt{selfdestruct} or through minery, then the balance will exceed the checking amount in the following transaction. In this case, the failed assertion will halt all future transactions, causing the victim contract to be in a deadlock state, as happened in an incident with Gridlock \cite{gridLock}. \textit{Balance comparison} is established to recognize this issue and asserts that balance is checked with a greater than or equal operator ‘>=’.
\subsubsection{Conditional Independence}
\label{P_conditionalInd}
If a conditional statement (if, for, while) calls an external contract and depends on the returned value to proceed with the execution, it might never proceed because the invoked contract might fail (throw or revert) permanently \cite{lu2021neucheck, arganaraz2020detection}. The resulting vulnerability is called \textit{DoS due to external contracts}. This vulnerability can be addressed by establishing \textit{conditional independence}, which asserts that conditional statements do not contain calls to external contracts; if values from external contracts are required, those should be obtained before executing the conditional statements. 
\subsubsection{Contract Addresses-related Properties}
\label{P_contractaddress}
As mentioned in Section~\ref{background}, each deployed contract on a blockchain has a unique address used to invoke its functions. A contract addresses-related property ensures proper uses of its address. Each of the following properties addresses specific vulnerabilities.
\begin{enumerate}[label=\Alph*]
 \item \textit{Address Size Soundness:} In Solidity, the function transferring ethers takes two arguments: the address of the receiving contract and the amount to be transferred. When a contract is compiled into the bytecode, the address is represented by a fixed 20-byte size followed by the amount of a fixed 32-byte size. EVM pads these arguments with zeros to fill the size; for instance, the amount has many leading zeros because it is a 256 bit integer (32 bytes). If the size of the receiver’s address (provided by the user) is less than 20 bytes, leading zeros are taken away from the amount and given to the shortened address. EVM will then automatically fill the missing bytes with zeros making the amount much larger. The resulting vulnerability is called \textit{address shortening} vulnerability \cite{nelaturu2020verified, ashouri2020etherolic, xing2020new}. \textit{Address size soundness} asserts that a contract that transfers funds accepts precisely twenty bytes for the address of the contract that receives funds.
 \item \textit{Address Integrity:} There is no readily available check to verify if external contract addresses refer to the intended contract in Solidity. Vulnerable \textit{external contract referencing} happens when a contract allows an arbitrary address to be used to refer to a contract irrespective of whether the address represents the intended contract \cite{nelaturu2020verified}. Moreover, in EOSIO smart contracts, not checking the transaction notification to ensure that the transaction's recipient is a correct contract address can lead to a \textit{forged transfer notification or fake receipt} \cite{huang2020eosfuzzer, he2021eosafe, li2022eosioanalyzer, Wasai2022, Li2022}. To detect these issues, \textit{address integrity} asserts that \romannumeral 1) hard-coded addresses are used if known before deployment to call external contracts, \romannumeral 2) the 'new' keyword is used for newly created contract instances to explicitly declare that these contracts are not yet live on the chain, and \romannumeral 3) the address of the transaction recipient is always checked when a transfer notification is received.
 \item \textit{Address Parameter Avoidance:} Solidity allows programmers to use any address to refer to a specific contract, which enables attacker contracts to hide malicious code by passing an incorrect address to a smart contract constructor as a parameter. This leads to \textit{misuse of address parameters in the constructor} \cite{praitheeshan2021solguard}. \textit{Address parameter avoidance} asserts that addresses are not taken as parameters in the constructor function. 
 \item \textit{Validity of Transfer Address:} In a call chain, contracts call other contracts, which further call more contracts. The environment variable \texttt{tx.origin} contains the address of the call chain originator and the variable \texttt{msg.sender} contains the address of the immediate invoker for any given contract in the call chain. Both values contain valid addresses; however, while authorizing a transaction, \textit{using \texttt{tx.origin} instead of \texttt{msg.sender}} is a vulnerability because the contract corresponding to \texttt{msg.sender} invoked the contract under consideration and requested this particular transfer and thus is eligible to receive the transfer \cite{lu2021neucheck, arganaraz2020detection, kalra2018zeus, tikhomirov2018smartcheck, brent2018vandal, EtherGIS2020, DefectChecker2022, Yao2022,ATL2022}. However, using \texttt{msg.sender} might not be correct in all cases; for example, if a contract invokes another to refund a transaction to a special investor account specified in parameters \cite{zhdarkin2021development}. In this case, using \texttt{msg.sender} entails that the contract that called the function could receive its deposit, whereas the right recipient is specified in the parameters. Furthermore, a recipient specified in the parameters can also be inaccurate; specifically, if an arbitrary contract can call a transaction, it may provide an address to a contract that never submitted any funds beforehand and is therefore not eligible to receive refunds. The contract, in this case, is called a “prodigal contract” that sends refunds to arbitrary contracts \cite{praitheeshan2019security, tang2021vulnerabilities}. These issues can be addressed by establishing \textit{validity of transfer address}, which asserts that while authenticating a transaction, \texttt{tx.origin} is not used, and if recipient addresses are provided in parameters, they are checked against (a list of) eligible contracts; otherwise, \texttt{msg.sender} is used.
\end{enumerate}
\subsubsection{Division-by-zero Guarding} 
\label{P_dividebyzero}
Values provided by contract users are often processed with arithmetic operations, making it possible for a denominator value to be zero. An \textit{unchecked division by zero} leads to errors \cite{so2020verismart, Khor2020improved, Osiris2018}. \textit{Division-by-zero guarding} asserts that the contract checks denominators before each integer division to avoid any possibility of division by zero. 
\subsubsection{Effective Callback Freeness}
\label{P_effectivecallback}
A contract can call external contracts, which can call it back. A contract may encounter a \textit{reentrancy} vulnerability if such a callback can result in an invalid state. Proving that every execution with a callback can also be simulated without callbacks can ensure that a contract is effectively callback free. \textit{Effective callback freeness} asserts that \textit{'for every execution trace with a callback, there exists an equivalent callback free execution trace yielding the same effect on the contract state'} \cite{albert2020taming, grossman2017online}.
\subsubsection{Exception and Return Value Handling}
\label{P_exceptions}
Some low-level operations in Solidity, such as the \texttt{send()}, and \texttt{call()} methods, do not throw an exception upon failure and instead return a boolean value. Because exceptions thrown by the callee contract cannot be propagated to the caller, missing checks for return values from such calls can cause several kinds of vulnerabilities relating to \textit{unchecked return value bug, non-verified external call, mishandled/unhandled exceptions or exceptions disorders} \cite{zhang2020smartshield, nelaturu2020verified, ashouri2020etherolic, arganaraz2020detection, semantics2020Jiao, wei2020smart, Oyente2016, kalra2018zeus, tikhomirov2018smartcheck, grishchenko2018semantic, Zhou2022, MANDO2022, DefectChecker2022, Yang2022, Yao2022}. More vulnerabilities also arise with unhandled exceptions and return values, including:
\begin{enumerate}[label=\Alph*]
 \item \textit{Call stack related issues:} the depth of the call stack (EVM stack) may go beyond the threshold (e.g., 1024) when a contract invokes other contracts via instructions such as \texttt{send()}, \texttt{call()}, and \texttt{delegatecall()}. Since these instructions do not throw an exception, unchecked return values can result in \textit{call stack depth attack vulnerability} \cite{wang2020contractward}.
    \item \textit{Out-of-gas exception-related issues:} When funds are transferred to a contract using \texttt{send()} or \texttt{transfer()} without specifying the target function, or when no function matches the given function signature in the target contract, the default fallback function -- of which only one is allowed per Solidity contract -- is automatically executed. The fallback function has to be marked payable to receive funds and can use at most 2300 units of gas only. If the gas allocated by the sender contract is insufficient to execute a costly fallback function, it sends an out-of-gas exception to the sender contract. If the sender does not check the exception properly, it will not realize the unsuccessful transfer, resulting in a \textit{gasless send} \cite{nguyen2020sfuzz, teng2021SmartGift, jiang2018contractfuzzer}. 
\end{enumerate}
These vulnerabilities can be detected by establishing \textit{exception and return value handling}. This property asserts that a caller contract always checks exceptions and returned values from low- and high-level calls to handle transaction failures and out-of-gas exceptions thrown by the callee contract.
\subsubsection{Explicit Access Control Modifiers Usage}
\label{P_explicitaccescontrol}
A contract is susceptible to an \textit{unspecified visibility level} vulnerability when functions or variables are not declared using proper access modifiers, such as public, private, and other keywords \cite{lu2021neucheck, tikhomirov2018smartcheck}. Also, \textit{access control-related} vulnerabilities arise when the access control modifier of the constructor is not defined, making it public and accessible to arbitrary users \cite{lu2021neucheck}. These vulnerabilities can be detected through \textit{explicit access modifier usage}, which asserts that a proper access modifier is explicitly specified while defining the constructor, functions, and variables of a contract.
\subsubsection{Gas-related Properties}
\label{P_gasrelated}
Gas-related properties address its consumption \cite{Correas2021static, grech2020madmax, arganaraz2020detection, reis2020tezla, nguyen2020sfuzz, albert2021don}. The following instances are addressed in the literature:
\begin{enumerate}[label=\Alph*]
\item \textit{Boundedness of Loops:}\label{P_boundednessLoops} In a loop's termination condition, an unchecked input from a user or a comparison of integers of different sizes can cause the loop to exceed the block gas limit or be too expensive to complete and result in transaction failures with \textit{non-terminating loops or unbounded mass operations} errors \cite{grech2020madmax, lu2021neucheck, nassirzadeh2021gas, Correas2021static}. Because Solidity admits type inference, not explicitly declaring variable types can lead to a failed comparison between integers of different sizes in a loop's termination condition. Such comparisons lead to \textit{unsafe type declaration or insecure inference} \cite{lu2021neucheck, arganaraz2020detection, tikhomirov2018smartcheck}. To ensure that loops will terminate, \textit{boundedness of loops} asserts that bounded values and variables with explicit types determine the number of iterations in the loops.
    \item \textit{Gas Transfer Viability:}\label{P_gasTransferViability}
    Solidity provides three built-in functions to transfer ethers, \texttt{transfer(x)}, \texttt{send(x)}, and \texttt{call(x)}. Along with \texttt{x} ethers, \texttt{transfer(x)} and \texttt{send(x)} transfer (the Ethereum-specified minimum required) 2300 units of gas for the receiver contract to process the transaction. The \texttt{call()} function is the one recommended by Solidity but not providing parameters with the \texttt{addr.call.value(x)()} redirects all the gas that was provided with the transaction to the recipient, resulting in \textit{transfer of all the gas} vulnerability \cite{arganaraz2020detection, tikhomirov2018smartcheck, aidee2021vulnerability}. \textit{Gas transfer viability} asserts that if a specific gas amount is required to be transferred, then \texttt{addr.call.value(x)()} is substituted with that amount, otherwise \texttt{transfer(x)} or preferably \texttt{addr.call.value(x)} is used.
\end{enumerate}
\subsubsection{Information Flow-related Properties}
\label{P_informationflow}
An information flow-related property asserts that information such as user-provided input and values in state variables flows between functions as intended and is not observable by arbitrary users \cite{brent2020ethainter, zhang2020ethploit}. The following instances arise in the literature:
\begin{enumerate}[label=\Alph*]
\item \textit{Guarding of \texttt{delegatecall}:} The \texttt{delegatecall} in Solidity calls external contracts and allows callee contracts to make state changes to the caller's contract, including sending funds to any address and destroying the contract. If a user-provided (nontrusted) value flows into the contract address invoked through a \texttt{delegatecall}, a non-trusted contract could be invoked, resulting in a high-risk security vulnerability called \textit{tainted or dangerous delegatecall} \cite{brent2020ethainter, wei2020smart, teng2021SmartGift, andesta2020testing, jiang2018contractfuzzer, EtherGIS2020, zhang2022smart,semantics2020Jiao}. This can be addressed by establishing \textit{guarding of \texttt{delegatecall}}, which asserts that functions with \texttt{delegatecall} are either private or accessible from public functions through specific checks only and if the target address in a \texttt{delegatecall} is derived from user-provided input, it is checked against trusted contract addresses  \cite{swcRegistry}, and it explicitly flows into arguments of a \texttt{delegatecall} instruction \cite{ILF2019}.
    \item \textit{Information Secrecy:} \label{P_informationSecrecy} A secret value is stored in a state variable and is used to gain rewards in gambling, quizzes, and gifts-based smart contracts. Although the secret setter function stores a hash value of such secret values, the plain text of the secret response is used in the transaction argument, which is recorded publicly in the blockchain. Since blockchain is an open platform, attackers can inspect such secret values by tracking previous transactions of a smart contract, resulting in an \textit{exposed secret} vulnerability \cite{zhang2020ethploit, lu2021neucheck}. \textit{Information secrecy} asserts that the secret values are transmitted in encrypted form so they do not appear in the contract execution path as plain text.
    \item \textit{Preservation of Owner Information:} \label{P_ownerInfo}
    A memory reference that holds the credentials of a contract owner is called an owner variable. When a contract terminates, the owner receives the remaining tokens or balance managed by the contract.
    If an owner variable is not initialized properly, for example, with the right credentials of the owner, money or tokens will be locked in the contract \cite{seijas2020efficient, Seijas2020Marlowe}. Similarly, if the owner variable becomes accessible to arbitrary users or values from arbitrary users can flow into the owner variable, it results in a \textit{tainted owner variable} vulnerability, which allows the arbitrary users to gain control of the contract along with the balance stored in the contract \cite{brent2020ethainter}.
    This vulnerability can arise by inadvertently exposing a constructor, for example, either by using public access control modifier (Sec. \ref{P_explicitaccescontrol}) or by misspelling its name \cite{swcRegistry} as happened in the notorious Parity Wallet hack \cite{ParityWallet, lu2021neucheck}. If an arbitrary user can become an owner of a contract, she gains access to sensitive operations, such as:
    \begin{enumerate}
        \item The \texttt{selfdestruct} instruction \label{P_selfdestruct}: In Solidity, this instruction deactivates a contract and transfers all available funds in the contract to the provided address. An \textit{accessible \texttt{selfdestruct}} means access control to a \texttt{selfdestruct} call can be bypassed due to the authentication mechanism being inadequate \cite{brent2020ethainter}. In this case, arbitrary access to \texttt{selfdestruct} results in vulnerabilities including \textit{unprotected \texttt{selfdestruct} or suicidal contract} \cite{aidee2021vulnerability, huang2021precise, EtherGIS2020}, \textit{vulnerable access control} \cite{zhang2020ethploit, kalra2018zeus}, and \textit{unprotected suicide and deadlock state} \cite{nelaturu2020verified}. A deadlock state arises when accessible \texttt{selfdestruct} allows the destruction of the wallet library, causing all contracts depending on it to be in a \textit{deadlock state}, meaning they can no longer perform funds transactions. These issues can be addressed by establishing \textit{guarding of \texttt{selfdestruct}}, which asserts that the \texttt{selfdestruct} call is accessible only to authorized users.
        \item The currency transfer instructions: Arbitrary access to currency transfer functions creates several vulnerabilities, including \textit{unfair payment} \cite{li2020protect}, \textit{leaking Ether} \cite{torres2021confuzzius, huang2021precise}, \textit{unprotected ether withdrawal and vulnerable access control} \cite{zhang2020ethploit, stephens2021smartpulse, tsankov2018securify, he2022tokencat, Cui2022}.        %
    \end{enumerate}
    These issues can be addressed by establishing the \textit{preservation of owner information} property. For Solidity contracts, it asserts that the owner variable is only initialized in the constructor, which itself is declared as private and has the same name as the contract. This way, user-provided (non-trusted) values cannot flow into the owner variable. For Marlowe contracts, it asserts that the owner variable is initialized with the credentials of the right owner. Together with the termination property (Sec. \ref{P_eventualTermination}), this property ensures that when a Marlowe contract terminates, the rightful owner will receive the tokens, and no money or tokens will be locked in the contract \cite{seijas2020efficient, Seijas2020Marlowe}.
    \item \textit{Preservation of Values in \texttt{staticcall}:} The EVM instruction \texttt{staticcall} calls a smart contract and uses the output of the called contract as its input while disallowing any modifications to the state of the caller contract during the call. 
    The output of a \texttt{staticcall} invocation must overwrite its input. In a problematic \texttt{staticcall}, the returned values do not overwrite the current memory buffer used for passing and returning data to CALL instructions, and the user-provided input is read as output (of the called contracts)  \cite{brent2020ethainter}. \textit{Preservation of values in \texttt{staticcall}} asserts that the information flows to/from \texttt{staticcall} operations as intended; for example, return values from such operations overwrite the current memory buffer.
    \item \textit{Preservation of \texttt{selfdestruct}:}  If input from an arbitrary user can flow into the target address used in the \texttt{selfdestruct} to transfer funds to, the resulting  vulnerability is a \textit{tainted \texttt{selfdestruct}} \cite{brent2020ethainter}. \textit{Preservation of \texttt{selfdestruct}} asserts that the receiver address used in the \texttt{selfdestruct} call is not updated with user-provided addresses.
\end{enumerate}
\subsubsection{Integer Division Guarding} 
\label{P_floatingpoint}
Solidity does not admit floating-point or decimal types, and integer division is always rounded towards zero \cite{arganaraz2020detection, tikhomirov2018smartcheck}. This choice of rounding method can produce results other than what some users might expect.
For example, -5/3 results in -1 instead of -2.
Guarding can be used to mitigate this problem when calculating the number of ethers or tokens.
\textit{Integer division guarding} asserts that the contract checks integer divisions whether they result in values according to the intent of the (developer of the) contract \cite{Cui2022}.
An example of such guarding is to check whether the numerator or denominator is negative in a division operation, and if such is the case, then to require explicit rounding.
\subsubsection{Integer over/underflow Guarding}
\label{P_integeroverunderflow}
An \textit{integer over/underflow} occurs when an arithmetic operation results in a value that exceeds the maximum or minimum range of integer representation \cite{ma2021pluto, ashouri2020etherolic, sun2020formal, so2020verismart, lai2020static, xing2020new, semantics2020Jiao, nguyen2020sfuzz, Khor2020improved, wang2020contractward, lu2021neucheck, yang2020hybrid, ding2021hfcontractfuzzer, andesta2020testing,stephens2021smartpulse, kalra2018zeus,mossberg2019manticore, Zhou2022, MANDO2022, wang2022gvd, DefectChecker2022, Cui2022, Son2022, Yao2022, Liu2022, liu2022learning, zhang2022smart}. \textit{Integer over/underflow guarding} asserts that valid defense mechanisms are placed with all occurrences of arithmetic operations to ensure that over/underflow does not occur.
\subsubsection{Isolation of External Calls} 
\label{P_isolationofcalls}
An external call can influence the caller: The call may include a complex or failing transaction that causes an out-of-gas exception. Such cases can lead to a \textit{wallet griefing} vulnerability that can result in an out-of-gas exception in the caller \cite{grech2020madmax}. A \textit{DoS due to unexpected revert or due to a failed call} occurs if a contract's execution is halted because the callee either deliberately or unexpectedly reverted due to a failure \cite{samreen2021smartscan, stephens2021smartpulse,tikhomirov2018smartcheck}. 
Issues relating to external calls can be minimized by allowing external contracts to initiate transactions instead of calling them. For example, it is better to let users withdraw funds rather than push funds to them automatically, thus reducing external calls. This way, external calls are isolated into separate transactions. Thus, \textit{isolation of external calls} asserts that the external calls are isolated in their own separate transactions so that reverted or failed transactions, out-of-gas exceptions, and possible DoS events in externally called contracts do not affect the execution or gas consumption of the caller.
\subsubsection{Map Disjointness}
\label{P_hashcollision}
Solidity's map data structure (similar to a hash table or a dictionary) uses a hash function to compute an index into an array of slots. If two maps are used in a contract, the sha3 function may produce the same hash value for two different memory locations, one in each map, creating a collision across maps. Lu et al. \cite{lu2021neucheck} refer to this issue as a \textit{hash collision}. In our survey, we propose a new terminology \textit{map overlap} for this vulnerability since hash collision \cite{maurer1975hash} is a standard term for a collision of hash values within one hash table, is part of the standard operation of a hash table and is, by itself, not a vulnerability. This vulnerability has occurred in a contract that used two maps, one for admins of the contract and the other for registered users, as arguments of a function that checked whether a given address belonged to admins or regularly registered users \cite{hashCollisionIncident}. An attacker, in this case, can manipulate the position of elements while registering as a regular user in such a way that the hash function returns the same hash results for certain addresses in the admins and regular users, thus bypassing authorization. This vulnerability can be addressed by establishing \textit{map disjointness}, which asserts that the contract does not use more than one map data structure.
\subsubsection{Memory References Distinction}
\label{memoryref}
Variables or memory references can be unintendedly overridden in inheriting contracts if they declare variables with the same names as their inherited contracts. The resulting vulnerability is called \textit{tainted memory (or shadow memory)} \cite{ashouri2020etherolic}. These vulnerabilities can be addressed by establishing  \textit{memory references distinction}, which asserts that the contract contains distinctive and non-overlapping variable names and memory references.
\subsubsection{Ownership of Assets}
\label{P_assetsownership}
A contract that handles assets such as cryptocurrencies can become susceptible to vulnerabilities caused by improper handling of assets. An example of improper handling is transferring assets without their owner's consent. \textit{Ownership of assets} asserts the ownership privileges of asset owners. Examples of ownership privileges include that \romannumeral 1) assets/resources have an owner, \romannumeral 2) they cannot be taken away from their current owner without their consent, and \romannumeral 3) an owner can only transfer assets that they own \cite{bram2021rich}.
\subsubsection{Preservation of Assets} 
\label{P_assetspreservation}
Losing references to objects that represent assets can mean the accidental loss of these assets. \textit{Preservation of assets} or asset retention asserts that assets (such as cryptocurrencies, tokens, and other resources) cannot be arbitrarily created, duplicated, or accidentally lost/destroyed \cite{coblenz2020obsidian, bram2021rich, blackshear2020resources}.
%
\subsubsection{Preservation of Assets in Transactions}
\label{P_assetsvalue}
Contracts can implement the exchange of tokens or digital assets, whereby certain tokens can be bought, sold, or traded. Many blockchains, like Bitcoin and Ethereum, enable the scarcity of their cryptocurrencies through the burn and mint mechanism (BME), whereby cryptocurrencies are burnt (destroyed) each time they are traded, and new units of cryptocurrencies are created (minted) to replace the burnt ones. On the other hand, Algorand utilizes smart contracts written in the Transaction Execution Approval Language (TEAL) to burn and mint its assets (such as Algos). Therefore, Algorand smart contracts must retain the constant amount of their digital assets until they are eventually burnt \cite{bartoletti2021formal}. Besides preserving the amount of assets, smart contracts must not allow unfair and duplicate trades of digital assets. A \textit{double-spending} vulnerability arises when the same transaction occurs more than once without reducing the balance \cite{bartoletti2021formal}. The vulnerability called \textit{unfair payment} arises due to a costless trade, which happens when a user can exchange the same token more than once, whereas once that token is spent, the user is no longer the owner of that token and thus not eligible to spend it again without getting it back first \cite{li2020protect}. \textit{Preservation of assets in transactions} ensures the preservation of assets and avoids duplicate trades by asserting that the amount of digital assets remains constant, and the same transaction or token cannot be issued/exchanged more than once.
\subsubsection{Preservation of State}
\label{P_statepreservation}
We recall from Property \ref{P_effectivecallback} that \textit{reentrancy} is a vulnerability that occurs when a contract calling another contract can be called back (reentered) before completing the original internal transaction; that is, before changing its state \cite{ma2021pluto, wustholz2020harvey, xue2020cross, chinen2020ra, schneidewind2020ethor, arganaraz2020detection, qian2020towards, nelaturu2020verified, ashouri2020etherolic, li2020protect, semantics2020Jiao, samreen2020reentrancy, ji2021security, lu2021neucheck, britten2021using, wang2020contractward, wei2020smart, teng2021SmartGift, andesta2020testing, stephens2021smartpulse, ribeiro2020formal, Oyente2016,tikhomirov2018smartcheck, mossberg2019manticore, grishchenko2018semantic, Zhou2022, MANDO2022, SAILFish2022, EtherGIS2020, DefectChecker2022, YanXingrun2022, Yang2022, Yao2022, zhang2022reentrancy, zhang2022smart, bang2020verification,ATL2022,grishchenko2018ethertrust}. An example of unsafe state change is if a state variable is changed after calling the external contract; in this case, the callee contract can call back, and the caller contract will be in its previous/old state. \textit{Preservation of state} asserts that state variables are never updated after calling an external contract, and they must be private; that is, they cannot be directly modified by functions in other smart contracts \cite{bram2021rich}.
\subsubsection{State Variable Declaration Ordering}
\label{P_statevariables}
When a contract uses \texttt{delegatecall} to invoke a function of another contract, all changes made by the callee contract to the state variables will affect the caller contract’s storage instead of the callee contract. If both contracts use state variables in different orders and unmatched types, attackers (acting as callee in this case) can compromise the caller contract by setting their address as the owner, resulting in gaining the contract in the attacker’s control. \textit{State variable declaration ordering} asserts that if \texttt{delegatecall} is used to invoke a contract’s function, the order and the variable type of all the state variables declared in the target contract should be identical to the caller contract \cite{praitheeshan2021solguard}.
\subsubsection{Transaction Cost Boundedness}
\label{P_transactionCost}
A \textit{DoS due to expensive transactions} is possible if a contract takes too long to execute and consequently requires more gas than the block gas limit \cite{chen2020gaschecker}. \textit{Transaction cost boundedness} asserts that the transactions of the contract under consideration do not involve expensive operations, such as repeated computations inside loops, unreachable code segments, and redundant operations \cite{li2022smartfast, SmartMixModel2022} and will complete using finite gas bounds (often determined by block gas limit). 
\subsubsection{Transaction Order Independence}
\label{P_transactionsordering}
In Ethereum, miners schedule a set of transactions for execution in each block. These sets can contain transactions in any order, mostly affected by user-specified prices per gas unit. If such transactions write to and read from the same storage variable, a malicious miner may schedule his transaction in such an order that produces the desired output. Varying the output of the contract by manipulating the transaction sequences results in \textit{transaction ordering dependency, read-after-write(RAW), or race conditions} \cite{torres2021confuzzius, ashouri2020etherolic, wang2020contractward, Oyente2016, kalra2018zeus,tsankov2018securify, grishchenko2018semantic, MANDO2022, SAILFish2022, EtherGIS2020, Yao2022}. These vulnerabilities can be detected by establishing \textit{transaction order independence}, which asserts that executing a set of transactions of the same contract in any order does not affect the final output.
%
%
\subsubsection{Transfer Amount Boundedness}
\label{P_amountBoundedness}
Marlowe contracts handle separate balances for contracts and accounts. \textit{Transfer amount boundedness} asserts that the contract will never spend more money or tokens than there are in an account, even if more is available in the contract \cite{seijas2020efficient, Seijas2020Marlowe}.
%
\subsubsection{Transfer Function Viability}
\label{P_transferfunctionviability}
 The choice of ether transfer function affects not only the transfer of gas (Sec. 
\ref{P_gasrelated} (\ref{P_gasTransferViability})) but also the execution of the caller (sender) contract because \texttt{call.value(x)} and \texttt{send(x)} do not return an exception in case of a failed transfer. In contrast, \texttt{transfer(x)} does throw an exception if the transaction is unsuccessful. 
This makes the \textit{use of \texttt{send} instead of \texttt{transfer}} \cite{arganaraz2020detection, tikhomirov2018smartcheck} a vulnerability. Even using \texttt{call.value(x)} can be problematic when the recipient function is not explicitly specified in the arguments of the \texttt{call}. By default, \texttt{call} invokes the default fallback function of the target contract so that the receiving contract can process the transferred balance. If a malicious contract deliberately does not implement a fallback function, any deposit made to it will not succeed, and the sender contract will not be able to function as intended due to failed transfers. This vulnerability is called \textit{missed to implement a fallback function}, which can lead to DoS issues in sender contract \cite{praitheeshan2021solguard}. These issues can be addressed by checking \textit{Transfer function viability}, which asserts that \texttt{transfer(x)} is used to send Ether, and \texttt{call} is used only when a specific gas amount is required to be transferred or when the name of the target function (in the receiving contract) can be provided along with transfer of ethers.
\subsubsection{Validity of Deposits}
\label{P_depositValidity}
Valid tokens are implemented using technical standards such as ERC-20 and eosio.token (Sec. \ref{blockchainsTokens}). While ERC-20 provides interfaces for implementing tokens, it does not specify the implementation details. For example, if a contract uses ERC-20 to implement a token, an exception should be thrown if there are insufficient tokens in the contract’s balance to spend. If a contract does not do that, it can deposit invalid tokens, resulting in a \textit{fake deposit} vulnerability \cite{ji2020deposafe}. During a transfer, if an EOSIO contract does not check that the EOS (or token) was generated using standard code (eosio.token), it results in the generation of \textit{fake EOS} \cite{huang2020eosfuzzer, he2021eosafe, li2022eosioanalyzer, Wasai2022, Li2022}. \textit{Validity of deposits} addresses these issues by asserting that a contract implements tokens using technical standards only and does not deposit invalid tokens into the exchange.
\subsubsection{Validity of Transfer Amount} 
\label{P_transfervalue}
Improper handling of transfer amounts and balances is a significant vulnerability because many contracts handle transactions of a large sum of cryptocurrencies and tokens. 
Adversaries can gain \textit{unlimited illicit gains}, for example, by manipulating the rewards calculations in gaming contracts \cite{zhang2020ethploit}.
If a contract does not contain checks for the amount to be transferred, or the checks are written poorly, it can have \textit{unchecked send bugs} \cite{yang2020hybrid, stephens2021smartpulse}. Accurate calculations of sums of balances are also important to ensure the integrity of cryptocurrencies and transactions \cite{elad2021summing, ahrendt2020functional}. \textit{Validity of transfer amount} asserts that any amount to be transferred between contracts or wallet accounts is checked so that the sum of balances is preserved during a transfer operation.
\subsubsection{Value Unpredictability}
\label{P_predictable}
In Solidity, block variables provide information about the current block. For example, \romannumeral 1) \texttt{block.timestamp} contains a Unix timestamp value (in UTC) of when the block was created/mined, \romannumeral 2) \texttt{block.number} contains the current block number or height, \romannumeral 3) \texttt{block.gaslimit} restricts maximum gas consumption for transactions within the block, and \romannumeral 4) \texttt{block.coinbase} represents the address of the miner who mined the current block. These values are often used in pseudo-random number generators or ether transfer functions. Unfortunately, since they are predictable, a miner holding a stake in a contract could gain an advantage by guessing a random number or by choosing a suitable timestamp for a block she is mining \cite{atzei2017survey, jiang2018contractfuzzer}. 
Similarly, in EOSIO smart contracts, block information can be obtained through \texttt{tapos\_block\_prefix()} or \texttt{tapos\_block\_num()} and used as the judgment condition of control flow \cite{li2022eosioanalyzer}. Using such variables can cause \textit{dependence on predictable variables, bad-randomness, block number, and timestamp dependency} \cite{lu2021neucheck, ma2021pluto, nguyen2020sfuzz, wang2020contractward, ashouri2020etherolic, wei2020smart, teng2021SmartGift, Oyente2016, kalra2018zeus, grishchenko2018semantic, li2022eosioanalyzer, Wasai2022, Zhou2022, MANDO2022, EtherGIS2020, DefectChecker2022, Yang2022, Li2022, zhang2022smart,grishchenko2018ethertrust}. \textit{Value unpredictability} asserts that predictable variables are not used in security-critical operations, such as ether transfer and random number generation.
\subsection{User-specified Properties}
\label{P_userspecified}
A user-specified property is one provided by contract designers or developers. This is usually done in the form of in-code annotations \cite{da2020whylson}. The following are examples of such a property,
\begin{enumerate}[label=\Alph*]
\item An \textit{Assertion} is a property that the author of the contract expects to be true. \textit{Assertion satisfaction} is the property that during the execution of a contract, it should never be possible to execute an instruction that is restricted by assertions \cite{schneidewind2020ethor, hajdu2019solc, hajdu2020formal, grieco2020echidna, torres2021confuzzius}. For example, an assertion in Solidity that restricts the balance of the contract to never be negative is written as \texttt{assert(balance >= 0)}.
\item \textit{A Contract-specific Invariant} is a property that holds under specific conditions \cite{zhong2020move, nelaturu2020verified, grieco2020echidna, hajdu2019solc, ahrendt2020functional}, and confirms the intent of the code \cite{marescotti2020accurate, SOLAR2018LiAo}. For example, an invariant in ERC20 \cite{ERC20Token} states that the sum of the account balances of all users is always equal to the total supply \cite{liu2022invcon, wang2019vultron}. Some examples of contract-specific invariants are as follows:
\begin{enumerate} \item \textit{Transaction invariants} are distinctive properties of smart contracts that hold under arbitrary interleaving transactions; for example, computing the total balance must not cause integer overflow at any state \cite{so2020verismart}. 
        \item \textit{Strong data integrity} is an invariant that forms a relation between internal data fields and the history of payments, which should not change for any interaction of the contract. For example, in an \texttt{Auction} contract, a data field \texttt{balances} is meant to store the accumulated funds sent by each bidder, minus anything that has been sent back \cite{ahrendt2020functional}. 
        \item \textit{Conservation invariant} is a safety invariant ensuring the integrity of cryptocurrencies. It asserts that \textit{the sum of the 'value' fields of all the 'Coin' objects in the system must be equal to the \texttt{'total\_value'} field of the Info object stored at the \texttt{ADMIN} address} \cite{patrignani2021robust}.
    \end{enumerate}
\item \textit{Frame Annotations} assert \textit{‘what a smart contract function cannot [or] will not do’} \cite{beckert2020specifying}. To formally verify smart contracts, such framing conditions are provided with formal specifications of contracts.
\item \textit{Functional Annotations}\label{P_functionalAnno} are functional pre-and post-conditions that specify what conditions must hold before and after a function executes \cite{schiffl2021towards, hajdu2019solc, antonino2020formalising}. Such annotations are defined with each function according to its specific functionality. An example of a precondition is to assert that the contract assumes that the sum of individual balances is equal to the total balance \cite{hajdu2019solc}, or that the caller must have access rights to the function \cite{schiffl2021towards}. An example of a post-condition is that a storage variable must not be modified unless certain preconditions are satisfied \cite{schiffl2021towards}.
\end{enumerate}
%
\section{Properties and Analysis Methods}
\label{methods}
In this section, we review the pre-deployment analysis methods that have been used to analyze smart contracts. We classify such methods into four categories:
\begin{itemize}
    \item Static analysis, including static type checking, abstract interpretation, control flow analysis, taint analysis, symbolic execution, and pattern matching-based analysis,
    \item Formal verification, including theorem proving and model checking,
    \item Dynamic analysis, including concolic testing and fuzzing,
    \item Machine learning methods, including supervised classification and deep neural networks.
\end{itemize}
These categories and their sub-categories are not mutually exclusive. For instance, taint analysis can be realized using abstract interpretation, and abstract interpretation can be seen as a type of formal verification. Because abstract interpretation is considered to be a framework for static analysis \cite{cousot1977abstract, cousot1979systematic, amadini2020abstract}, we choose to include it here under static analysis. 
Similarly, symbolic execution can be used for verification, testing, and debugging. We categorize always terminating algorithms as static analysis and allow possibly non-terminating to be categorized as formal methods (Sec. \ref{background}). We categorize static symbolic execution analysis as static analysis because it explores only paths up to a certain length (Sec. \ref{SymbolicExecution}) and is therefore terminating.
\subsection{Static Analysis}
A static analysis is an always-terminating algorithm that examines a program to determine whether it satisfies a particular property \cite{cousot2010gentle}. It can be used for a variety of applications, including  program optimization, correctness, and developer experience. 
A fundamental limitation of this approach is that statically proving that a program written in a Turing complete language has a nontrivial property is, in general,  undecidable (See, for example, Rice's theorem \cite{Kozen1977}). This means that for any interesting property, a static analysis that recognizes precisely the set of programs that satisfy this property may not always exist. Nevertheless, as the successful examples mentioned earlier illustrate, there are ample examples of useful static analyses that approximate such a set. 
Often, these conservative approximations are formalized in the abstract interpretation literature \cite{cousot1977abstract}. It is an approximation in that some details are ignored; for example, we may approximate with the set of Integers a set which includes only even Integers or one which simply includes the value 7. It is conservative in the sense that it cannot lead to incorrect conclusions. For example, any approximation of a set must include all of its elements -- if our goal is to reason about what the set may include (but not if we want to reason about what the set must include).\\
We now turn to the review of the static analysis methods used to analyze smart contracts as well as the properties established by these methods.
Table~\ref{tab:SAmethods} presents an overview of what we will cover.
The table is constructed in a manner that we will also follow for the other categories of analysis methods. The goal of this organization is to facilitate the comparison between methods by highlighting what set of properties, if any, is a good match for the method. The table is constructed as follows: First, methods (columns) are listed in increasing order of the total number of properties that each cover. Second, properties (rows) are sorted by the order that fills up the leftmost column with references to work where the method is used to address that property. Within that set of properties, they are sorted in increasing order of the number of other methods addressing this property.
 \input{T5_PropertiesStaticAnalysis.tex}
\subsubsection{Static Type Checking}
\label{M_typeChecking}
A type system for a programming language is a collection of rules used to assign specific types to various constructs, such as variables, expressions, functions, and modules in the program in the context of assumed types for the environment. Type checking ensures that all program statements and expressions adhere to such these rules; for instance, arithmetic and logical operators can only be performed on compatible types \cite{agrawal1991static}. Static type checkers check typing at compile-time only but can verify that the checked conditions hold for all possible executions of the program. Some examples of statically typed languages are C, C++, C\#, Java, and Haskell. Our review pointed to seven smart contract domain-specific languages (DSLs) that use static typing to establish certain safety guarantees,
\begin{itemize}
    \item Albert \cite{10.1007/978-3-030-54455-3_41} uses a linear type system to statically check the duplication and destruction of resources (Sec. \ref{P_assetspreservation}).
    \item Lolisa \cite{yang2020lolisa} uses generalized algebraic data types (GADTs) \cite{xi2003guarded}, which allows it to have a stronger static type system than Solidity. The formal syntax and semantics of Lolisa ensure that all expressions and values in Lolisa are deterministic. Lolisa extends the notion of type safety initially reinforced by Solidity; as such, it does not allow constructing ill-typed terms. 
    \item Mini-Michelson \cite{nishida2022helmholtz} uses a refinement type system to statically check specifications written in the form of refinement types and checks Functional Annotations (Sec. \ref{P_userspecified} (\ref{P_functionalAnno})) of contracts written in a statically typed stack-based language, Michelson \cite{goodmanimichelson}.
    \item Nomos \cite{das2021resource} uses a linear type system \cite{GIRARD19871} to prevent duplication or deletion of assets (Sec. \ref{P_assetspreservation}) and session types to prevent re-entrance into a contract in an inconsistent state (Sec. \ref{P_assetspreservation}).
    \item Obsidian’s \cite{coblenz2020obsidian} type system uses typestate and linear types to statically ensure that objects are manipulated correctly and safely according to their current states and checks the Preservation of Assets (Sec. \ref{P_assetspreservation}).
    \item Zkay \cite{zkay1.0, baumann2020zkay} introduces privacy types defining owners of private values to prevent unintended information leaks and performs type checks for Information flow-related properties (Sec. \ref{P_informationflow}). 
    \item Zeestar \cite{steffen2022zeestar} extends Zkay to expand its expressiveness. The privacy annotations of ZeeStar and zkay are identical, along with the privacy type analysis. However, ZeeStar allows for foreign expressions disallowed in zkay and treats binary operations differently.
\end{itemize}
Papers on Static Typing appear to play an interesting role for this survey: Even though there is only a small number of properties addressed by Static Typing, and also only recently, these properties appear fundamental for the domain of smart contracts. These properties are Preservation of Assets (Sec. \ref{P_assetspreservation}) \cite{10.1007/978-3-030-54455-3_41, das2021resource, coblenz2020obsidian}, Information Secrecy (Sec. \ref{P_informationflow}(\ref{P_informationSecrecy})) \cite{zkay1.0, baumann2020zkay}, and Preservation of Owner Information (Sec. \ref{P_informationflow}(\ref{P_ownerInfo})) \cite{zkay1.0, baumann2020zkay}. Noteworthy is that no other type of static analysis addresses the first property, even though it seems to be a highly general and desirable property for smart contracts. At the other extreme, properties, such as the Preservation of State, have been addressed extensively by all other static analysis methods as well.\\
As such, work on Static Typing seems to point to (and possibly, in some cases, to help formulate) properties that are fundamental to the domain.
%
\subsubsection{Abstract interpretation}
Abstract interpretation is a general methodology for sound approximation of formal semantics of a programming language and for the design of decidable approximations of potentially undecidable properties \cite{cousot1977abstract, cousot1979systematic}.\\
The most notable characteristic of works that use Abstract Interpretation is that they seem to address properties that focus on -- or are related to -- language constructs. The specific instances we find in the literature are Boundedness of Loops (Sec. \ref{P_gasrelated}(\ref{P_boundednessLoops})) \cite{grech2020madmax}, Exception handling (Sec. \ref{P_isolationofcalls}) \cite{tsankov2018securify, smaragdakis2021symbolic}, Guarding of \texttt{selfdestruct} (Sec. \ref{P_informationflow}(\ref{P_selfdestruct})) \cite{smaragdakis2021symbolic}, and Integer over/underflow guarding (Sec. \ref{P_integeroverunderflow}) \cite {Michelson2022}.
One plausible explanation is that being built systematically on top of a particular formal semantics, which is often itself driven by the formal types and syntax of the language, Abstract Interpretation is a natural tool to deal with issues intrinsic to specific language constructs.\\
That said, it should be noted that Abstract Interpretation has also been used to address two of the four fundamental domain-specific properties addressed by static typing, namely Preservation of Owner Information (Sec. \ref{P_informationflow}(\ref{P_ownerInfo})) \cite{tsankov2018securify, smaragdakis2021symbolic}, and Preservation of State (Sec. \ref{P_statepreservation}) \cite{tsankov2018securify, smaragdakis2021symbolic,schneidewind2020ethor}. It has also been used to address four somewhat more complex, domain-specific properties, namely, Isolation of External Calls (Sec. \ref{P_isolationofcalls}) \cite{grech2020madmax}, and Transaction Order Independence (Sec. \ref{P_transactionsordering}), Argument Precondition Validation (Sec. \ref{P_argumentsValidation}), and Withdrawal Acceptance (Sec. \ref{P_withdrawal}) \cite{tsankov2018securify}.\\
As such, among static analysis methods, Abstract Interpretation appears to have the flexibility to address properties based on both deep domain-specific concepts as well as language constructs.
\subsubsection{Control Flow Analysis}
Control flow analysis considers all execution paths that will be traversed at runtime \cite{allen1970control}.\\
It has been used to address three of the four properties that we see as fundamental to the domain. In terms of properties based on language constructs, works using Control Flow Analysis addressed all the ones that have been addressed using Abstract Interpretation. Furthermore, there is one other such property; namely, Division-by-zero Guarding (Sec. \ref{P_dividebyzero}) \cite{peng2019sif}. Moreover, it appears to lend itself naturally to addressing several properties that are based on specific op-codes of the virtual machine underlying the contract execution model, namely \texttt{delegatecall}, \texttt{selfdestruct}, \texttt{staticcall}, \cite{brent2020ethainter} and \texttt{transfer} \cite{brent2018vandal}.\\
As such, among static analysis methods, Control Flow Analysis seems to have something to offer for properties based on both domain-specific concepts and language constructs, and add to that the ability to express constructs based on virtual machine op-codes, which themselves can be viewed as a finer type of properties based on language constructs.
\subsubsection{Taint Analysis (Dataflow Analysis)}
\label{M_taint}
Taint analysis tracks the flow of data between input from a user (source) to program points that it can reach (sinks) \cite{tripp2009taj}. Examples of taint sources are inputs from unauthorized users, and sinks are sensitive instructions reachable by tainted input \cite{lerch2014flowtwist}. Taint analysis has been used to address two kinds of properties, integrity-related properties, for instance, \textit{"can a piece of user-provided information propagate to the internal file system?"} and confidentiality-related properties, for instance, \textit{"can private information become publicly observable?"} In practice, taint analysis is also referred to as information flow analysis that tracks data flow between pre-specified locations in the program.\\
As the table shows, Taint Analysis has been used to address the properties that Control Flow has been used for, with the exception of Argument Precondition Validation (Sec. \ref{P_argumentsValidation}), Boundedness of Loops (Sec. \ref{P_boundednessLoops}), and Isolation of External Calls (Sec. \ref{P_isolationofcalls}). It is plausible that these three properties are not addressed because they involve constraints over a set of actions that have to be done in a particular sequence, which may not be something that can be naturally done by tracking the flow of certain data values. Among properties that are also checked by previously reviewed static analysis methods, some point to a particular class of properties, namely (non-)determinism, that dataflow analysis appears to be useful in addressing \cite{NPChecker2019}. Regarding what has been addressed by Taint Analysis but not the other methods already reviewed so far, there are only two properties: Balance Comparison (Sec. \ref{P_balance}) \cite{ali2021sescon} and Value Unpredictability (Sec. \ref{P_predictable}) \cite{ali2021sescon, li2022smartfast, Mythril}. The most interesting characteristic that these two properties have is that all three works addressing them use Taint Analysis in conjunction with another analysis method.\\
As such, Taint Analysis may be an enabler/catalyst for other types of analysis, with the exception of a small number of works, namely \cite{feist2019slither, NPChecker2019}.
\subsubsection{Symbolic Execution}
\label{SymbolicExecution}
Symbolic execution uses symbolic expressions instead of concrete values to explore the possible program paths and to reason about the conditions under which the program execution will branch in a specific way \cite{king1976symbolic}. Symbolic execution-based approaches are usually fully automated, use a set of properties, and automatically build a model for the system based on an input (like the source code or bytecode). These approaches explore paths up to a certain length, usually defined through abstract interpretation and partial-order reduction, and utilize off-the-shelf SMT solvers (e.g., z3 \cite{de2008z3}) to explore whether or not these properties hold.\\
Symbolic Execution-based approaches detect two of four liveness properties, Eventual Contract Termination (Sec. \ref{P_eventualTermination}) \cite{seijas2020efficient, Seijas2020Marlowe}, and Withdrawal Acceptance (Sec. \ref{P_withdrawal}) \cite{MAIAN}, among which the former is only checked using Symbolic Execution. The properties checked using Symbolic Execution but not by other static analysis methods we have covered so far are Address Integrity (Sec. \ref{P_contractaddress}) \cite{he2021eosafe}, and Validity of Deposits (Sec. \ref{P_depositValidity}) \cite{he2021eosafe, ji2020deposafe}. The works addressing these properties only use Symbolic Execution \cite{he2021eosafe, ji2020deposafe}, among which one work \cite{he2021eosafe} addresses both properties.
As such, among static analysis methods, Symbolic Execution is the only method that has been used to check these domain-specific properties.
\subsubsection{Pattern Matching-based Analysis}
\label{M_pattern}
Pattern matching-based analysis, also referred to as rule-based analysis, scans the source code and checks it against a set of rules or patterns \cite{christodorescu2006static}. Patterns are written in an intermediate representation such as XML or datalog-based specifications~\cite{jeffrey1988principles_datalog}. This method uses an abstract syntax tree (AST) representation of the program, converts it to an intermediate representation, for example, an XML parse tree, and scans it to match against predefined patterns, such as XPath patterns. Pattern matching-based analyzers contain libraries or repositories of rules of predefined known vulnerabilities patterns, which can be further complemented with new patterns to establish custom properties, assertions, and incode-annotations \cite{tikhomirov2018smartcheck, samreen2021smartscan}.\\ 
 Pattern Matching-based Analysis has been applied to a wide range of properties, although it is not necessarily able to provide strong guarantees. A symptom of this is that deeply semantic properties such as Eventual Contract Termination (Sec. \ref{P_eventualTermination}) are not addressed by any works focusing on Pattern Matching-based Analysis. That said, it is practical to codify programming anti-patterns that practitioners constantly identify as they gain experience working in any domain.
%
\subsection{Formal Verification Methods}
Formal verification methods can be used to mechanically build or check proofs that a program satisfies a particular property. This section reviews formal verification methods used to analyze smart contracts, categorizing such methods into either theorem proving or model checking. Table~\ref{tab:FVmethods} gives an overview of the papers that fall in this category and its subcategories. This table is organized similarly to table~\ref{tab:SAmethods}.
\subsubsection{Theorem Proving}
\input{T6_PropertiesFormalMethods.tex}
Theorem proving \cite{cook1971complexity} methods can be used to prove a property of mathematical objects, including programs and their semantics. Theorem provers that need human interaction to help in building proofs are called semi-automatic/interactive theorem provers or proof-assistants. Commonly used proof assistants for smart contracts are Coq \cite{bertot2013interactive} or Isabelle/HOL \cite{nipkow2002isabelle} (see examples \cite{ ribeiro2020formal, arusoaie2021certifying, annenkov2020concert, annenkov2022extracting, milo2022finding}). 
Such proof assistants have also been used after the source code of smart contracts is translated into a formal intermediate representation of a specific verification language, such as WhyML \cite{da2020whylson, 9284726}, IELE \cite{kasampalis2018iele},  Scilla \cite{sergey2018temporal, Scilla2018, Scilla2019}, Boogie \cite{wang2019formal}, F* \cite{bhargavan2016formal}, and SPARK \cite{SPARK2019}.\\ 
Theorem Proving has been used to check two of four liveness properties of smart contracts, namely, Deposit Acceptance (Sec. \ref{P_deposit}) \cite{park2020end}, and Eventual Contract Termination (Sec. \ref{P_eventualTermination}) \cite{genet2020termination}. Among verification methods, the two properties are addressed only by Theorem Proving and not Model Checking. Not only that, among all papers reviewed in this survey, Deposit Acceptance has only been addressed through Theorem Proving. It will be interesting to see in future work if Theorem Proving or verification methods, in general, provide a particular advantage for establishing these properties in particular or liveness properties in general.\\
As such, while several properties have been addressed by Theorem Proving and Model Checking, the key property type addressed by verification methods and not by other methods is Functional Annotations and Assertions (Sec. \ref{P_userspecified}). Such user-specified properties are the natural strength of Theorem Proving. In exchange for this expressive power, more work is typically needed to address useability issues, such as error messages that may come up if a property is not satisfied or the method is unable to establish the property even though it holds.
\subsubsection{Model Checking}
Model checking \cite{baier2008principles} uses a finite-state model of a system to exhaustively check whether or not a given property holds for all possible instances of this model. Bounded model checking explores the system up to a bound $k$ where $k$ is the number of transitions taken from some initial state for violations of a given property \cite{antonino2020formalising}. If a certain property holds for the model within that bound, that is reported; if not, a counterexample is reported to help the user identify the mistake and correct bugs.\\
Model Checking has been used to check the two other liveness properties, namely, Eventual Contract Removal (Sec. \ref{P_wallet}) \cite{nelaturu2020verified} and Withdrawal Acceptance (Sec. \ref{P_withdrawal}) \cite{nelaturu2020verified, stephens2021smartpulse}. Among all papers reviewed in this survey, Model Checking is the only method that has been used to establish Eventual Contract Removal. It is also useful to note that even though Withdrawal Acceptance (Sec. \ref{P_withdrawal}) appears to be one of the most widely addressed properties by the methods surveyed, the variants addressed by non-verification methods do not appear to guarantee liveness.\\
As such, on the whole, formal verification, having been used to check all four liveness properties, with two of them not addressed by any of the other works addressed in this survey, appears to provide an advantage in terms of proving liveness properties.
\subsection{Pre-deployment Dynamic Analysis}
Dynamic analysis \cite{ball1999concept} analyzes properties of a program while it is executing \cite{dynamicAnalysisSurvey2015}. For smart contracts, dynamic analysis is commonly used to simulate attack scenarios to expose exploitable vulnerabilities. This section reviews the most commonly adopted types of dynamic analysis performed on smart contracts, including concolic testing and fuzzing, as presented in Table~\ref{tab:DAmethods} with connection to smart contract properties (presented in rows) that have been established through these methods. 
This table is organized similarly to table~\ref{tab:SAmethods}.
\subsubsection{Concolic Testing}
\input{T7_PropertiesDynamicAnalysis.tex}
Dynamic symbolic execution, also known as concolic testing \cite{concolicTesting}, is a hybrid method that combines CONCrete and symbOLIC execution. It executes the target program symbolically, in a forward manner, and systematically explores its feasible paths by using concrete inputs to improve code coverage. It is commonly used as a test case generation technique to aid in fuzzing techniques that can prioritize executions of interest \cite{parvez2016combining}.\\ 
Concolic Testing has been used to check two user-specified properties, namely, Contract-specific Invariants \cite{SOLAR2018LiAo} and Assertion Satisfaction (Sec. \ref{P_userspecified}) \cite{Annotary2019}. Among other properties that have been checked using Concolic Testing, one is a relatively complex domain-specific property, Preservation of Assets in Transactions (Sec. \ref{P_assetsvalue}) \cite{li2020protect}, which is only checked by one other method, that is, Theorem Proving.\\
As such, Concolic Testing seems to provide flexibility in checking user-specified and complex domain-specific properties. 
\subsubsection{Fuzzing}
Fuzzing \cite{sutton2007fuzzing} feeds random or systematically generated inputs to the program at runtime to explore vulnerable program parts. It is usually an automated process that is commonly performed to find bugs, performance issues, zero-day attacks, and exploits in the program. The fuzzing-based analysis applies two kinds of test case generation approaches; the grammar-based approaches \cite{10.1145/1379022.1375607} generate test cases from a user-specified model following the program's input format, whereas mutation-based approaches \cite{van2005fuzzing} randomly mutate inputs to cause crashes in the program.\\
Fuzzing has been used to check user-specified custom properties and Assertions \cite{torres2021confuzzius, grieco2020echidna}. Other properties addressed using Fuzzing are two language constructs-related properties, namely, Guarding of \texttt{selfdestruct} (Sec. \ref{P_informationflow}(\ref{P_selfdestruct})) \cite{torres2021confuzzius} and Exception Handling (Sec. \ref{P_exceptions}) \cite{torres2021confuzzius, teng2021SmartGift, nguyen2020sfuzz, wang2020oracle, wei2020smart, jiang2018contractfuzzer, 8859497, ILF2019}. Fuzzing has also been used to check fundamental as well as complex domain-specific properties, as shown in table~\ref{tab:SAmethods}. \\
As such, Fuzzing appears to be a useful tool for falsifying (i.e., finding counterexamples) for a wide range of properties, including fundamental, user-specified, and complex domain-specific ones.
%
\subsection{Machine Learning Methods}
\label{M_Machinelearning}
Recently, several works investigated using machine learning methods to analyze smart contracts \cite{MLmodel2019, qian2020towards, zhuang2020smart, hao2020scscan, wang2020contractward, xing2020new, ashizawa2021eth2vec, huang2021hunting, liu2021combining, xu2021novel, GUPTA2021107583, narayana2021automation, mi2021vscl, Zhang2022, zhang2022smart, zhang2022reentrancy, Nha2022, Yao2022, Yang2022, YanXingrun2022}. Machine learning methods learn from data and build predictions. For example, in supervised learning, labeled data is used to train a model to produce a function to predict outputs for new inputs \cite{sen2020supervised}. Machine learning methods do not require predefined patterns for specific properties or vulnerabilities. Experts can label a number of contracts to train a model that can automatically predict whether a contract has a specific type of vulnerability \cite{liu2021combining}. \textit{A supervised classification algorithm} can learn from such training sets and then assigns new contracts to a particular class, for example, vulnerable or not vulnerable. Using labeled contracts, \textit{deep learning algorithms (neural networks)} can automatically construct important features, thereby obviating the need for manual patterns specification \cite{mudgal2018deep}. Table~\ref{tab:MLmethods} gives an overview of the papers that fall in this category and its subcategories.
\input{T8_PropertiesMachineLearning.tex}
\\
%
%
Classification algorithm-based approaches have shown an increased detection performance compared to approaches using traditional analyses based on various metrics. For example, the supervised classification model presented by Momeni et al. \cite{MLmodel2019} shows high accuracy and precision for vulnerabilities corresponding to Integer over/underflow Guarding (Sec. \ref{P_integeroverunderflow}), Preservation of State (Sec. \ref{P_statepreservation}), and Guarding of \texttt{selfdestruct} (Sec. \ref{P_informationflow}(\ref{P_selfdestruct})). Similarly, SCScan \cite{hao2020scscan} shows a 100\% recognition rate (the number of identified contracts with vulnerabilities divided by the total number of vulnerable contracts) for vulnerabilities corresponding to the Preservation of State, Integer over/underflow Guarding, Transfer Function Viability (Sec. \ref{P_transferfunctionviability}), and Value Unpredictability (Sec. \ref{P_predictable}).\\ 
Approaches using deep learning algorithms have also shown better detection performance (with higher accuracy and precision) than state-of-the-art static analysis approaches. For example, the approach of Qian et al. \cite{qian2020towards}, although targeted toward detecting reentrancy only, is reported to have higher precision and recall than four state-of-the-art static analysis-based tools, namely, Oyente \cite{Oyente2016}, Securify \cite{tsankov2018securify}, SmartCheck \cite{tikhomirov2018smartcheck}, and Mythril \cite{Mythril}. Those tools are also evaluated with the approach of Zhuang et al. \cite {zhuang2020smart}, which is reported to provide better accuracy and precision for the Preservation of State, Value Unpredictability, and Boundedness of Loops (Sec. \ref{P_boundednessLoops}). These properties and the same set of static tools and Slither \cite{feist2019slither} are evaluated with the approach of Liu et al. \cite{liu2021combining}, which shows improved accuracy than those tools for those properties. Similarly, SVChecker \cite{SVChecker2022} is reported to outperform Oyente, Securify, Slither, and SmartCheck, giving as high as a 100\% detection rate for two vulnerabilities, namely, reentrancy and unchecked low-level calls (Sec. \ref{P_exceptions}). Eth2Vec \cite{ashizawa2021eth2vec} is reported to outperform the (SVM classification-based) model by Momeni et al. \cite{MLmodel2019} in terms of precision, recall, and F1 score. \\
As such, machine learning methods, particularly deep learning algorithms, seem to have covered a wide range of properties, including liveness and complex domain-specific properties, and have shown increased detection performance over state-of-the-art static analysis tools.
\section{Emerging Directions and Opportunities}
\label{discussion}
In this section, we discuss emerging research directions and opportunities based on our review of the literature in this survey.\\
\textbf{Findings on pre-deployment analysis of smart contracts. } A high-level trend seems to be that works addressing liveness properties appear primarily after 2020. This may indicate that the community is shifting attention from safety properties to the -- often more challenging to establish -- liveness properties. In particular, one liveness property, Eventual contract termination (Sec. \ref{P_eventualTermination}), is commonly checked for Ethereum and Cardano blockchains, indicating growing research on properties of smart contracts of platforms beyond Ethereum. In terms of challenges, we observe that with few exceptions (cf. \cite{xue2020cross, ma2021pluto, chinen2020ra, Annotary2019}), most methods analyze individual contracts in isolation, that is, without considering calls to external contracts. So, the contracts with external calls, for instance, to contracts whose source is not available for analysis or to off-chain services, remain vulnerable to issues that only arise when they interact with other contracts or services. Modular methods can accommodate this challenge by allowing the user to express assumptions about called methods, but it is conceivable that the ideal solution here may need to come from new protocols for interaction between contracts to limit the extent to which one contract can unilaterally affect the resources of another contract. Another challenge is that analysis tools can generate too many unnecessary alarms by detecting vulnerabilities that are not exploitable in practice. For example, many contracts were deemed vulnerable by various analysis tools, but not all were found exploitable in practice \cite{perez2021smart}. Such findings call for exploring the vulnerabilities from multiple perspectives, including the possibilities of their occurrences and exploitations in live contracts.  
Similarly, as languages for writing smart contracts evolve, some properties may be addressed by dynamic checks. For example, the Solidity 0.8.0+ compilers check divisions by zero (Sec. \ref{P_dividebyzero}) and integer over/underflow (Sec. \ref{P_integeroverunderflow}). This may alleviate some of the need for addressing this kind of property through pre-deployment analysis, shifting the concern from being a security concern to a performance or cost concern.\\
\textbf{{Findings on commonly investigated properties. }}
Among the properties reviewed in this survey (Sec. \ref{properties}), the Preservation of State (Sec. \ref{P_statepreservation}) is the most extensively researched. Moreover, certain types of analysis are entirely dedicated to checking this property \cite{xue2020cross, albert2020taming, schneidewind2020ethor, li2020protect, wang2021mar, wustholz2020harvey, qian2020towards}. A possible reason behind the attention given to this property is that its violation resulted in a notorious reentrancy vulnerability, which was the culprit behind the famous DAO attack \cite{DAOAttack}, which eventually led to a hard fork of Ethereum. Unfortunately, this and similar incidents did not stop with the DAO attack as 'reentrancy' is often exploited in live contracts; for example, in an incident in which an attacker stole tokens worth USD 2 Million \cite{ReentrancyIncident}.
Another frequently studied property is Integer over/underflow Guarding (Sec. \ref{P_integeroverunderflow}). As noted previously, this and similar other properties (such as Sec. \ref{P_dividebyzero}) may cease to be a pain point in the future due to evolving language support through dynamic checking, but that would not eliminate the value of static checking. For example, Solidity's checked arithmetic uses more gas, which can also be unnecessary in some cases, such as with the case of overflow checks in a loop over a fixed-size array. Besides the commonly investigated properties, our investigation reveals that only one property, Transfer amount boundedness (Sec. \ref{P_amountBoundedness}), is checked exclusively in contracts written in Marlowe for Cardano blockchain. This property is unique to Cardano contracts as they handle separate balances for contracts and accounts.\\
\textbf{{Findings on formalization and rigorous treatment of properties.}} 
Out of the 35 properties we reviewed, only 16 have been formalized. This shows that there are ample opportunities for formalization and rigorous treatment of properties. We add a “$^\star$” symbol in Table~\ref{tab:propDetails} next to the properties that have been formalized and the studies that have formalized them to facilitate future efforts to formalize properties and vulnerabilities.\\
%
\textbf{{Findings on analysis Methods. }} 
Here we highlight some of the key insights from our review of analysis methods (Sec. \ref{methods}). Static Typing seems to point to (and possibly, in some cases, to help formulate) properties that are fundamental to the domain. Abstract Interpretation appears to have the flexibility to address properties based on both domain-specific concepts as well as language constructs. Control Flow methods seem to have something to offer for properties based on domain-specific concepts and language constructs, and add to that the ability to express constructs based on virtual machine op-codes, which themselves can be viewed as a finer type of properties based on language constructs. Taint analysis appears to be often an enabler/catalyst for other types of analysis, except for a small number of works, namely \cite{feist2019slither, NPChecker2019}. Symbolic Execution is the only method that has been used to check particular domain-specific properties, namely, Address Integrity (Sec. \ref{P_contractaddress}) and Validity of Deposits (Sec. \ref{P_depositValidity}).\\
Formal methods having been used to check all four liveness properties, with two not addressed by any of the other works addressed in this survey, appear to provide an advantage in proving liveness properties.\\
Among dynamic analysis methods reviewed in this survey, Concolic Testing seems to provide flexibility in checking user-specified as well as complex domain-specific properties. Fuzzing appears to be a useful tool for falsifying (finding counterexamples) various kinds of properties, such as fundamental and complex domain-specific properties and user-specified properties.\\
Machine learning methods, particularly deep learning algorithms, seem to have covered a wide range of properties, including liveness and complex domain-specific properties and have shown increased detection performance over state-of-the-art static analysis tools. Methodologically, through the notions of precision and recall, work on ML methods provides a fresh perspective on static analysis that may prove useful in honing in on methods with the highest level of usability. \\
 Lastly, a valuable observation made by one of the reviewers about analysis methods is that only Static Typing has been widely applied to non-Ethereum smart contracts compared to other traditional analysis methods. Ethereum contracts are primarily written in Solidity. In contrast, various DSLs are being developed for other platforms, for instance, Albert for Tezos, that aim to address some fundamental properties, such as asset linearizability (Sec. \ref{P_assetspreservation}) by using various type systems. These types of checks reinforce the security of these languages. Some platforms also incorporate languages that make specific types of analysis and guarantees about properties possible. For example, Michelson is a domain-specific language for the Tezos platform that has a formal verification process built into the language, which ensures that smart contracts are correct and secure before they are deployed to the blockchain. Also, other platforms are newer than Ethereum––while some traditional analyses have been applied to those platforms (cf. \cite{Michelson2022, seijas2020efficient, Seijas2020Marlowe, grishchenko2018semantic}), many approaches are still to be investigated for them.\\
\textbf{{Findings on soundness. }} 
The soundness of the methods has been established mathematically for only 11 out of 35 properties (Sec. \ref{P_deposit}, \ref{P_eventualTermination}, \ref{P_contractaddress}, \ref{P_exceptions}, \ref{P_informationflow}(\ref{P_selfdestruct}), \ref{P_integeroverunderflow}, \ref{P_assetspreservation}, \ref{P_statepreservation}, \ref{P_transactionsordering}, \ref{P_predictable}, \ref{P_userspecified} (A \& D)) and 17 approaches based on static type checking \cite{coblenz2020obsidian, nishida2022helmholtz, das2021resource}, abstract interpretation \cite{schneidewind2020ethor}, symbolic execution \cite{permenev2020verx}, model checking \cite{marescotti2020accurate, wesley2021compositional,kalra2018zeus,grishchenko2018ethertrust,grishchenko2018foundations}, and theorem proving \cite{park2020end, elad2021summing, ribeiro2020formal,annenkov2020concert,annenkov2022extracting,genet2020termination,grishchenko2018semantic}. This observation suggests that there are ample opportunities for further investigation of the utility of mathematical semantic analysis methods in this domain. Also, it would be particularly interesting to see if there are ML-based methods that can also provide strong guarantees about properties.
\\
\textbf{{Findings on benchmarking. }} Comparing the performance of analysis methods requires practical standards that can be used as benchmarks. The lack of benchmarks for comparison of tools as well as standardization for naming vulnerabilities has also been acknowledged before \cite{gupta2020insecurity}. With respect to benchmarking, assessing false negatives and positives reported by such tools requires non-vulnerable and vulnerable benchmark entries. Towards this end, we wish to acknowledge some efforts in this direction: Durieux et al. \cite{durieux2020empirical}, Ferreira et al. \cite{ferreira2020smartbugs}, and Torres et al. \cite{Osiris2018} have generously made classified datasets of smart contracts available to the public; which present an opportunity for the addition of sample contracts representing newer classes of vulnerabilities. Also, recent taxonomies \cite{atzei2017survey, khan2020survey, chen2020survey} have enabled a better understanding of different classes of vulnerabilities, which shows potential for further investigation of properties as well.

\bibliographystyle{ACM-Reference-Format-num}
\bibliography{
references,
references2016,
references2017,
references2018,
references2019,
references2020,
references2021,
references2022}

\appendix

\end{document}

%% file: T1_PropertiesSpecifications.tex
\begin{table}[!h]
 \small
\setlength\abovecaptionskip{-0.1\baselineskip}
\setlength\belowcaptionskip{-0.1\baselineskip}
  \caption{Variations of the same property across different studies}
  \label{tab:Specifications}
    \resizebox{\columnwidth}{!}{
  \begin{tabular}
  {| p{15mm} | p{24mm} | p {136mm}|}
    \hline
 \textbf{Study} & \textbf{Property Name} & \textbf{Property Specification} \\
 \hline
\cite{tolmach2020survey} & Single-entrancy & 
    “no writes after call” \cite{tsankov2018securify},
    “the contract cannot perform any more calls once it has been reentered” \cite{schneidewind2020ethor}\\
 \hline
\cite{schneidewind2020good} & Single-entrancy & 
“Single-entrancy captures that the reentering execution of a contract should not initiate any further internal transactions.”\\
 \hline
 
This Survey & \raggedright{Preservation of state (Sec. \ref{P_statepreservation})} & 
Preservation of state asserts that state variables are never updated after calling an external contract, and they must be private; that is, they cannot be directly modified by functions in other smart contracts.\\
  \hline
\end{tabular}
  }
 \vspace{-4mm}
\end{table}

%% file: T2_RelatedSurveys.tex
\begin{wraptable}{r}{9.9cm}
\tiny
\setlength\abovecaptionskip{-0.1\baselineskip}
\setlength\belowcaptionskip{-0.1\baselineskip}
  \caption{Related Studies}
  \label{tab:surveys}
  \begin{tabular} {| l | P{6mm} | P{7mm} | P{6mm} | P{12mm} | P{7mm} |}
  \hline
    \textbf{Study} 
    & \textbf{Proper-ties} 
    & \textbf{Vulner-abilities} 
    & \textbf{Methods} 
    & \textbf{Peer-reviewed publications}
    & \textbf{Multi-Platform} \\

\hline \hline

    This Survey &
    \raisebox{-.25\height}{ 
    \includegraphics [width=0.25cm]{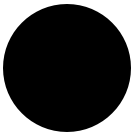}}
    &
    \raisebox{-.25\height}{ 
    \includegraphics [width=0.25cm]{yes}}
    &
    \raisebox{-.25\height}{ 
    \includegraphics [width=0.25cm]{yes}}
    &
    \raisebox{-.25\height}{ 
    \includegraphics [width=0.25cm]{yes}}
    &
    \raisebox{-.25\height}{ 
    \includegraphics [width=0.25cm]{yes}} \\ 

 \hline 
 \hline  
 \multicolumn{6}{| l | }{\textbf{Surveys that address smart contract properties, vulnerabilities, and methods}} \\
 \hline
    Tolmach et al., 2021 \cite{tolmach2020survey} 
    & 
    \raisebox{-.25\height}{ 
    \includegraphics [width=0.25cm]{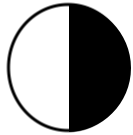}}
    &
    \raisebox{-.25\height}{ 
    \includegraphics [width=0.25cm]{somewhat}}
    &
    \raisebox{-.25\height}{ 
    \includegraphics [width=0.25cm]{yes}}
    &
    \raisebox{-.25\height}{ 
    \includegraphics [width=0.25cm]{yes}}
    &
    \raisebox{-.25\height}{ 
    \includegraphics [width=0.25cm]{yes}}\\
\hline
    \vtop{
    \hbox{\strut Schneidewind et al., 2020 \cite{schneidewind2020good}}
    \hbox{\strut Grishchenko et al. 2018 \cite{grishchenko2018foundations}}}
    &
    \raisebox{-.75\height}{ 
    \includegraphics [width=0.25cm]{somewhat}}
    &
    \raisebox{-.75\height}{
    \includegraphics [width=0.25cm]{somewhat}}
    &
    \raisebox{-.75\height}{
    \includegraphics [width=0.25cm]{yes}}
    &
    \raisebox{-.75\height}{
    \includegraphics [width=0.25cm]{yes}}
    &
    \raisebox{-.75\height}{
    \includegraphics [width=0.25cm]{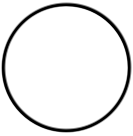}}\\

 \hline 
 \hline
 
  \multicolumn{6}{| l | }{\textbf{Surveys that address vulnerabilities and methods}} \\
  \hline
   Praitheeshan et al., 2019 \cite{praitheeshan2019security}&
    \raisebox{-.25\height}{ 
    \includegraphics [width=0.25cm]{no}}
    &
    \raisebox{-.25\height}{ 
    \includegraphics [width=0.25cm]{yes}}
    &
    \raisebox{-.25\height}{ 
    \includegraphics [width=0.25cm]{yes}}
    &
    \raisebox{-.25\height}{ 
    \includegraphics [width=0.25cm]{yes}}
    &
    \raisebox{-.25\height}{ 
    \includegraphics [width=0.25cm]{no}}\\

 \hline
    \vtop{
    \hbox{\strut Tang et al., 2021 \cite{tang2021vulnerabilities}, 
    Antonio et al., 2020 \cite{lopez2020smart}}
    \hbox{\strut Garfattaet et al., 2021 \cite{garfatta2021survey}}
    \hbox{\strut Kim \& Sukyoung, 2020 \cite{kim9230065}}}&
    \raisebox{-1.5\height}{
    \includegraphics [width=0.25cm]{no}}
    &
    \raisebox{-1.5\height}{
    \includegraphics [width=0.25cm]{yes}}
    &
    \raisebox{-1.5\height}{
    \includegraphics [width=0.25cm]{yes}}
    &
    \raisebox{-1.5\height}{
    \includegraphics [width=0.25cm]{somewhat}}
    &
    \raisebox{-1.5\height}{
    \includegraphics [width=0.25cm]{no}}\\

   \hline
    \vtop{
    \hbox{\strut Chen H. et al., 2020 \cite{chen2020survey}}
    \hbox{\strut Dika \& Nowostawski, 2018 \cite{dika2018security}}}&
    \raisebox{-.75\height}{
    \includegraphics [width=0.25cm]{no}}
    &
    \raisebox{-.75\height}{
    \includegraphics [width=0.25cm]{yes}}
    &
    \raisebox{-.75\height}{
    \includegraphics [width=0.25cm]{somewhat}}
    &
    \raisebox{-.75\height}{
    \includegraphics [width=0.25cm]{yes}}
    &
    \raisebox{-.75\height}{
    \includegraphics [width=0.25cm]{no}}\\
\hline
    \vtop{
    \hbox{\strut Gupta et al., 2020 \cite{gupta2020insecurity}, 
    Groce et al., 2020 \cite{groce2020actual}}
    \hbox{\strut Khan \& Namin, 2020 \cite{khan2020survey}}
    \hbox{\strut Mense \& Flastscher, 2018 \cite{menseVulnerabilities}}}&
    \raisebox{-1.5\height}{
    \includegraphics [width=0.25cm]{no}}
    &
    \raisebox{-1.5\height}{
    \includegraphics [width=0.25cm]{yes}}
    &
    \raisebox{-1.5\height}{
    \includegraphics [width=0.25cm]{somewhat}}
    &
    \raisebox{-1.5\height}{
    \includegraphics [width=0.25cm]{somewhat}}
    &
    \raisebox{-1.5\height}{
    \includegraphics [width=0.25cm]{no}}\\
  \hline
    Rahimian \& Clark, 2021 \cite{rahimian2021tokenhook} &
    \raisebox{-.25\height}{
    \includegraphics [width=0.25cm]{no}}&
    \raisebox{-.25\height}{
    \includegraphics [width=0.25cm]{yes}}&
    \raisebox{-.25\height}{
    \includegraphics [width=0.25cm]{somewhat}}&
    \raisebox{-.25\height}{
    \includegraphics [width=0.25cm]{no}}&
    \raisebox{-.25\height}{
    \includegraphics [width=0.25cm]{somewhat}}\\

\hline

    Yamashita et al., 2019 \cite{yamashitaPotentialRisks} &
    \raisebox{-.25\height}{
    \includegraphics [width=0.25cm]{no}}
    &
    \raisebox{-.25\height}{
    \includegraphics [width=0.25cm]{yes}}
    &
    \raisebox{-.25\height}{
    \includegraphics [width=0.25cm]{somewhat}}
    &
    \raisebox{-.25\height}{
    \includegraphics [width=0.25cm]{no}}
    &
    \raisebox{-.25\height}{
    \includegraphics [width=0.25cm]{no}}\\

 \hline 
 \hline
   \multicolumn{6}{| l | }{\textbf{Surveys that address vulnerabilities}} \\
   \hline
    \vtop{
    \hbox{\strut  Atzei et al., 2017 \cite{atzei2017survey}, 
    Ji et al., 2021 \cite{ji2021evaluating}}
    \hbox{\strut  Demir et al., 2020 \cite{demir2019security}}}& 
    \raisebox{-.75\height}{
    \includegraphics [width=0.25cm]{no}}
    &
    \raisebox{-.75\height}{
    \includegraphics [width=0.25cm]{yes}}
    &
    \raisebox{-.75\height}{
    \includegraphics [width=0.25cm]{no}}
    &
    \raisebox{-.75\height}{
    \includegraphics [width=0.25cm]{yes}}
    &
    \raisebox{-.75\height}{
    \includegraphics [width=0.25cm]{no}}\\

\hline
    \vtop{
    \hbox{\strut Krupa et al., 2021 \cite{krupa2021security}, 
    Daojing et al., 2020 \cite{He9143290}}
    \hbox{\strut Staderini et al., 2020 \cite{staderini2020classification}}}& 
    \raisebox{-.75\height}{
    \includegraphics [width=0.25cm]{no}}
    &
    \raisebox{-.75\height}{
    \includegraphics [width=0.25cm]{yes}}
    &
    \raisebox{-.75\height}{
    \includegraphics [width=0.25cm]{no}}
    &
    \raisebox{-.75\height}{
    \includegraphics [width=0.25cm]{somewhat}}
    &
    \raisebox{-.75\height}{
    \includegraphics [width=0.25cm]{no}}\\
 
 \hline
 
    Zhu et al., 2018 \cite{zhu2018research} &
    \raisebox{-.25\height}{
    \includegraphics [width=0.25cm]{no}}
    &
    \raisebox{-.25\height}{
    \includegraphics [width=0.25cm]{yes}}
    &
    \raisebox{-.25\height}{
    \includegraphics [width=0.25cm]{no}}
    &
    \raisebox{-.25\height}{
    \includegraphics [width=0.25cm]{no}}
    &
    \raisebox{-.25\height}{
    \includegraphics [width=0.25cm]{somewhat}}\\

\hline

    Chen et al., 2020 \cite{chen2020Defects}, Li et al., 2020 \cite{li2020research} &
    \raisebox{-.25\height}{
    \includegraphics [width=0.25cm]{no}}
    &
    \raisebox{-.25\height}{
    \includegraphics [width=0.25cm]{yes}}
    &
    \raisebox{-.25\height}{
    \includegraphics [width=0.25cm]{no}}
    &
    \raisebox{-.25\height}{
    \includegraphics [width=0.25cm]{no}}
    &
    \raisebox{-.25\height}{
    \includegraphics [width=0.25cm]{no}}\\
    
\hline 
\hline

     \multicolumn{6}{| l | }{\textbf{Surveys that address analysis methods}} \\
     \hline
    \vtop{
    \hbox{\strut Wang et al., 2021 \cite{wang2021security}, Singh et al., 2020 \cite{singh2020blockchain}
    }
    \hbox{\strut Almakhour et al., 2020 \cite{almakhour2020verification}}
    \hbox{\strut Liu J. \& Liu Z., 2019 \cite{liu2019survey}}} &
    \raisebox{-1.5\height}{ 
    \includegraphics [width=0.25cm]{no}}
    &
    \raisebox{-1.5\height}{
    \includegraphics [width=0.25cm]{no}}
    &
    \raisebox{-1.5\height}{
    \includegraphics [width=0.25cm]{yes}}
    &
    \raisebox{-1.5\height}{
    \includegraphics [width=0.25cm]{yes}}
    &
    \raisebox{-1.5\height}{
    \includegraphics [width=0.25cm]{somewhat}}\\
 \hline 
 \hline
  \multicolumn{6}{| l | }{\textbf{Surveys that have limited coverage of properties, vulnerabilities, and methods }} \\
  \hline
     \vtop{
     \hbox{\strut Sayeed et al., 2020 \cite{sayeed2020Attacks},
     Saad et al., 2019 \cite{saad2019exploring}
     }
     \hbox{\strut Perez \& Livshits, 2021 \cite{perez2021smart}}
     \hbox{\strut Kushawa et al., 2022 \cite{9667515}, Rameder et al., 2022 \cite{rameder2022review}}
     \hbox{\strut Piantadosi et al., 2022 \cite{piantadosi2022detecting}}
     }&   
     \raisebox{-2\height}{
     \includegraphics [width=0.25cm]{no}}
     &
     \raisebox{-2\height}{
     \includegraphics [width=0.25cm]{somewhat}}
     &
     \raisebox{-2\height}{
     \includegraphics [width=0.25cm]{somewhat}}
     &
     \raisebox{-2\height}{
     \includegraphics [width=0.25cm]{somewhat}}
     &
     \raisebox{-2\height}{
     \includegraphics [width=0.25cm]{no}}\\

\hline
     \vtop{
     \hbox{\strut Praitheeshan et al., 2020 \cite{praitheeshan2020security}}
     \hbox{\strut 
     Vacca et al., 2020 \cite{vacca2020systematic}
     }}
     &
     \raisebox{-.75\height}{
     \includegraphics [width=0.25cm]{no}}
     &
     \raisebox{-.75\height}{
     \includegraphics [width=0.25cm]{somewhat}}
     &
     \raisebox{-.75\height}{
     \includegraphics [width=0.25cm]{no}}
     &
     \raisebox{-.75\height}{
     \includegraphics [width=0.25cm]{yes}}
     &
     \raisebox{-.75\height}{
     \includegraphics [width=0.25cm]{yes}}\\

\hline

    Alharby \& Moorsel, 2018 \cite{alharby2018blockchain} &
    \raisebox{-.25\height}{
    \includegraphics [width=0.25cm]{no}}
    &
    \raisebox{-.25\height}{
    \includegraphics [width=0.25cm]{somewhat}}
    &
    \raisebox{-.25\height}{
    \includegraphics [width=0.25cm]{no}}
    &
    \raisebox{-.25\height}{
    \includegraphics [width=0.25cm]{somewhat}}
    &
    \raisebox{-.25\height}{
    \includegraphics [width=0.25cm]{yes}}\\

\hline

    Tantikul \& Ngamsuriyaroj, 2020  \cite{tantikul2020exploring} &
    \raisebox{-.25\height}{
    \includegraphics [width=0.25cm]{no}}
    &
    \raisebox{-.25\height}{
    \includegraphics [width=0.25cm]{somewhat}}
    &
    \raisebox{-.25\height}{
    \includegraphics [width=0.25cm]{no}}
    &
    \raisebox{-.25\height}{
    \includegraphics [width=0.25cm]{somewhat}}
    &
    \raisebox{-.25\height}{
    \includegraphics [width=0.25cm]{no}}\\
    
  \hline
\end{tabular}
\end{wraptable}

%% file: T_Blockchains.tex
\begin{wraptable}{r}{10cm}
\tiny
\setlength\abovecaptionskip{-0.1\baselineskip}
 \setlength\belowcaptionskip{-0.1\baselineskip}
  \caption{Blockchains, their supported languages, crypto-assets, and technical standards}
  \label{tab:Blockchains}
  \begin{tabular} {| l | l |l| c| l |}
    \hline
    \textbf{Blockchain} & \textbf{Languages} & \textbf{Native (fungible) Cryptocurrency} & \textbf{NFTs} & \textbf{Technical Standards} \\
    \hline
    Ethereum & Solidity, Vyper 
    & Ether(ETH) &  
    \includegraphics [width=0.2cm]{yes}
    & ERC-20, ERC721
    \\
    \hline
    Bitcoin & Script & Bitcoin(BTC) &  \includegraphics [width=0.2cm]{no} & - \\
    \hline
    EOSIO & C++, WebAssembly (WASM) & EOS & \includegraphics [width=0.2cm]{yes} & eosio.token 
    \\
    \hline
    Hyperledger Fabric
    & Go, JavaScript(node.js) 
    & None & \includegraphics [width=0.2cm]{no} & - \\
    \hline
    Algorand
    & TEAL, Python (PyTeal) & Algo, Algorand Standard Assets (ASA) & \includegraphics [width=0.2cm]{no} & ASA \\
    \hline
    Cardano
    & Marlowe, Plutus, Haskell & ADA, Cardano Tokens & \includegraphics [width=0.2cm]{no} & - \\ 
    \hline
    Tezos
    & Michelson, SmartPy, Ligo, Fi & tez(XTZ) & \includegraphics [width=0.2cm]{yes} & FA1.2, FA2, TZIP-16\\
  \hline
\end{tabular}
\end{wraptable}

%% file: T4_PropertiesDetails.tex
\begin{table}
\tiny
\setlength\abovecaptionskip{-0.1\baselineskip}
 \setlength\belowcaptionskip{-0.2\baselineskip}
  \caption{Smart contract properties with corresponding vulnerabilities. $^\star$ marks the properties that have been formalized and studies that formalize properties. Languages: \textbf{V}yper, \textbf{S}olidity,  \textbf{-}any language supported by the platform.}
  \label{tab:propDetails}
\begin{tabular}{| p{31mm} | p{31mm} | p{7mm} | c | p {55mm} |}
    \hline
   \textbf{Property} & \textbf{Vulnerabilities} & \textbf{Platform} & \textbf{Language} & \textbf{Studies} \\
    \hline
    \multicolumn{5}{| l | }{\textbf{Liveness}} \\
    \hline
    Deposit acceptance 
        &  Failing to accept new deposit (\ref{P_deposit}) & Ethereum & V0.1.0B13 & \cite{park2020end} \\
    \hline
    Eventual contract removal$^\star$ & References to destroyed contracts (\ref{P_wallet}) & Ethereum & S & \cite{nelaturu2020verified}$^\star$\\
       \hline
    \multirow{2}{45mm}{Eventual contract termination$^\star$} 
        & Non-terminating contracts (\ref{P_eventualTermination})
        & Ethereum & - & \cite{genet2020termination}$^\star$  \\
        \cline{3-5}
        & & Cardano & Marlowe & \cite{seijas2020efficient,Seijas2020Marlowe} \\
    \hline
    Withdrawal acceptance$^\star$  
        & Locking, locked or freezing Ether (\ref{P_withdrawal}) & Ethereum & - &
        \cite{lu2021neucheck, ashouri2020etherolic, xing2020new, wei2020smart, Li2021Clue, tikhomirov2018smartcheck,jiang2018contractfuzzer, DefectChecker2022}\cite{nelaturu2020verified,stephens2021smartpulse}$^\star$\\
    \hline
    \hline
    \multicolumn{5}{| l | }{\textbf{Safety}} \\
    \hline
    \multirow{2}{45mm}{Address integrity$^\star$}
        & Forged notification, Fake receipt (\ref{P_contractaddress}) & EOSIO & - & \cite{huang2020eosfuzzer, he2021eosafe, li2022eosioanalyzer,Wasai2022, Li2022} \\
        \cline{2-5}
        & External contract referencing (\ref{P_contractaddress}) & Ethereum & S & \cite{nelaturu2020verified}$^\star$ \\
    \hline
    Address parameter avoidance 
        & Misuse of address parameters (\ref{P_contractaddress})  & Ethereum & S & \cite{praitheeshan2021solguard}\\
    \hline
    Address size soundness$^\star$ 
        & Address shortening (\ref{P_contractaddress}) & Ethereum & S & \cite{ashouri2020etherolic, xing2020new}\cite{nelaturu2020verified}$^\star$\\
    \hline
    Argument precondition validation & Use of untrusted arguments (\ref{P_argumentsValidation}) & Ethereum & - & \cite{tsankov2018securify} \\
    \hline
    Balance comparison 
        & Strict equality of balance (\ref{P_balance}) &  Ethereum & S < 0.8.0 & \cite{lu2021neucheck, arganaraz2020detection,tikhomirov2018smartcheck, DefectChecker2022} \\
    \hline      
    \multirow{3}{45mm}{Boundedness of loops}    
        & Un-bounded mass operations (\ref{P_gasrelated}) & Ethereum & S & \cite{grech2020madmax, lu2021neucheck, nassirzadeh2021gas, Correas2021static}\\
        & Unsafe type declaration (\ref{P_gasrelated}) &  & & \cite{lu2021neucheck, arganaraz2020detection, tikhomirov2018smartcheck} \\
      \hline 
       Conditional independence  
        & DoS due to external contracts (\ref{P_conditionalInd}) & Ethereum & - & \cite{arganaraz2020detection, lu2021neucheck}\\
    \hline
    Division by zero guarding 
        & Unchecked division by zero (\ref{P_dividebyzero})& Ethereum & S < 0.8.0 & \cite{so2020verismart, Khor2020improved, Osiris2018}\\ 
    \hline 
    Effective callback freeness & Reentrancy (\ref{P_effectivecallback}) & Ethereum & S
        &\cite{albert2020taming, grossman2017online}\\    
    \hline 
    \multirow{3}{40mm}{Exception \& return value handling$^\star$}    
        & Callstack depth issue (\ref{P_exceptions}) &   & & \cite{wang2020contractward}\\
        & Gasless send (\ref{P_exceptions}) &  Ethereum & - & \cite{nguyen2020sfuzz, teng2021SmartGift, jiang2018contractfuzzer} \\
        & Mishandled/unhandled exceptions (\ref{P_exceptions}) &  & & \cite{zhang2020smartshield, ashouri2020etherolic, arganaraz2020detection, nguyen2020sfuzz, wei2020smart, Oyente2016, tikhomirov2018smartcheck, Zhou2022, MANDO2022, Yang2022, Yao2022, DefectChecker2022}\cite{nelaturu2020verified,semantics2020Jiao,kalra2018zeus,grishchenko2018semantic}$^\star$\\  
    \hline 
    \multirow{2}{40mm}{Explicit access control modifiers usage}
        & Vulnerable access (\ref{P_explicitaccescontrol}) & Ethereum & S & \cite{lu2021neucheck} \\
        & Unspecified visibility level (\ref{P_explicitaccescontrol}) &  &  & \cite{lu2021neucheck, tikhomirov2018smartcheck} \\
    \hline
    Gas transfer viability 
        & Transfer of all the gas (\ref{P_gasrelated}) & Ethereum & S & \cite{arganaraz2020detection, tikhomirov2018smartcheck, aidee2021vulnerability}\\
    \hline
    Guarding of \texttt{delegatecall}$^\star$ 
        & Tainted/dangerous \texttt{delegatecall} (\ref{P_informationflow}) & Ethereum & S & \cite{brent2020ethainter, wei2020smart, teng2021SmartGift, andesta2020testing, jiang2018contractfuzzer, EtherGIS2020}\cite{semantics2020Jiao}$^\star$ \\  
    \hline
     Information secrecy 
        &  Exposed secret (\ref{P_informationflow}) & Ethereum & - & \cite{zhang2020ethploit, lu2021neucheck}\\
    \hline 
    Integer division guarding 
        & Unexpected results from divisions
        & Ethereum & S
        & \cite{arganaraz2020detection, tikhomirov2018smartcheck}\\
        \cline{3-5}
        &   (\ref{P_floatingpoint}) & Solana & Rust & \cite{Cui2022}\\
    \hline
    Integer over/underflow guarding$^\star$  
        &  Integer over/underflow  (\ref{P_integeroverunderflow}) & Ethereum & S < 0.8.0
        & \cite{ma2021pluto, ashouri2020etherolic, sun2020formal, so2020verismart, lai2020static, xing2020new, nguyen2020sfuzz, Khor2020improved, wang2020contractward, lu2021neucheck, andesta2020testing,mossberg2019manticore, Zhou2022, MANDO2022, wang2022gvd, Son2022, Yao2022, Liu2022, liu2022learning, zhang2022smart}\cite{semantics2020Jiao,stephens2021smartpulse,kalra2018zeus,yang2020hybrid}$^\star$ \\ 
        \cline{3-5}
        &  & H. Fabric & Golang & \cite{ding2021hfcontractfuzzer}\\
\cline{3-5}
        &  & Solana & Rust & \cite{Cui2022}\\   
    \hline
    \multirow{2}{45mm}{Isolation of external calls$^\star$}
        & DoS due to external contracts (\ref{P_isolationofcalls}) & Ethereum & - & \cite{samreen2021smartscan,tikhomirov2018smartcheck}\cite{stephens2021smartpulse}$^\star$\\
        &  Wallet griefing (\ref{P_isolationofcalls}) &   & & \cite{grech2020madmax} \\
    \hline
    Map disjointness 
        & Map overlap (\ref{P_hashcollision}) &  Ethereum & S & \cite{lu2021neucheck}\\
    \hline
    Memory references distinction 
    & Tainted or shadow memory (\ref{memoryref}) & Ethereum & S & \cite{ashouri2020etherolic}\\
    \hline 
    Ownership of assets 
        & Improper handling of assets (\ref{P_assetsownership})
        & Ethereum & V &\cite{bram2021rich} \\
    \hline
    Preservation of assets & Improper handling of assets (\ref{P_assetspreservation}) & Libra/Diem & Move & \cite{blackshear2020resources}\\
    \cline{3-5}
    & & H. Fabric & Obsidian & \cite{coblenz2020obsidian}\\
    \cline{3-5}
    & & Ethereum & V & \cite{bram2021rich} \\
    \hline
    \multirow{2}{45mm}{Preservation of assets in transactions$^\star$}
        & Unfair payment (\ref{P_assetsvalue}) & Ethereum & S & \cite{li2020protect} \\ 
        \cline{2-5}
        &  Double-spending (\ref{P_assetsvalue}) & Algorand & TEAL & \cite{bartoletti2021formal}$^\star$\\
    \hline
        Preservation of \texttt{selfdestruct}
        & Tainted \texttt{selfdestruct} (\ref{P_informationflow}) &  Ethereum & S & \cite{brent2020ethainter}
        \\
     \hline
    Preservation of state$^\star$  
        & Reentrancy (\ref{P_statepreservation}) & Ethereum & - &
        \cite{ma2021pluto, wustholz2020harvey, xue2020cross, chinen2020ra, schneidewind2020ethor, arganaraz2020detection, qian2020towards, ashouri2020etherolic, bang2020verification, li2020protect, samreen2020reentrancy, ji2021security, lu2021neucheck, britten2021using, wang2020contractward, wei2020smart, teng2021SmartGift, andesta2020testing, ribeiro2020formal, Oyente2016,tikhomirov2018smartcheck, mossberg2019manticore, Zhou2022, MANDO2022, SAILFish2022, EtherGIS2020, DefectChecker2022, YanXingrun2022, Yang2022, bram2021rich, zhang2022smart, zhang2022reentrancy, Yao2022}\cite{nelaturu2020verified,semantics2020Jiao,stephens2021smartpulse,ATL2022,grishchenko2018ethertrust,grishchenko2018semantic}$^\star$
        \\
     \hline
     \multirow{8}{40mm}{Preservation of owner information$^\star$} 
        & Locked amount (\ref{P_informationflow}) & Cardano & Marlowe & \cite{seijas2020efficient,Seijas2020Marlowe} \\ 
        \cline{2-5}
        & Deadlock state, Unprotected suicide (\ref{P_informationflow}) &   &  & \cite{nelaturu2020verified}$^\star$\\ 
        & Leaking Ether (\ref{P_informationflow}) &  &  & \cite{torres2021confuzzius, huang2021precise} \\
        & Tainted owner variable (\ref{P_informationflow}) &  &  & \cite{brent2020ethainter}\\
        & Unfair payment (\ref{P_informationflow}) & Ethereum & - & \cite{li2020protect} \\ 
        & Unprotected/accessible \texttt{selfdestruct}\ref{P_informationflow} &  &  & \cite{brent2020ethainter, aidee2021vulnerability, huang2021precise, EtherGIS2020} \\ 
        & Vulnerable access control (\ref{P_informationflow}) & & & \cite{zhang2020ethploit}\cite{kalra2018zeus}$^\star$ \\
        & Unprotected withdrawal (\ref{P_informationflow}) &  &  & \cite{zhang2020ethploit, tsankov2018securify}\cite{stephens2021smartpulse}$^\star$ \\
         \cline{2-5}
         & Vulnerable access control (\ref{P_informationflow}) & Solana & Rust & \cite{Cui2022}\\
    \hline
    Preservation of \texttt{staticcall} 
        & Problematic \texttt{staticcall} (\ref{P_informationflow}) & Ethereum &  S & \cite{brent2020ethainter}\\
    \hline
    State variables declaration ordering
        &  Improper variable declaration (\ref{P_statevariables}) & Ethereum & S & \cite{praitheeshan2021solguard}\\
    \hline  
    Transaction cost boundedness
        &  DoS from expensive transactions (\ref{P_transactionCost}) & Ethereum &  - & \cite{chen2020gaschecker, li2022smartfast, SmartMixModel2022}\\
    \hline
    Transaction order independence$^\star$  
        &  Transaction ordering dependency (\ref{P_transactionsordering}) & Ethereum & - &
        \cite{torres2021confuzzius, ashouri2020etherolic, wang2020contractward, Oyente2016,tsankov2018securify, MANDO2022, SAILFish2022, EtherGIS2020}\cite{kalra2018zeus,grishchenko2018semantic}$^\star$\\
    \hline
    Transfer amount boundedness  
        & Unrestricted amount transfer (\ref{P_amountBoundedness}) & Cardano & Marlowe
        & \cite{seijas2020efficient, Seijas2020Marlowe} \\
    \hline
    \multirow{2}{45mm}{Transfer function viability}
        & Missed to implement fallback (\ref{P_transferfunctionviability}) &  Ethereum & S & \cite{praitheeshan2021solguard}\\
        &  Use of send instead of transfer (\ref{P_transferfunctionviability}) &  & & \cite{arganaraz2020detection, tikhomirov2018smartcheck} \\
    \hline 
    \multirow{2}{45mm}{Validity of deposits}
        & Fake deposit (\ref{P_depositValidity}) & Ethereum & S & \cite{ji2020deposafe} \\
        \cline{2-5}
        & Fake EOS (\ref{P_depositValidity}) & EOSIO & - & \cite{huang2020eosfuzzer, he2021eosafe, li2022eosioanalyzer, Wasai2022, Li2022} \\
    \hline 
    Validity of transfer amount$^\star$ 
       & Unchecked send bugs (\ref{P_transfervalue}) &  Ethereum & S & \cite{yang2020hybrid, stephens2021smartpulse}$^\star$\\
       & Unlimited illicit gains (\ref{P_transfervalue}) & & &  \cite{zhang2020ethploit}\\
       
    \hline 
        \multirow{2}{45mm}{Validity of transfer address$^\star$}
        & Misuse of \texttt{msg.sender} (\ref{P_contractaddress})&   & & \cite{zhdarkin2021development}\\
        & Misuse of \texttt{tx.origin} (\ref{P_contractaddress}) & Ethereum & - & \cite{lu2021neucheck, arganaraz2020detection, tikhomirov2018smartcheck, brent2018vandal, Son2022, EtherGIS2020, DefectChecker2022, Yao2022}\cite{kalra2018zeus,ATL2022}$^\star$\\
        & Prodigal contract (\ref{P_contractaddress}) & & & \cite{praitheeshan2019security, tang2021vulnerabilities}\\
    \hline
    
    Value unpredictability$^\star$  
        & Dependence on predictable variables  &  Ethereum & S & \cite{lu2021neucheck, ma2021pluto, nguyen2020sfuzz, wang2020contractward,ashouri2020etherolic, wei2020smart, teng2021SmartGift, Oyente2016, Zhou2022, MANDO2022, EtherGIS2020, DefectChecker2022, Yang2022}\cite{kalra2018zeus,grishchenko2018ethertrust,grishchenko2018semantic}$^\star$\\
    \cline{3-5}
        & (\ref{P_predictable}) & EOSIO & - & \cite{li2022eosioanalyzer, Li2022, Wasai2022}\\
    \hline
\end{tabular}
\end{table}

%% file: T5_PropertiesStaticAnalysis.tex
\begin{table*}
\tiny
\setlength\abovecaptionskip{-0.1\baselineskip}
\setlength\belowcaptionskip{-0.1\baselineskip}
\caption{Properties addressed in the papers adopting static analysis methods. Liveness properties are appended with $^\ast$.}
\label{tab:SAmethods}
\begin{tabular}{
| l |p{9mm} | p{10mm} |p{11mm} | p{17mm} | p{22mm} | p{23mm} |}
    \hline
    \textbf{Property} & 
    \textbf{Static Type Checking} &
    \textbf{Abstract Interpretation} &
    \textbf{Control Flow Analysis} &
    \textbf{Taint (Dataflow) Analysis}  &
    \textbf{Symbolic Execution} &
    \textbf{Pattern-based Analysis}
     \\
    \hline
    Preservation of assets  
        & \cite{10.1007/978-3-030-54455-3_41, das2021resource, coblenz2020obsidian}
        &
        &
        &
        &
        & \\
    \hline
    Information secrecy  
        & \cite{zkay1.0, baumann2020zkay}
        &
        & \cite{zhang2020ethploit}
        & \cite{zhang2020ethploit}
        &
        & \cite{lu2021neucheck}
         \\  
    \hline 
    Preservation of owner info.  
        & \cite{zkay1.0, baumann2020zkay}
        & \cite{tsankov2018securify, smaragdakis2021symbolic}
        & \cite{brent2020ethainter}
        & \cite{brent2020ethainter}
        & \cite{smaragdakis2021symbolic, seijas2020efficient, Seijas2020Marlowe, he2022tokencat}
        & \cite{tsankov2018securify, Cui2022}
        \\
    \hline
    Preservation of state 
        &  \cite{das2021resource}
        & \cite{schneidewind2020ethor, tsankov2018securify, smaragdakis2021symbolic}
        & \cite{xue2020cross, brent2018vandal}
        & \cite{feist2019slither, NPChecker2019, li2022smartfast, xue2020cross, ali2021sescon, SAILFish2022}
        & \cite{chinen2020ra, smaragdakis2021symbolic, ma2021pluto, Oyente2016, MPro2019, Liu2022, DefectChecker2022}
        & \cite{li2022smartfast, lu2021neucheck, arganaraz2020detection, tikhomirov2018smartcheck, tsankov2018securify, brent2018vandal, SolChecker2022, ali2021sescon} 
        \\
    \hline
    Argument precondition valid.
        &
        & \cite{tsankov2018securify} 
        & \cite{tsankov2018securify}
        &
        &
        & \cite{tsankov2018securify} 
        \\
    \hline  
    Isolation of external calls
        & 
        & \cite{grech2020madmax} 
        & \cite{grech2020madmax}
        &
        &
        & \cite{tikhomirov2018smartcheck, samreen2021smartscan} 
        \\
    \hline
    Boundedness of loops
        & 
        & \cite{grech2020madmax} 
        & \cite{nassirzadeh2021gas}
        &
        & \cite{seijas2020efficient, Seijas2020Marlowe}
        & \cite{lu2021neucheck}
        \\
    \hline
    Exception handling  
        & 
        & \cite{tsankov2018securify, smaragdakis2021symbolic}
        & \cite{brent2018vandal} 
        & \cite{NPChecker2019, ali2021sescon, Mythril}
        & \cite{smaragdakis2021symbolic, ma2021pluto, Oyente2016, Mythril, Liu2022, DefectChecker2022}
        & \cite{lu2021neucheck, arganaraz2020detection, tikhomirov2018smartcheck, tsankov2018securify, brent2018vandal, SolChecker2022, ali2021sescon}
        \\
    \hline
    Guarding of \texttt{selfdestruct} 
        & 
        & \cite{smaragdakis2021symbolic}
        & \cite{brent2018vandal, zhang2020ethploit}
        & \cite{zhang2020ethploit, ali2021sescon, Mythril}
        & \cite{smaragdakis2021symbolic, Mythril, nikolic2018finding}
        & \cite{aidee2021vulnerability, brent2018vandal, ali2021sescon}
        \\
    \hline
    Integer over/underflow guard. 
        & 
        & \cite {Michelson2022}
        & \cite{peng2019sif}
        & \cite{ali2021sescon, Osiris2018, Mythril} 
        & \cite{permenev2020verx, Osiris2018, Mythril, Liu2022}
        & \cite{lai2020static, lu2021neucheck, arganaraz2020detection, tikhomirov2018smartcheck, Cui2022}
        \\
    \hline
    Transaction order independ.
        & 
        & \cite{tsankov2018securify}
        & \cite{tsankov2018securify}
        & \cite{NPChecker2019, ali2021sescon, SAILFish2022}
        & \cite{Oyente2016}
        & \cite{tsankov2018securify, ali2021sescon}
        \\
    \hline
    Withdrawal acceptance$^\ast$
        & 
        & \cite{tsankov2018securify}
        & \cite{tsankov2018securify}
        & \cite{li2022smartfast}
        & \cite{nikolic2018finding, DefectChecker2022}
        & \cite{li2022smartfast, lu2021neucheck, arganaraz2020detection, 
        tikhomirov2018smartcheck, tsankov2018securify}
        \\
  \hline
      Preservation of \texttt{selfdestruct} \& \texttt{staticcall}
        & 
        & 
        & \cite{brent2020ethainter} 
        & \cite{brent2020ethainter}
        &
        & 
        \\
    \hline
    Validity of transfer amount  
        &
        & 
        & \cite{zhang2020ethploit}
        & \cite{zhang2020ethploit}
        & 
        & 
        \\
    \hline
    Guarding of \texttt{delegatecall}
        & 
        &
        & \cite{brent2020ethainter} 
        & \cite{brent2020ethainter, ali2021sescon}
        &
        & \cite{ali2021sescon}
        \\
    \hline
    Division by zero guarding 
        &
        & 
        & \cite{peng2019sif}
        & \cite{Osiris2018}
        & \cite{Osiris2018}
        & \cite{zhou2018security}
        \\
    \hline
    Validity of transfer address
        & 
        &
        & \cite{brent2018vandal}
        & \cite{li2022smartfast}
        & \cite{nikolic2018finding, zhdarkin2021development, DefectChecker2022}
        & \cite{li2022smartfast, lu2021neucheck, arganaraz2020detection, tikhomirov2018smartcheck, brent2018vandal, zhou2018security}
        \\
    \hline
    Balance comparison  
        &
        &
        &
        & \cite{ali2021sescon}
        & \cite{DefectChecker2022}
        & \cite{lu2021neucheck, arganaraz2020detection, tikhomirov2018smartcheck, SolChecker2022, ali2021sescon}
        \\
    \hline
    Value unpredictability 
        &
        &
        &
        & \cite{ali2021sescon, li2022smartfast,Mythril}
        & \cite{yang2020seraph, ma2021pluto, Oyente2016, Mythril, DefectChecker2022}
        & \cite{li2022smartfast, li2022eosioanalyzer, lu2021neucheck, arganaraz2020detection, tikhomirov2018smartcheck, zhou2018security, SolChecker2022, ali2021sescon}
         \\
    \hline
    Eventual contract termination$^\ast$ 
        & 
        & 
        & 
        & 
        & \cite{seijas2020efficient, Seijas2020Marlowe}
        & 
        \\
    \hline
    Address integrity 
        &
        &
        &
        &  
        & \cite{he2021eosafe}
        & \cite{li2022eosioanalyzer} 
        \\
    \hline
    Validity of deposits 
        & 
        &
        &
        &
        & \cite{he2021eosafe, ji2020deposafe}
        & \cite{li2022eosioanalyzer} 
        \\
    \hline
    Address parameter av. \&  State variable decl.
        &
        &
        &
        &
        &
        & \cite{praitheeshan2021solguard} 
        \\
    \hline
    Conditional independence
        &
        &
        &
        &
        &
        & \cite{lu2021neucheck, arganaraz2020detection}
        \\
    \hline 
    Explicit access control modifier. 
        &
        &
        &  
        &
        &
        & \cite{lu2021neucheck, tikhomirov2018smartcheck} 
        \\
    \hline
    Gas transfer viability
        &
        &
        &
        & 
        &
        & \cite{arganaraz2020detection, tikhomirov2018smartcheck, aidee2021vulnerability}
        \\
    \hline
    Integer division guarding
        &
        &
        &
        &
        &
        & \cite{arganaraz2020detection, tikhomirov2018smartcheck, Cui2022} 
         \\
    \hline    
    Map Disjointness  
        &
        &
        &
        & 
        &
        & \cite{lu2021neucheck}
        \\
    \hline
    Transfer function viability  
        &
        &
        &
        & 
        &
        & \cite{praitheeshan2021solguard, arganaraz2020detection, tikhomirov2018smartcheck}
        \\
    \hline
\end{tabular}
\end{table*}

%% file: T6_PropertiesFormalMethods.tex
\begin{wraptable}{r}{7.7cm}
 \tiny
\setlength\abovecaptionskip{-0.1\baselineskip}
\setlength\belowcaptionskip{-0.1\baselineskip}
\caption{Properties addressed in the papers adopting formal verification methods. Liveness properties are appended with $^\ast$.}
\label{tab:FVmethods}
\begin{tabular}{ | l | l | l |}
\hline
    \textbf{Property} & 
    \textbf{Theorem Proving} &
    \textbf{Model Checking} \\
    \hline
    
    Contract-specific invariants
        & \cite{Mi-Cho-Coq2020, da2020whylson}
        & \\
    \hline

    Deposit acceptance$^\ast$
        & \cite{park2020end}
        & \\
    \hline
        
     Eventual contract termination$^\ast$ 
        & \cite{genet2020termination}
        & \\
    \hline
    Preservation of assets in transactions
        &  \cite{bartoletti2021formal}
        & \\
    \hline    
    
    Exception \& return value handling  
        &  \cite{grishchenko2018semantic}
        & \cite{nelaturu2020verified, kalra2018zeus}
        \\
    \hline
 
    Functional annotations \& Assertions
        & \cite{zhong2020move, beckert2018formal, da2020whylson, 9284726}
        & \cite{SAFEVM2019, duo2020formal, marescotti2020accurate, hajdu2019solc, hajdu2020formal, SolCMC2022, Son2022, wang2019formal}
        \\
    \hline
    
    Integer over/underflow guarding  
        & \cite{sun2020formal, yang2020hybrid, FSPVME2019, Daejun2018FormalVerification}
        & \cite{kalra2018zeus, stephens2021smartpulse, hajdu2019solc, hajdu2020formal, SolCMC2022, Son2022}
        \\
    \hline
    
    Preservation of state  
        & \cite{britten2021using, ahrendt2020functional, grishchenko2018semantic, SPARK2019} 
        & \cite{kalra2018zeus, nelaturu2020verified, duo2020formal, ATL2022, hajdu2019solc, hajdu2020formal, grishchenko2018ethertrust, grishchenko2018foundations, SolCMC2022}
        \\
    \hline

    Transaction order independence 
        & \cite{grishchenko2018semantic}
        & \cite{kalra2018zeus, duo2020formal}
        \\
    \hline 

    Validity of transfer amount  
        &  \cite{yang2020hybrid, FSPVME2019}
        & \cite{stephens2021smartpulse}
        \\
    \hline
    
    Value unpredictability  
        & \cite{grishchenko2018semantic}
        & \cite{kalra2018zeus, duo2020formal, grishchenko2018ethertrust, grishchenko2018foundations}
        \\
    \hline   
    
     Address size soundness
        &
        & \cite{nelaturu2020verified}
        \\
    \hline    

    Eventual contract removal$^\ast$
      &
      & \cite{nelaturu2020verified}
    \\
    \hline
    
    Guarding of \texttt{selfdestruct}
        &
        & \cite{kalra2018zeus, nelaturu2020verified}
        \\
    \hline    

    Isolation of external calls 
        &
        & \cite{stephens2021smartpulse}
        \\
    \hline
    
    Validity of transfer address  
        &
        & \cite{kalra2018zeus, ATL2022, Son2022}
         \\
    \hline 
    
    Withdrawal acceptance$^\ast$ 
        &
        & \cite{nelaturu2020verified, stephens2021smartpulse} \\ 
    \hline
\end{tabular}
\end{wraptable}

%% file: T7_PropertiesDynamicAnalysis.tex
\begin{wraptable}{r}{8.5cm}
\tiny
\setlength\abovecaptionskip{-0.1\baselineskip}
\setlength\belowcaptionskip{-0.1\baselineskip}
\caption{Properties addressed in the papers adopting dynamic analysis methods. Liveness properties are appended with $^\ast$.}
\label{tab:DAmethods}
\begin{tabular}{| l | l | l | }
\hline
    \textbf{Property} & 
    \textbf{Concolic Testing} &
    \textbf{Fuzzing}  \\
    \hline  
    Contract-specific invariants
        & \cite{SOLAR2018LiAo}
        & \\
    \hline
    Preservation of assets in transactions
    & \cite{li2020protect}
    &
    \\
    \hline
    Validity of transfer address
        & \cite{SKLEE2022}
        & \\
    \hline
    Assertion satisfaction
        & \cite{Annotary2019}
        & \cite{torres2021confuzzius, grieco2020echidna}\\
    \hline
    Exception \& return value handling  
        & \cite{SKLEE2022}
        & \cite{torres2021confuzzius, teng2021SmartGift, nguyen2020sfuzz, wang2020oracle, wei2020smart, jiang2018contractfuzzer, 8859497, ILF2019}\\
    \hline
    Guarding of \texttt{selfdestruct}  
        & \cite{huang2021precise, SKLEE2022}
        & \cite{torres2021confuzzius}\\   
    \hline
    Integer over/underflow guarding  
        & \cite{mossberg2019manticore, huang2021precise, SKLEE2022}
        & \cite{torres2021confuzzius, teng2021SmartGift, nguyen2020sfuzz, wang2020oracle, wei2020smart, Soliaudit2019, ding2021hfcontractfuzzer} \\  
    \hline
    Preservation of state  
        & \cite{wang2021mar, li2020protect, mossberg2019manticore, SKLEE2022}
        &  \cite{wustholz2020harvey,torres2021confuzzius, nguyen2020sfuzz,teng2021SmartGift,jiang2018contractfuzzer, 8859497, wei2020smart, wang2020oracle, Soliaudit2019}\\
    \hline
      Transaction order independence 
        & \cite{SKLEE2022}
        & \cite{torres2021confuzzius}\\
    \hline
    Value unpredictability
        &  \cite{SKLEE2022, huang2021precise}
        & \cite{torres2021confuzzius, teng2021SmartGift, nguyen2020sfuzz, wang2020oracle, wei2020smart, jiang2018contractfuzzer, 8859497, huang2020eosfuzzer, Wasai2022, ILF2019, Li2022}\\
    \hline  
     Withdrawal acceptance$^\ast$ 
        & \cite{huang2021precise, SKLEE2022}
        & \cite{ILF2019, torres2021confuzzius, wei2020smart, wang2020oracle, jiang2018contractfuzzer, 8859497} \\
    \hline
    Address integrity 
        & 
        & \cite{huang2020eosfuzzer, Wasai2022, Li2022} \\
    \hline
    Owner information preservation
        & 
        & \cite{ILF2019, torres2021confuzzius} \\ 
    \hline
    Guarding of \texttt{delegatecall}
        & 
        & \cite{jiang2018contractfuzzer, 8859497, torres2021confuzzius, wei2020smart, wang2020oracle, teng2021SmartGift, ILF2019} \\
      \hline  
     Validity of deposits
        & 
        & \cite{huang2020eosfuzzer, Wasai2022, Li2022}\\  
\hline
\end{tabular}
\end{wraptable}

%% file: T8_PropertiesMachineLearning.tex
\begin{wraptable}{r}{10cm}
\tiny
\setlength\abovecaptionskip{-0.1\baselineskip}
\setlength\belowcaptionskip{-0.1\baselineskip}
\caption{Properties addressed in the papers adopting machine learning methods. Liveness properties are appended with $^\ast$.}
\label{tab:MLmethods}
\begin{tabular}{|l | l | l |}
    \hline
    \textbf{Property} & 
    \textbf{Classification Algos} & 
    \textbf{Neural Network Algorithms} \\
    \hline
     Guarding of \texttt{selfdestruct}  
        & \cite{MLmodel2019}
        & \cite{EtherGIS2020}\\
    \hline
    Boundedness of loops
        & \cite{xu2021novel}
        & \cite{zhuang2020smart, liu2021combining, narayana2021automation}
        \\
    \hline
    Exception handling
        & \cite{wang2020contractward, xu2021novel}
        & \cite{SVChecker2022, Yao2022, Yang2022, MANDO2022, Zhou2022}
         \\
    \hline 
    Integer over/underflow guarding
        & \cite{MLmodel2019, wang2020contractward, xu2021novel, hao2020scscan, wang2022gvd}
        & \cite{ashizawa2021eth2vec, xing2020new, SVChecker2022, liu2022learning, zhang2022smart,Yao2022, MANDO2022, Zhou2022}
        \\
     \hline
    Preservation of state
        & \cite{wang2020contractward, xu2021novel, MLmodel2019, hao2020scscan, YanXingrun2022}
        & \cite{ashizawa2021eth2vec, zhuang2020smart, liu2021combining, qian2020towards, mi2021vscl, SVChecker2022, HAMBiLSTM2022, zhang2022smart, zhang2022reentrancy, Yao2022, Yang2022, EtherGIS2020, MANDO2022, Zhou2022}\\
     \hline
    Transaction order independence
        & \cite{wang2020contractward, xu2021novel}
        & \cite{SVChecker2022, Yao2022, EtherGIS2020, MANDO2022}
         \\
    \hline
    Transfer function viability
        & \cite{hao2020scscan}
        & \cite{GUPTA2021107583, narayana2021automation}
       \\
    \hline
    Value unpredictability
        & \cite{wang2020contractward, xu2021novel, hao2020scscan}
        & \cite{ashizawa2021eth2vec, zhuang2020smart, liu2021combining, zhang2022smart,Yao2022, Yang2022, EtherGIS2020, MANDO2022, Zhou2022}
        \\
     \hline
    Address size soundness
        & 
        & \cite{xing2020new}
        \\
    \hline
    Balance comp. \& Transaction cost bound.
        &
        & \cite{SmartMixModel2022}
    \\  
    \hline
    Explicit access cont. \& Validity of deposits 
        &
        & \cite{ashizawa2021eth2vec}
        \\
    \hline 
      Guarding of \texttt{delegatecall}
        & 
        & \cite{zhang2022smart, EtherGIS2020} \\
      \hline 
        Validity of transfer address
        &
        & \cite{GUPTA2021107583, narayana2021automation, SmartMixModel2022,Yao2022, EtherGIS2020}
         \\
    \hline
    Withdrawal acceptance$^\ast$ 
        &
        & \cite{SmartMixModel2022}
    \\  
    \hline
\end{tabular}
\end{wraptable}

%% file: manuscript.bbl

\begin{thebibliography}{268}


\ifx \showCODEN    \undefined \def \showCODEN     #1{\unskip}     \fi
\ifx \showDOI      \undefined \def \showDOI       #1{#1}\fi
\ifx \showISBNx    \undefined \def \showISBNx     #1{\unskip}     \fi
\ifx \showISBNxiii \undefined \def \showISBNxiii  #1{\unskip}     \fi
\ifx \showISSN     \undefined \def \showISSN      #1{\unskip}     \fi
\ifx \showLCCN     \undefined \def \showLCCN      #1{\unskip}     \fi
\ifx \shownote     \undefined \def \shownote      #1{#1}          \fi
\ifx \showarticletitle \undefined \def \showarticletitle #1{#1}   \fi
\ifx \showURL      \undefined \def \showURL       {\relax}        \fi
\providecommand\bibfield[2]{#2}
\providecommand\bibinfo[2]{#2}
\providecommand\natexlab[1]{#1}
\providecommand\showeprint[2][]{arXiv:#2}

\bibitem[\protect\citeauthoryear{??}{ama}{[n. d.]}]%
        {amadini2020abstract}
 \bibinfo{year}{[n. d.]}\natexlab{}.
\newblock


\bibitem[\protect\citeauthoryear{Agrawal, DeMichiel, et~al\mbox{.}}{Agrawal
  et~al\mbox{.}}{1991}]%
        {agrawal1991static}
\bibfield{author}{\bibinfo{person}{Rakesh Agrawal}, \bibinfo{person}{Linda~G
  DeMichiel}, {et~al\mbox{.}}} \bibinfo{year}{1991}\natexlab{}.
\newblock \showarticletitle{Static type checking of multi-methods}.
\newblock \bibinfo{journal}{\emph{ACM SIGPLAN Notices}} \bibinfo{volume}{26},
  \bibinfo{number}{11} (\bibinfo{year}{1991}), \bibinfo{pages}{113--128}.
\newblock


\bibitem[\protect\citeauthoryear{Ahrendt and Bubel}{Ahrendt and Bubel}{2020}]%
        {ahrendt2020functional}
\bibfield{author}{\bibinfo{person}{Wolfgang Ahrendt} {and}
  \bibinfo{person}{Richard Bubel}.} \bibinfo{year}{2020}\natexlab{}.
\newblock \showarticletitle{Functional Verification of Smart Contracts via
  Strong Data Integrity}. In \bibinfo{booktitle}{\emph{Leveraging Applications
  of Formal Methods, Verification and Validation: Applications}}.
  \bibinfo{publisher}{Springer International Publishing},
  \bibinfo{address}{Cham}, \bibinfo{pages}{9--24}.
\newblock
\showISBNx{978-3-030-61467-6}


\bibitem[\protect\citeauthoryear{Aiello, Kanig, et~al\mbox{.}}{Aiello
  et~al\mbox{.}}{2020}]%
        {SPARK2019}
\bibfield{author}{\bibinfo{person}{M.~Anthony Aiello},
  \bibinfo{person}{Johannes Kanig}, {et~al\mbox{.}}}
  \bibinfo{year}{2020}\natexlab{}.
\newblock \showarticletitle{Call Me Back, I Have a Type Invariant}. In
  \bibinfo{booktitle}{\emph{Formal Methods. FM 2019 International Workshops}}.
  \bibinfo{publisher}{Springer International Publishing},
  \bibinfo{address}{Cham}, \bibinfo{pages}{325--336}.
\newblock
\showISBNx{978-3-030-54994-7}


\bibitem[\protect\citeauthoryear{Albert, Correas, et~al\mbox{.}}{Albert
  et~al\mbox{.}}{2019}]%
        {SAFEVM2019}
\bibfield{author}{\bibinfo{person}{Elvira Albert}, \bibinfo{person}{Jes\'{u}s
  Correas}, {et~al\mbox{.}}} \bibinfo{year}{2019}\natexlab{}.
\newblock \showarticletitle{SAFEVM: A Safety Verifier for Ethereum Smart
  Contracts}. In \bibinfo{booktitle}{\emph{Proceedings of the 28th ACM SIGSOFT
  International Symposium on Software Testing and Analysis}}.
  \bibinfo{publisher}{Association for Computing Machinery},
  \bibinfo{address}{New York, NY, USA}, \bibinfo{pages}{386–389}.
\newblock
\showISBNx{9781450362245}


\bibitem[\protect\citeauthoryear{Albert, Correas, et~al\mbox{.}}{Albert
  et~al\mbox{.}}{2021}]%
        {albert2021don}
\bibfield{author}{\bibinfo{person}{Elvira Albert}, \bibinfo{person}{Jes{\'u}s
  Correas}, {et~al\mbox{.}}} \bibinfo{year}{2021}\natexlab{}.
\newblock \showarticletitle{Don’t run on fumes—parametric gas bounds for
  smart contracts}.
\newblock \bibinfo{journal}{\emph{Journal of Systems and Software}}
  \bibinfo{volume}{176} (\bibinfo{year}{2021}), \bibinfo{pages}{110923}.
\newblock


\bibitem[\protect\citeauthoryear{Albert, Grossman, et~al\mbox{.}}{Albert
  et~al\mbox{.}}{2020}]%
        {albert2020taming}
\bibfield{author}{\bibinfo{person}{Elvira Albert}, \bibinfo{person}{Shelly
  Grossman}, {et~al\mbox{.}}} \bibinfo{year}{2020}\natexlab{}.
\newblock \showarticletitle{Taming callbacks for smart contract modularity}.
\newblock \bibinfo{journal}{\emph{Proceedings of the ACM on Programming
  Languages}} \bibinfo{volume}{4}, \bibinfo{number}{OOPSLA}
  (\bibinfo{year}{2020}), \bibinfo{pages}{1--30}.
\newblock


\bibitem[\protect\citeauthoryear{Alharby, Aldweesh, et~al\mbox{.}}{Alharby
  et~al\mbox{.}}{2018}]%
        {alharby2018blockchain}
\bibfield{author}{\bibinfo{person}{Maher Alharby}, \bibinfo{person}{Amjad
  Aldweesh}, {et~al\mbox{.}}} \bibinfo{year}{2018}\natexlab{}.
\newblock \showarticletitle{Blockchain-based smart contracts: A systematic
  mapping study of academic research (2018)}. In \bibinfo{booktitle}{\emph{2018
  International Conference on Cloud Computing, Big Data and Blockchain
  (ICCBB)}}. IEEE, \bibinfo{pages}{1--6}.
\newblock


\bibitem[\protect\citeauthoryear{Ali, Abideen, et~al\mbox{.}}{Ali
  et~al\mbox{.}}{2021}]%
        {ali2021sescon}
\bibfield{author}{\bibinfo{person}{Amir Ali}, \bibinfo{person}{Zain~Ul
  Abideen}, {et~al\mbox{.}}} \bibinfo{year}{2021}\natexlab{}.
\newblock \showarticletitle{SESCon: Secure Ethereum Smart Contracts by
  Vulnerable Patterns’ Detection}.
\newblock \bibinfo{journal}{\emph{Security and Communication Networks}}
  \bibinfo{volume}{2021} (\bibinfo{year}{2021}).
\newblock


\bibitem[\protect\citeauthoryear{Allen}{Allen}{1970}]%
        {allen1970control}
\bibfield{author}{\bibinfo{person}{Frances~E Allen}.}
  \bibinfo{year}{1970}\natexlab{}.
\newblock \showarticletitle{Control flow analysis}.
\newblock \bibinfo{journal}{\emph{ACM Sigplan Notices}} \bibinfo{volume}{5},
  \bibinfo{number}{7} (\bibinfo{year}{1970}), \bibinfo{pages}{1--19}.
\newblock


\bibitem[\protect\citeauthoryear{Almakhour, Sliman, et~al\mbox{.}}{Almakhour
  et~al\mbox{.}}{2020}]%
        {almakhour2020verification}
\bibfield{author}{\bibinfo{person}{Mouhamad Almakhour}, \bibinfo{person}{Layth
  Sliman}, {et~al\mbox{.}}} \bibinfo{year}{2020}\natexlab{}.
\newblock \showarticletitle{On the Verification of Smart Contracts: A
  Systematic Review}. In \bibinfo{booktitle}{\emph{International Conference on
  Blockchain}}. Springer, \bibinfo{pages}{94--107}.
\newblock


\bibitem[\protect\citeauthoryear{Alpern and Schneider}{Alpern and
  Schneider}{1987}]%
        {alpern1987recognizing}
\bibfield{author}{\bibinfo{person}{Bowen Alpern} {and} \bibinfo{person}{Fred~B
  Schneider}.} \bibinfo{year}{1987}\natexlab{}.
\newblock \showarticletitle{Recognizing safety and liveness}.
\newblock \bibinfo{journal}{\emph{Distributed computing}} \bibinfo{volume}{2},
  \bibinfo{number}{3} (\bibinfo{year}{1987}), \bibinfo{pages}{117--126}.
\newblock


\bibitem[\protect\citeauthoryear{Alt, Blicha, et~al\mbox{.}}{Alt
  et~al\mbox{.}}{2022}]%
        {SolCMC2022}
\bibfield{author}{\bibinfo{person}{Leonardo Alt}, \bibinfo{person}{Martin
  Blicha}, {et~al\mbox{.}}} \bibinfo{year}{2022}\natexlab{}.
\newblock \showarticletitle{SolCMC: Solidity Compiler's Model Checker}. In
  \bibinfo{booktitle}{\emph{Computer Aided Verification}},
  \bibfield{editor}{\bibinfo{person}{Sharon Shoham} {and}
  \bibinfo{person}{Yakir Vizel}} (Eds.). \bibinfo{publisher}{Springer
  International Publishing}, \bibinfo{address}{Cham},
  \bibinfo{pages}{325--338}.
\newblock
\showISBNx{978-3-031-13185-1}


\bibitem[\protect\citeauthoryear{Ammin}{Ammin}{2022}]%
        {introSC}
\bibfield{author}{\bibinfo{person}{Samaj Ammin}.}
  \bibinfo{year}{2022}\natexlab{}.
\newblock \bibinfo{title}{INTRODUCTION TO SMART CONTRACTS}.
\newblock
\newblock
\urldef\tempurl%
\url{https://ethereum.org/en/developers/docs/smart-contracts/}
\showURL{%
Retrieved April 26, 2022 from \tempurl}


\bibitem[\protect\citeauthoryear{Andesta, Faghih, et~al\mbox{.}}{Andesta
  et~al\mbox{.}}{2020}]%
        {andesta2020testing}
\bibfield{author}{\bibinfo{person}{Erfan Andesta}, \bibinfo{person}{Fathiyeh
  Faghih}, {et~al\mbox{.}}} \bibinfo{year}{2020}\natexlab{}.
\newblock \showarticletitle{Testing smart contracts gets smarter}. In
  \bibinfo{booktitle}{\emph{2020 10th International Conference on Computer and
  Knowledge Engineering (ICCKE)}}. IEEE, \bibinfo{pages}{405--412}.
\newblock


\bibitem[\protect\citeauthoryear{Annenkov, Milo, et~al\mbox{.}}{Annenkov
  et~al\mbox{.}}{2022}]%
        {annenkov2022extracting}
\bibfield{author}{\bibinfo{person}{Danil Annenkov}, \bibinfo{person}{Mikkel
  Milo}, {et~al\mbox{.}}} \bibinfo{year}{2022}\natexlab{}.
\newblock \showarticletitle{Extracting functional programs from Coq, in Coq}.
\newblock \bibinfo{journal}{\emph{Journal of Functional Programming}}
  \bibinfo{volume}{32} (\bibinfo{year}{2022}), \bibinfo{pages}{e11}.
\newblock


\bibitem[\protect\citeauthoryear{Annenkov, Nielsen, et~al\mbox{.}}{Annenkov
  et~al\mbox{.}}{2020}]%
        {annenkov2020concert}
\bibfield{author}{\bibinfo{person}{Danil Annenkov},
  \bibinfo{person}{Jakob~Botsch Nielsen}, {et~al\mbox{.}}}
  \bibinfo{year}{2020}\natexlab{}.
\newblock \showarticletitle{ConCert: a smart contract certification framework
  in Coq}. In \bibinfo{booktitle}{\emph{Proceedings of the 9th ACM SIGPLAN
  International Conference on Certified Programs and Proofs}}.
  \bibinfo{pages}{215--228}.
\newblock


\bibitem[\protect\citeauthoryear{Antonino and Roscoe}{Antonino and
  Roscoe}{2020}]%
        {antonino2020formalising}
\bibfield{author}{\bibinfo{person}{Pedro Antonino} {and} \bibinfo{person}{AW
  Roscoe}.} \bibinfo{year}{2020}\natexlab{}.
\newblock \showarticletitle{Formalising and verifying smart contracts with
  Solidifier: a bounded model checker for Solidity}.
\newblock  (\bibinfo{year}{2020}).
\newblock
\showeprint[arxiv]{2002.02710}


\bibitem[\protect\citeauthoryear{Arga{\~n}araz, Ber{\'o}n,
  et~al\mbox{.}}{Arga{\~n}araz et~al\mbox{.}}{2020}]%
        {arganaraz2020detection}
\bibfield{author}{\bibinfo{person}{Mauro Arga{\~n}araz}, \bibinfo{person}{Mario
  Ber{\'o}n}, {et~al\mbox{.}}} \bibinfo{year}{2020}\natexlab{}.
\newblock \showarticletitle{Detection of vulnerabilities in smart contracts
  specifications in ethereum platforms}. In \bibinfo{booktitle}{\emph{9th
  Symposium on Languages, Applications and Technologies (SLATE 2020)}},
  Vol.~\bibinfo{volume}{83}. \bibinfo{pages}{1--16}.
\newblock


\bibitem[\protect\citeauthoryear{Arrojado~da Horta, Santos~Reis,
  et~al\mbox{.}}{Arrojado~da Horta et~al\mbox{.}}{2020}]%
        {9284726}
\bibfield{author}{\bibinfo{person}{Luís~Pedro Arrojado~da Horta},
  \bibinfo{person}{João Santos~Reis}, {et~al\mbox{.}}}
  \bibinfo{year}{2020}\natexlab{}.
\newblock \showarticletitle{A tool for proving Michelson Smart Contracts in
  WHY3}. In \bibinfo{booktitle}{\emph{2020 IEEE International Conference on
  Blockchain (Blockchain)}}. \bibinfo{pages}{409--414}.
\newblock


\bibitem[\protect\citeauthoryear{Arusoaie}{Arusoaie}{2021}]%
        {arusoaie2021certifying}
\bibfield{author}{\bibinfo{person}{Andrei Arusoaie}.}
  \bibinfo{year}{2021}\natexlab{}.
\newblock \showarticletitle{Certifying Findel derivatives for blockchain}.
\newblock \bibinfo{journal}{\emph{Journal of Logical and Algebraic Methods in
  Programming}}  \bibinfo{volume}{121} (\bibinfo{year}{2021}),
  \bibinfo{pages}{100665}.
\newblock


\bibitem[\protect\citeauthoryear{Ashizawa, Yanai, et~al\mbox{.}}{Ashizawa
  et~al\mbox{.}}{2021}]%
        {ashizawa2021eth2vec}
\bibfield{author}{\bibinfo{person}{Nami Ashizawa}, \bibinfo{person}{Naoto
  Yanai}, {et~al\mbox{.}}} \bibinfo{year}{2021}\natexlab{}.
\newblock \showarticletitle{Eth2Vec: Learning contract-wide code
  representations for vulnerability detection on ethereum smart contracts}. In
  \bibinfo{booktitle}{\emph{Proceedings of the 3rd ACM International Symposium
  on Blockchain and Secure Critical Infrastructure}}. \bibinfo{pages}{47--59}.
\newblock


\bibitem[\protect\citeauthoryear{Ashouri}{Ashouri}{2020}]%
        {ashouri2020etherolic}
\bibfield{author}{\bibinfo{person}{Mohammadreza Ashouri}.}
  \bibinfo{year}{2020}\natexlab{}.
\newblock \showarticletitle{Etherolic: a practical security analyzer for smart
  contracts}. In \bibinfo{booktitle}{\emph{Proceedings of the 35th Annual ACM
  Symposium on Applied Computing}}. \bibinfo{pages}{353--356}.
\newblock


\bibitem[\protect\citeauthoryear{Atzei, Bartoletti, et~al\mbox{.}}{Atzei
  et~al\mbox{.}}{2017}]%
        {atzei2017survey}
\bibfield{author}{\bibinfo{person}{Nicola Atzei}, \bibinfo{person}{Massimo
  Bartoletti}, {et~al\mbox{.}}} \bibinfo{year}{2017}\natexlab{}.
\newblock \showarticletitle{A survey of attacks on ethereum smart contracts
  (sok)}. In \bibinfo{booktitle}{\emph{International conference on principles
  of security and trust}}. Springer, \bibinfo{pages}{164--186}.
\newblock


\bibitem[\protect\citeauthoryear{Baier and Katoen}{Baier and Katoen}{2008}]%
        {baier2008principles}
\bibfield{author}{\bibinfo{person}{Christel Baier} {and}
  \bibinfo{person}{Joost-Pieter Katoen}.} \bibinfo{year}{2008}\natexlab{}.
\newblock \bibinfo{booktitle}{\emph{Principles of model checking}}.
\newblock \bibinfo{publisher}{MIT press}.
\newblock


\bibitem[\protect\citeauthoryear{Ball}{Ball}{1999}]%
        {ball1999concept}
\bibfield{author}{\bibinfo{person}{Thomas Ball}.}
  \bibinfo{year}{1999}\natexlab{}.
\newblock \showarticletitle{The concept of dynamic analysis}. In
  \bibinfo{booktitle}{\emph{Software Engineering—ESEC/FSE’99}}. Springer,
  \bibinfo{pages}{216--234}.
\newblock


\bibitem[\protect\citeauthoryear{Bang, Nguyen, et~al\mbox{.}}{Bang
  et~al\mbox{.}}{2020}]%
        {bang2020verification}
\bibfield{author}{\bibinfo{person}{Tam Bang}, \bibinfo{person}{Hoang~H Nguyen},
  {et~al\mbox{.}}} \bibinfo{year}{2020}\natexlab{}.
\newblock \showarticletitle{Verification of Ethereum Smart Contracts: A Model
  Checking Approach}.
\newblock \bibinfo{journal}{\emph{International Journal of Machine Learning and
  Computing}} \bibinfo{volume}{10}, \bibinfo{number}{4} (\bibinfo{year}{2020}).
\newblock


\bibitem[\protect\citeauthoryear{Bartoletti, Bracciali,
  et~al\mbox{.}}{Bartoletti et~al\mbox{.}}{2021}]%
        {bartoletti2021formal}
\bibfield{author}{\bibinfo{person}{Massimo Bartoletti}, \bibinfo{person}{Andrea
  Bracciali}, {et~al\mbox{.}}} \bibinfo{year}{2021}\natexlab{}.
\newblock \showarticletitle{A formal model of Algorand smart contracts}. In
  \bibinfo{booktitle}{\emph{International Conference on Financial Cryptography
  and Data Security}}. Springer, \bibinfo{pages}{93--114}.
\newblock


\bibitem[\protect\citeauthoryear{Bau, Min\'{e}, et~al\mbox{.}}{Bau
  et~al\mbox{.}}{2022}]%
        {Michelson2022}
\bibfield{author}{\bibinfo{person}{Guillaume Bau}, \bibinfo{person}{Antoine
  Min\'{e}}, {et~al\mbox{.}}} \bibinfo{year}{2022}\natexlab{}.
\newblock \showarticletitle{Abstract Interpretation of Michelson
  Smart-Contracts}. In \bibinfo{booktitle}{\emph{Proceedings of the 11th ACM
  SIGPLAN International Workshop on the State Of the Art in Program Analysis}}.
  \bibinfo{publisher}{Association for Computing Machinery},
  \bibinfo{address}{New York, NY, USA}, \bibinfo{pages}{36–43}.
\newblock
\showISBNx{9781450392747}


\bibitem[\protect\citeauthoryear{Baumann, Steffen, et~al\mbox{.}}{Baumann
  et~al\mbox{.}}{2020}]%
        {baumann2020zkay}
\bibfield{author}{\bibinfo{person}{Nick Baumann}, \bibinfo{person}{Samuel
  Steffen}, {et~al\mbox{.}}} \bibinfo{year}{2020}\natexlab{}.
\newblock \showarticletitle{zkay v0.2: Practical Data Privacy for Smart
  Contracts}.
\newblock  (\bibinfo{year}{2020}).
\newblock
\showeprint[arxiv]{2009.01020}


\bibitem[\protect\citeauthoryear{Beckert, Herda, et~al\mbox{.}}{Beckert
  et~al\mbox{.}}{2018}]%
        {beckert2018formal}
\bibfield{author}{\bibinfo{person}{Bernhard Beckert}, \bibinfo{person}{Mihai
  Herda}, {et~al\mbox{.}}} \bibinfo{year}{2018}\natexlab{}.
\newblock \showarticletitle{Formal specification and verification of
  Hyperledger Fabric chaincode}. In \bibinfo{booktitle}{\emph{3rd Symposium on
  Distributed Ledger Technology (SDLT-2018) co-located with ICFEM}}.
  \bibinfo{pages}{44--48}.
\newblock


\bibitem[\protect\citeauthoryear{Beckert and Schiffl}{Beckert and
  Schiffl}{2020}]%
        {beckert2020specifying}
\bibfield{author}{\bibinfo{person}{Bernhard Beckert} {and}
  \bibinfo{person}{Jonas Schiffl}.} \bibinfo{year}{2020}\natexlab{}.
\newblock \showarticletitle{Specifying Framing Conditions for Smart Contracts}.
  In \bibinfo{booktitle}{\emph{International Symposium on Leveraging
  Applications of Formal Methods}}. Springer, \bibinfo{pages}{43--59}.
\newblock


\bibitem[\protect\citeauthoryear{Bernardo, Cauderlier, et~al\mbox{.}}{Bernardo
  et~al\mbox{.}}{2020a}]%
        {Mi-Cho-Coq2020}
\bibfield{author}{\bibinfo{person}{Bruno Bernardo},
  \bibinfo{person}{Rapha{\"e}l Cauderlier}, {et~al\mbox{.}}}
  \bibinfo{year}{2020}\natexlab{a}.
\newblock \showarticletitle{Making Tezos Smart Contracts More Reliable with
  Coq}. In \bibinfo{booktitle}{\emph{Leveraging Applications of Formal Methods,
  Verification and Validation: Applications}}. \bibinfo{publisher}{Springer
  International Publishing}, \bibinfo{pages}{60--72}.
\newblock
\showISBNx{978-3-030-61467-6}


\bibitem[\protect\citeauthoryear{Bernardo, Cauderlier, et~al\mbox{.}}{Bernardo
  et~al\mbox{.}}{2020b}]%
        {10.1007/978-3-030-54455-3_41}
\bibfield{author}{\bibinfo{person}{Bruno Bernardo},
  \bibinfo{person}{Rapha{\"e}l Cauderlier}, {et~al\mbox{.}}}
  \bibinfo{year}{2020}\natexlab{b}.
\newblock \showarticletitle{Albert, An Intermediate Smart-Contract Language for
  the Tezos Blockchain}. In \bibinfo{booktitle}{\emph{Financial Cryptography
  and Data Security}}. \bibinfo{publisher}{Springer International Publishing},
  \bibinfo{address}{Cham}, \bibinfo{pages}{584--598}.
\newblock
\showISBNx{978-3-030-54455-3}


\bibitem[\protect\citeauthoryear{Bertot and Cast{\'e}ran}{Bertot and
  Cast{\'e}ran}{2013}]%
        {bertot2013interactive}
\bibfield{author}{\bibinfo{person}{Yves Bertot} {and} \bibinfo{person}{Pierre
  Cast{\'e}ran}.} \bibinfo{year}{2013}\natexlab{}.
\newblock \bibinfo{booktitle}{\emph{Interactive theorem proving and program
  development: Coq’Art: the calculus of inductive constructions}}.
\newblock \bibinfo{publisher}{Springer Science \& Business Media}.
\newblock


\bibitem[\protect\citeauthoryear{Bhargavan, Delignat-Lavaud,
  et~al\mbox{.}}{Bhargavan et~al\mbox{.}}{2016}]%
        {bhargavan2016formal}
\bibfield{author}{\bibinfo{person}{Karthikeyan Bhargavan},
  \bibinfo{person}{Antoine Delignat-Lavaud}, {et~al\mbox{.}}}
  \bibinfo{year}{2016}\natexlab{}.
\newblock \showarticletitle{Formal verification of smart contracts: Short
  paper}. In \bibinfo{booktitle}{\emph{Proceedings of the 2016 ACM Workshop on
  Programming Languages and Analysis for Security}}. \bibinfo{pages}{91--96}.
\newblock


\bibitem[\protect\citeauthoryear{Blackshear, Dill, et~al\mbox{.}}{Blackshear
  et~al\mbox{.}}{2020}]%
        {blackshear2020resources}
\bibfield{author}{\bibinfo{person}{Sam Blackshear}, \bibinfo{person}{David~L.
  Dill}, {et~al\mbox{.}}} \bibinfo{year}{2020}\natexlab{}.
\newblock \bibinfo{title}{Resources: A Safe Language Abstraction for Money}.
\newblock
\newblock
\showeprint[arxiv]{cs.PL/2004.05106}


\bibitem[\protect\citeauthoryear{Bose, Das, et~al\mbox{.}}{Bose
  et~al\mbox{.}}{2022}]%
        {SAILFish2022}
\bibfield{author}{\bibinfo{person}{Priyanka Bose}, \bibinfo{person}{Dipanjan
  Das}, {et~al\mbox{.}}} \bibinfo{year}{2022}\natexlab{}.
\newblock \showarticletitle{SAILFISH: Vetting Smart Contract
  State-Inconsistency Bugs in Seconds}. In \bibinfo{booktitle}{\emph{2022 IEEE
  Symposium on Security and Privacy (SP)}}. \bibinfo{pages}{161--178}.
\newblock
\urldef\tempurl%
\url{https://doi.org/10.1109/SP46214.2022.9833721}
\showDOI{\tempurl}


\bibitem[\protect\citeauthoryear{Br{\"a}m, Eilers, et~al\mbox{.}}{Br{\"a}m
  et~al\mbox{.}}{2021}]%
        {bram2021rich}
\bibfield{author}{\bibinfo{person}{Christian Br{\"a}m}, \bibinfo{person}{Marco
  Eilers}, {et~al\mbox{.}}} \bibinfo{year}{2021}\natexlab{}.
\newblock \showarticletitle{Rich specifications for Ethereum smart contract
  verification}.
\newblock \bibinfo{journal}{\emph{Proceedings of the ACM on Programming
  Languages}} \bibinfo{volume}{5}, \bibinfo{number}{OOPSLA}
  (\bibinfo{year}{2021}), \bibinfo{pages}{1--30}.
\newblock


\bibitem[\protect\citeauthoryear{Brent, Grech, et~al\mbox{.}}{Brent
  et~al\mbox{.}}{2020}]%
        {brent2020ethainter}
\bibfield{author}{\bibinfo{person}{Lexi Brent}, \bibinfo{person}{Neville
  Grech}, {et~al\mbox{.}}} \bibinfo{year}{2020}\natexlab{}.
\newblock \showarticletitle{Ethainter: a smart contract security analyzer for
  composite vulnerabilities.}. In \bibinfo{booktitle}{\emph{PLDI}}.
  \bibinfo{pages}{454--469}.
\newblock


\bibitem[\protect\citeauthoryear{Brent, Jurisevic, et~al\mbox{.}}{Brent
  et~al\mbox{.}}{2018}]%
        {brent2018vandal}
\bibfield{author}{\bibinfo{person}{Lexi Brent}, \bibinfo{person}{Anton
  Jurisevic}, {et~al\mbox{.}}} \bibinfo{year}{2018}\natexlab{}.
\newblock \showarticletitle{Vandal: A scalable security analysis framework for
  smart contracts}.
\newblock  (\bibinfo{year}{2018}).
\newblock


\bibitem[\protect\citeauthoryear{Britten, Sj{\"o}berg, et~al\mbox{.}}{Britten
  et~al\mbox{.}}{2021}]%
        {britten2021using}
\bibfield{author}{\bibinfo{person}{Daniel Britten}, \bibinfo{person}{Vilhelm
  Sj{\"o}berg}, {et~al\mbox{.}}} \bibinfo{year}{2021}\natexlab{}.
\newblock \showarticletitle{Using Coq to Enforce the
  Checks-Effects-Interactions Pattern in DeepSEA Smart Contracts (Short
  Paper)}. In \bibinfo{booktitle}{\emph{3rd International Workshop on Formal
  Methods for Blockchains (FMBC 2021)}}. Schloss Dagstuhl-Leibniz-Zentrum
  f{\"u}r Informatik.
\newblock


\bibitem[\protect\citeauthoryear{Chakravarty, Chapman,
  et~al\mbox{.}}{Chakravarty et~al\mbox{.}}{2020}]%
        {chakravarty2020extended}
\bibfield{author}{\bibinfo{person}{Manuel~MT Chakravarty},
  \bibinfo{person}{James Chapman}, {et~al\mbox{.}}}
  \bibinfo{year}{2020}\natexlab{}.
\newblock \showarticletitle{The extended UTXO model}. In
  \bibinfo{booktitle}{\emph{International Conference on Financial Cryptography
  and Data Security}}. Springer, \bibinfo{pages}{525--539}.
\newblock


\bibitem[\protect\citeauthoryear{Chen, Pendleton, et~al\mbox{.}}{Chen
  et~al\mbox{.}}{2020b}]%
        {chen2020survey}
\bibfield{author}{\bibinfo{person}{Huashan Chen}, \bibinfo{person}{Marcus
  Pendleton}, {et~al\mbox{.}}} \bibinfo{year}{2020}\natexlab{b}.
\newblock \showarticletitle{A survey on ethereum systems security:
  Vulnerabilities, attacks, and defenses}.
\newblock \bibinfo{journal}{\emph{ACM Computing Surveys (CSUR)}}
  \bibinfo{volume}{53}, \bibinfo{number}{3} (\bibinfo{year}{2020}),
  \bibinfo{pages}{1--43}.
\newblock


\bibitem[\protect\citeauthoryear{Chen, Xia, et~al\mbox{.}}{Chen
  et~al\mbox{.}}{2020c}]%
        {chen2020Defects}
\bibfield{author}{\bibinfo{person}{Jiachi Chen}, \bibinfo{person}{Xin Xia},
  {et~al\mbox{.}}} \bibinfo{year}{2020}\natexlab{c}.
\newblock \showarticletitle{Defining Smart Contract Defects on Ethereum}.
\newblock \bibinfo{journal}{\emph{IEEE Transactions on Software Engineering}}
  (\bibinfo{year}{2020}), \bibinfo{pages}{1--1}.
\newblock


\bibitem[\protect\citeauthoryear{Chen, Xia, et~al\mbox{.}}{Chen
  et~al\mbox{.}}{2022b}]%
        {DefectChecker2022}
\bibfield{author}{\bibinfo{person}{Jiachi Chen}, \bibinfo{person}{Xin Xia},
  {et~al\mbox{.}}} \bibinfo{year}{2022}\natexlab{b}.
\newblock \showarticletitle{DefectChecker: Automated Smart Contract Defect
  Detection by Analyzing EVM Bytecode}.
\newblock \bibinfo{journal}{\emph{IEEE Transactions on Software Engineering}}
  \bibinfo{volume}{48}, \bibinfo{number}{7} (\bibinfo{year}{2022}),
  \bibinfo{pages}{2189--2207}.
\newblock
\urldef\tempurl%
\url{https://doi.org/10.1109/TSE.2021.3054928}
\showDOI{\tempurl}


\bibitem[\protect\citeauthoryear{Chen, Feng, et~al\mbox{.}}{Chen
  et~al\mbox{.}}{2020a}]%
        {chen2020gaschecker}
\bibfield{author}{\bibinfo{person}{Ting Chen}, \bibinfo{person}{Youzheng Feng},
  {et~al\mbox{.}}} \bibinfo{year}{2020}\natexlab{a}.
\newblock \showarticletitle{Gaschecker: Scalable analysis for discovering
  gas-inefficient smart contracts}.
\newblock \bibinfo{journal}{\emph{IEEE Transactions on Emerging Topics in
  Computing}} (\bibinfo{year}{2020}).
\newblock


\bibitem[\protect\citeauthoryear{Chen, Sun, et~al\mbox{.}}{Chen
  et~al\mbox{.}}{2022a}]%
        {Wasai2022}
\bibfield{author}{\bibinfo{person}{Weimin Chen}, \bibinfo{person}{Zihan Sun},
  {et~al\mbox{.}}} \bibinfo{year}{2022}\natexlab{a}.
\newblock \showarticletitle{WASAI: Uncovering Vulnerabilities in Wasm Smart
  Contracts}. \bibinfo{publisher}{Association for Computing Machinery},
  \bibinfo{address}{New York, NY, USA}, \bibinfo{pages}{703–715}.
\newblock
\showISBNx{9781450393799}


\bibitem[\protect\citeauthoryear{Chinen, Yanai, et~al\mbox{.}}{Chinen
  et~al\mbox{.}}{2020}]%
        {chinen2020ra}
\bibfield{author}{\bibinfo{person}{Yuchiro Chinen}, \bibinfo{person}{Naoto
  Yanai}, {et~al\mbox{.}}} \bibinfo{year}{2020}\natexlab{}.
\newblock \showarticletitle{RA: Hunting for Re-Entrancy Attacks in Ethereum
  Smart Contracts via Static Analysis}. In \bibinfo{booktitle}{\emph{2020 IEEE
  International Conference on Blockchain (Blockchain)}}. IEEE,
  \bibinfo{pages}{327--336}.
\newblock


\bibitem[\protect\citeauthoryear{Christidis and Devetsikiotis}{Christidis and
  Devetsikiotis}{2016}]%
        {blockchain_Smartcontracts}
\bibfield{author}{\bibinfo{person}{Konstantinos Christidis} {and}
  \bibinfo{person}{Michael Devetsikiotis}.} \bibinfo{year}{2016}\natexlab{}.
\newblock \showarticletitle{Blockchains and Smart Contracts for the Internet of
  Things}.
\newblock \bibinfo{journal}{\emph{IEEE Access}}  \bibinfo{volume}{4}
  (\bibinfo{year}{2016}).
\newblock


\bibitem[\protect\citeauthoryear{Christodorescu and Jha}{Christodorescu and
  Jha}{2006}]%
        {christodorescu2006static}
\bibfield{author}{\bibinfo{person}{Mihai Christodorescu} {and}
  \bibinfo{person}{Somesh Jha}.} \bibinfo{year}{2006}\natexlab{}.
\newblock \bibinfo{booktitle}{\emph{Static analysis of executables to detect
  malicious patterns}}.
\newblock \bibinfo{type}{{T}echnical {R}eport}. \bibinfo{institution}{Wisconsin
  Univ-Madison Dept of Computer Sciences}.
\newblock


\bibitem[\protect\citeauthoryear{Coblenz, Oei, et~al\mbox{.}}{Coblenz
  et~al\mbox{.}}{2020}]%
        {coblenz2020obsidian}
\bibfield{author}{\bibinfo{person}{Michael Coblenz}, \bibinfo{person}{Reed
  Oei}, {et~al\mbox{.}}} \bibinfo{year}{2020}\natexlab{}.
\newblock \showarticletitle{Obsidian: Typestate and Assets for Safer Blockchain
  Programming}.
\newblock  \bibinfo{volume}{42}, \bibinfo{number}{3} (\bibinfo{year}{2020}).
\newblock
\showISSN{0164-0925}
\urldef\tempurl%
\url{https://doi.org/10.1145/3417516}
\showURL{%
\tempurl}


\bibitem[\protect\citeauthoryear{CoinMarketCap.com}{CoinMarketCap.com}{2022}]%
        {cryptocurrencies}
\bibfield{author}{\bibinfo{person}{CoinMarketCap.com}.}
  \bibinfo{year}{2022}\natexlab{}.
\newblock \bibinfo{title}{All Cryptocurrencies}.
\newblock
\newblock
\urldef\tempurl%
\url{https://coinmarketcap.com/all/views/all/}
\showURL{%
Retrieved Oct 11, 2021 from \tempurl}


\bibitem[\protect\citeauthoryear{Coinranking.Com}{Coinranking.Com}{2020}]%
        {Coinranking}
\bibfield{author}{\bibinfo{person}{Coinranking.Com}.}
  \bibinfo{year}{2020}\natexlab{}.
\newblock \bibinfo{title}{Coinranking - Crypto market overview and total market
  cap}.
\newblock
\newblock
\urldef\tempurl%
\url{https://coinranking.com/overview}
\showURL{%
Retrieved January 19, 2022 from \tempurl}


\bibitem[\protect\citeauthoryear{Community}{Community}{2020}]%
        {swcRegistry}
\bibfield{author}{\bibinfo{person}{Ethereum Community}.}
  \bibinfo{year}{2020}\natexlab{}.
\newblock \bibinfo{title}{SWC Registry - Smart Contract Weakness Classification
  and Test Cases}.
\newblock
\newblock
\urldef\tempurl%
\url{https://swcregistry.io/}
\showURL{%
Retrieved January 20, 2021 from \tempurl}


\bibitem[\protect\citeauthoryear{ConsenSys}{ConsenSys}{2021}]%
        {Mythril}
\bibfield{author}{\bibinfo{person}{ConsenSys}.}
  \bibinfo{year}{2021}\natexlab{}.
\newblock \bibinfo{title}{Mythril}.
\newblock \bibinfo{howpublished}{\url{https://github.com/ConsenSys/mythril}}.
\newblock


\bibitem[\protect\citeauthoryear{Cook}{Cook}{1971}]%
        {cook1971complexity}
\bibfield{author}{\bibinfo{person}{Stephen~A Cook}.}
  \bibinfo{year}{1971}\natexlab{}.
\newblock \showarticletitle{The complexity of theorem-proving procedures}. In
  \bibinfo{booktitle}{\emph{Proceedings of the third annual ACM symposium on
  Theory of computing}}. \bibinfo{pages}{151--158}.
\newblock


\bibitem[\protect\citeauthoryear{Correas, Gordillo, et~al\mbox{.}}{Correas
  et~al\mbox{.}}{2021}]%
        {Correas2021static}
\bibfield{author}{\bibinfo{person}{Jesús Correas}, \bibinfo{person}{Pablo
  Gordillo}, {et~al\mbox{.}}} \bibinfo{year}{2021}\natexlab{}.
\newblock \showarticletitle{Static Profiling and Optimization of Ethereum Smart
  Contracts Using Resource Analysis}.
\newblock \bibinfo{journal}{\emph{IEEE Access}}  \bibinfo{volume}{9}
  (\bibinfo{year}{2021}), \bibinfo{pages}{25495--25507}.
\newblock


\bibitem[\protect\citeauthoryear{Cousot and Cousot}{Cousot and Cousot}{1977}]%
        {cousot1977abstract}
\bibfield{author}{\bibinfo{person}{Patrick Cousot} {and}
  \bibinfo{person}{Radhia Cousot}.} \bibinfo{year}{1977}\natexlab{}.
\newblock \showarticletitle{Abstract interpretation: a unified lattice model
  for static analysis of programs by construction or approximation of
  fixpoints}. In \bibinfo{booktitle}{\emph{Proceedings of the 4th ACM
  SIGACT-SIGPLAN symposium on Principles of programming languages}}.
  \bibinfo{pages}{238--252}.
\newblock


\bibitem[\protect\citeauthoryear{Cousot and Cousot}{Cousot and Cousot}{1979}]%
        {cousot1979systematic}
\bibfield{author}{\bibinfo{person}{Patrick Cousot} {and}
  \bibinfo{person}{Radhia Cousot}.} \bibinfo{year}{1979}\natexlab{}.
\newblock \showarticletitle{Systematic design of program analysis frameworks}.
  In \bibinfo{booktitle}{\emph{Proceedings of the 6th ACM SIGACT-SIGPLAN
  symposium on Principles of programming languages}}.
  \bibinfo{pages}{269--282}.
\newblock


\bibitem[\protect\citeauthoryear{Cousot and Cousot}{Cousot and Cousot}{2010}]%
        {cousot2010gentle}
\bibfield{author}{\bibinfo{person}{Patrick Cousot} {and}
  \bibinfo{person}{Radhia Cousot}.} \bibinfo{year}{2010}\natexlab{}.
\newblock \showarticletitle{A gentle introduction to formal verification of
  computer systems by abstract interpretation}.
\newblock In \bibinfo{booktitle}{\emph{Logics and Languages for Reliability and
  Security}}. \bibinfo{publisher}{IOS Press}, \bibinfo{pages}{1--29}.
\newblock


\bibitem[\protect\citeauthoryear{Cui, Zhao, et~al\mbox{.}}{Cui
  et~al\mbox{.}}{2022}]%
        {Cui2022}
\bibfield{author}{\bibinfo{person}{Siwei Cui}, \bibinfo{person}{Gang Zhao},
  {et~al\mbox{.}}} \bibinfo{year}{2022}\natexlab{}.
\newblock \showarticletitle{VRust: Automated Vulnerability Detection for Solana
  Smart Contracts}. In \bibinfo{booktitle}{\emph{Proceedings of the 2022 ACM
  SIGSAC Conference on Computer and Communications Security}}
  \emph{(\bibinfo{series}{CCS '22})}. \bibinfo{publisher}{Association for
  Computing Machinery}, \bibinfo{address}{New York, NY, USA},
  \bibinfo{pages}{639–652}.
\newblock
\showISBNx{9781450394505}
\urldef\tempurl%
\url{https://doi.org/10.1145/3548606.3560552}
\showDOI{\tempurl}


\bibitem[\protect\citeauthoryear{da~Horta, Reis, et~al\mbox{.}}{da~Horta
  et~al\mbox{.}}{2020}]%
        {da2020whylson}
\bibfield{author}{\bibinfo{person}{Lu{\'\i}s Pedro~Arrojado da Horta},
  \bibinfo{person}{Jo{\~a}o~Santos Reis}, {et~al\mbox{.}}}
  \bibinfo{year}{2020}\natexlab{}.
\newblock \showarticletitle{WhylSon: Proving your Michelson Smart Contracts in
  Why3}.
\newblock  (\bibinfo{year}{2020}).
\newblock
\showeprint[arxiv]{2005.14650}


\bibitem[\protect\citeauthoryear{Dai, Yang, et~al\mbox{.}}{Dai
  et~al\mbox{.}}{2022}]%
        {dai2022superdetector}
\bibfield{author}{\bibinfo{person}{Meiyi Dai}, \bibinfo{person}{Zhe Yang},
  {et~al\mbox{.}}} \bibinfo{year}{2022}\natexlab{}.
\newblock \showarticletitle{SuperDetector: A Framework for Performance
  Detection on Vulnerabilities of Smart Contracts}. In
  \bibinfo{booktitle}{\emph{Journal of Physics: Conference Series}},
  Vol.~\bibinfo{volume}{2289}. IOP Publishing, \bibinfo{pages}{012010}.
\newblock


\bibitem[\protect\citeauthoryear{Das, Balzer, et~al\mbox{.}}{Das
  et~al\mbox{.}}{2021}]%
        {das2021resource}
\bibfield{author}{\bibinfo{person}{Ankush Das}, \bibinfo{person}{Stephanie
  Balzer}, {et~al\mbox{.}}} \bibinfo{year}{2021}\natexlab{}.
\newblock \showarticletitle{Resource-aware session types for digital
  contracts}. In \bibinfo{booktitle}{\emph{2021 IEEE 34th Computer Security
  Foundations Symposium (CSF)}}. IEEE, \bibinfo{pages}{1--16}.
\newblock


\bibitem[\protect\citeauthoryear{De~Moura and Bj{\o}rner}{De~Moura and
  Bj{\o}rner}{2008}]%
        {de2008z3}
\bibfield{author}{\bibinfo{person}{Leonardo De~Moura} {and}
  \bibinfo{person}{Nikolaj Bj{\o}rner}.} \bibinfo{year}{2008}\natexlab{}.
\newblock \showarticletitle{Z3: An efficient SMT solver}. In
  \bibinfo{booktitle}{\emph{International conference on Tools and Algorithms
  for the Construction and Analysis of Systems}}. Springer,
  \bibinfo{pages}{337--340}.
\newblock


\bibitem[\protect\citeauthoryear{Demir, Alalfi, et~al\mbox{.}}{Demir
  et~al\mbox{.}}{2019}]%
        {demir2019security}
\bibfield{author}{\bibinfo{person}{Mehmet Demir}, \bibinfo{person}{Manar
  Alalfi}, {et~al\mbox{.}}} \bibinfo{year}{2019}\natexlab{}.
\newblock \showarticletitle{Security smells in smart contracts}. In
  \bibinfo{booktitle}{\emph{2019 IEEE 19th International Conference on Software
  Quality, Reliability and Security Companion (QRS-C)}}. IEEE,
  \bibinfo{pages}{442--449}.
\newblock


\bibitem[\protect\citeauthoryear{Dika and Nowostawski}{Dika and
  Nowostawski}{2018}]%
        {dika2018security}
\bibfield{author}{\bibinfo{person}{Ardit Dika} {and} \bibinfo{person}{Mariusz
  Nowostawski}.} \bibinfo{year}{2018}\natexlab{}.
\newblock \showarticletitle{Security vulnerabilities in ethereum smart
  contracts}. In \bibinfo{booktitle}{\emph{2018 IEEE International Conference
  on Internet of Things and IEEE Green Computing and Communications and IEEE
  Cyber, Physical and Social Computing and IEEE Smart Data}}. IEEE,
  \bibinfo{pages}{955--962}.
\newblock


\bibitem[\protect\citeauthoryear{Ding, Li, et~al\mbox{.}}{Ding
  et~al\mbox{.}}{2021}]%
        {ding2021hfcontractfuzzer}
\bibfield{author}{\bibinfo{person}{Mengjie Ding}, \bibinfo{person}{Peiru Li},
  {et~al\mbox{.}}} \bibinfo{year}{2021}\natexlab{}.
\newblock \showarticletitle{HFContractFuzzer: Fuzzing Hyperledger Fabric Smart
  Contracts for Vulnerability Detection}.
\newblock In \bibinfo{booktitle}{\emph{Evaluation and Assessment in Software
  Engineering}}. \bibinfo{pages}{321--328}.
\newblock


\bibitem[\protect\citeauthoryear{Dong, Zhou, et~al\mbox{.}}{Dong
  et~al\mbox{.}}{2022}]%
        {SolChecker2022}
\bibfield{author}{\bibinfo{person}{Weiliang Dong}, \bibinfo{person}{Teng Zhou},
  {et~al\mbox{.}}} \bibinfo{year}{2022}\natexlab{}.
\newblock \showarticletitle{SolChecker: A Practical Static Analysis Framework
  for Ethereum Smart Contract}. In \bibinfo{booktitle}{\emph{2022 International
  Conference on Networks, Communications and Information Technology (CNCIT)}}.
  \bibinfo{pages}{179--186}.
\newblock


\bibitem[\protect\citeauthoryear{Duo, Xin, et~al\mbox{.}}{Duo
  et~al\mbox{.}}{2020}]%
        {duo2020formal}
\bibfield{author}{\bibinfo{person}{Wang Duo}, \bibinfo{person}{Huang Xin},
  {et~al\mbox{.}}} \bibinfo{year}{2020}\natexlab{}.
\newblock \showarticletitle{Formal analysis of smart contract based on colored
  petri nets}.
\newblock \bibinfo{journal}{\emph{IEEE Intelligent Systems}}
  \bibinfo{volume}{35}, \bibinfo{number}{3} (\bibinfo{year}{2020}),
  \bibinfo{pages}{19--30}.
\newblock


\bibitem[\protect\citeauthoryear{Durieux, Ferreira, et~al\mbox{.}}{Durieux
  et~al\mbox{.}}{2020}]%
        {durieux2020empirical}
\bibfield{author}{\bibinfo{person}{Thomas Durieux}, \bibinfo{person}{Jo{\~a}o~F
  Ferreira}, {et~al\mbox{.}}} \bibinfo{year}{2020}\natexlab{}.
\newblock \showarticletitle{Empirical review of automated analysis tools on
  47,587 Ethereum smart contracts}. In \bibinfo{booktitle}{\emph{Proceedings of
  the ACM/IEEE 42nd International Conference on Software Engineering}}.
  \bibinfo{pages}{530--541}.
\newblock


\bibitem[\protect\citeauthoryear{Elad, Rain, et~al\mbox{.}}{Elad
  et~al\mbox{.}}{2021}]%
        {elad2021summing}
\bibfield{author}{\bibinfo{person}{Neta Elad}, \bibinfo{person}{Sophie Rain},
  {et~al\mbox{.}}} \bibinfo{year}{2021}\natexlab{}.
\newblock \showarticletitle{Summing up Smart Transitions}. In
  \bibinfo{booktitle}{\emph{International Conference on Computer Aided
  Verification}}. Springer, \bibinfo{pages}{317--340}.
\newblock


\bibitem[\protect\citeauthoryear{Ethereum}{Ethereum}{2020}]%
        {EthereumDepositContract}
\bibfield{author}{\bibinfo{person}{Ethereum}.} \bibinfo{year}{2020}\natexlab{}.
\newblock \bibinfo{title}{Ethereum 2.0 Deposit Contract}.
\newblock
\newblock
\urldef\tempurl%
\url{https://github.com/ethereum/consensus-specs/blob/v0.11.2/deposit_contract/contracts/validator_registration.vy}
\showURL{%
Retrieved December 08, 2022 from \tempurl}


\bibitem[\protect\citeauthoryear{Feist, Grieco, et~al\mbox{.}}{Feist
  et~al\mbox{.}}{2019}]%
        {feist2019slither}
\bibfield{author}{\bibinfo{person}{Josselin Feist}, \bibinfo{person}{Gustavo
  Grieco}, {et~al\mbox{.}}} \bibinfo{year}{2019}\natexlab{}.
\newblock \showarticletitle{Slither: a static analysis framework for smart
  contracts}. In \bibinfo{booktitle}{\emph{2019 IEEE/ACM 2nd International
  Workshop on Emerging Trends in Software Engineering for Blockchain
  (WETSEB)}}. IEEE, \bibinfo{pages}{8--15}.
\newblock


\bibitem[\protect\citeauthoryear{Ferreira, Cruz, et~al\mbox{.}}{Ferreira
  et~al\mbox{.}}{2020}]%
        {ferreira2020smartbugs}
\bibfield{author}{\bibinfo{person}{Jo{\~a}o~F Ferreira}, \bibinfo{person}{Pedro
  Cruz}, {et~al\mbox{.}}} \bibinfo{year}{2020}\natexlab{}.
\newblock \showarticletitle{SmartBugs: a framework to analyze solidity smart
  contracts}. In \bibinfo{booktitle}{\emph{Proceedings of the 35th IEEE/ACM
  International Conference on Automated Software Engineering}}.
  \bibinfo{pages}{1349--1352}.
\newblock


\bibitem[\protect\citeauthoryear{Finance}{Finance}{2022}]%
        {ReentrancyIncident}
\bibfield{author}{\bibinfo{person}{Revest Finance}.}
  \bibinfo{year}{2022}\natexlab{}.
\newblock \bibinfo{title}{Revest Protocol Exploit Recovery Plan}.
\newblock
\newblock
\urldef\tempurl%
\url{https://revestfinance.medium.com/revest-protocol-exploit-recovery-plan-b06ca33fbdf5}
\showURL{%
Retrieved December 08, 2022 from \tempurl}


\bibitem[\protect\citeauthoryear{Forum}{Forum}{2020}]%
        {BlockchainStandards}
\bibfield{author}{\bibinfo{person}{World~Economic Forum}.}
  \bibinfo{year}{2020}\natexlab{}.
\newblock \bibinfo{booktitle}{\emph{Global Standards Mapping Initiative: An
  overview of blockchain technical standards}}.
\newblock \bibinfo{type}{{T}echnical {R}eport}.
\newblock
\urldef\tempurl%
\url{https://www3.weforum.org/docs/WEF_GSMI_Technical_Standards_2020.pdf}
\showURL{%
\tempurl}


\bibitem[\protect\citeauthoryear{Garfatta, Klai, et~al\mbox{.}}{Garfatta
  et~al\mbox{.}}{2021}]%
        {garfatta2021survey}
\bibfield{author}{\bibinfo{person}{Ikram Garfatta}, \bibinfo{person}{Kais
  Klai}, {et~al\mbox{.}}} \bibinfo{year}{2021}\natexlab{}.
\newblock \showarticletitle{A Survey on Formal Verification for Solidity Smart
  Contracts}. In \bibinfo{booktitle}{\emph{2021 Australasian Computer Science
  Week Multiconference}}. \bibinfo{pages}{1--10}.
\newblock


\bibitem[\protect\citeauthoryear{Genet, Jensen, et~al\mbox{.}}{Genet
  et~al\mbox{.}}{2020}]%
        {genet2020termination}
\bibfield{author}{\bibinfo{person}{Thomas Genet}, \bibinfo{person}{Thomas
  Jensen}, {et~al\mbox{.}}} \bibinfo{year}{2020}\natexlab{}.
\newblock \bibinfo{booktitle}{\emph{{Termination of Ethereum's Smart
  Contracts}}}.
\newblock \bibinfo{type}{Research Report}. \bibinfo{institution}{{Univ Rennes,
  Inria, CNRS, IRISA}}.
\newblock


\bibitem[\protect\citeauthoryear{Girard}{Girard}{1987}]%
        {GIRARD19871}
\bibfield{author}{\bibinfo{person}{Jean-Yves Girard}.}
  \bibinfo{year}{1987}\natexlab{}.
\newblock \showarticletitle{Linear logic}.
\newblock \bibinfo{journal}{\emph{Theoretical Computer Science}}
  \bibinfo{volume}{50}, \bibinfo{number}{1} (\bibinfo{year}{1987}),
  \bibinfo{pages}{1--101}.
\newblock
\showISSN{0304-3975}


\bibitem[\protect\citeauthoryear{Godefroid, Kiezun, et~al\mbox{.}}{Godefroid
  et~al\mbox{.}}{2008}]%
        {10.1145/1379022.1375607}
\bibfield{author}{\bibinfo{person}{Patrice Godefroid}, \bibinfo{person}{Adam
  Kiezun}, {et~al\mbox{.}}} \bibinfo{year}{2008}\natexlab{}.
\newblock \showarticletitle{Grammar-Based Whitebox Fuzzing}.
\newblock \bibinfo{journal}{\emph{SIGPLAN Not.}} \bibinfo{volume}{43},
  \bibinfo{number}{6} (\bibinfo{date}{jun} \bibinfo{year}{2008}),
  \bibinfo{pages}{206–215}.
\newblock
\showISSN{0362-1340}


\bibitem[\protect\citeauthoryear{Gosain and Sharma}{Gosain and Sharma}{2015}]%
        {dynamicAnalysisSurvey2015}
\bibfield{author}{\bibinfo{person}{Anjana Gosain} {and} \bibinfo{person}{Ganga
  Sharma}.} \bibinfo{year}{2015}\natexlab{}.
\newblock \showarticletitle{A Survey of Dynamic Program Analysis Techniques and
  Tools}. In \bibinfo{booktitle}{\emph{Proceedings of the 3rd International
  Conference on Frontiers of Intelligent Computing: Theory and Applications
  (FICTA) 2014}}. \bibinfo{publisher}{Springer International Publishing},
  \bibinfo{address}{Cham}, \bibinfo{pages}{113--122}.
\newblock


\bibitem[\protect\citeauthoryear{Grech, Kong, et~al\mbox{.}}{Grech
  et~al\mbox{.}}{2018}]%
        {grech2020madmax}
\bibfield{author}{\bibinfo{person}{Neville Grech}, \bibinfo{person}{Michael
  Kong}, {et~al\mbox{.}}} \bibinfo{year}{2018}\natexlab{}.
\newblock \showarticletitle{MadMax: Surviving out-of-Gas Conditions in Ethereum
  Smart Contracts}.
\newblock  \bibinfo{volume}{2}, \bibinfo{number}{OOPSLA}, Article
  \bibinfo{articleno}{116} (\bibinfo{date}{Oct.} \bibinfo{year}{2018}),
  \bibinfo{numpages}{27}~pages.
\newblock


\bibitem[\protect\citeauthoryear{Grieco, Song, et~al\mbox{.}}{Grieco
  et~al\mbox{.}}{2020}]%
        {grieco2020echidna}
\bibfield{author}{\bibinfo{person}{Gustavo Grieco}, \bibinfo{person}{Will
  Song}, {et~al\mbox{.}}} \bibinfo{year}{2020}\natexlab{}.
\newblock \showarticletitle{Echidna: effective, usable, and fast fuzzing for
  smart contracts}. In \bibinfo{booktitle}{\emph{Proceedings of the 29th ACM
  SIGSOFT International Symposium on Software Testing and Analysis}}.
  \bibinfo{pages}{557--560}.
\newblock


\bibitem[\protect\citeauthoryear{Grishchenko, Maffei,
  et~al\mbox{.}}{Grishchenko et~al\mbox{.}}{2018a}]%
        {grishchenko2018ethertrust}
\bibfield{author}{\bibinfo{person}{Ilya Grishchenko}, \bibinfo{person}{Matteo
  Maffei}, {et~al\mbox{.}}} \bibinfo{year}{2018}\natexlab{a}.
\newblock \showarticletitle{Ethertrust: Sound static analysis of ethereum
  bytecode}.
\newblock \bibinfo{journal}{\emph{Technische Universit{\"a}t Wien, Tech. Rep}}
  (\bibinfo{year}{2018}), \bibinfo{pages}{1--41}.
\newblock


\bibitem[\protect\citeauthoryear{Grishchenko, Maffei,
  et~al\mbox{.}}{Grishchenko et~al\mbox{.}}{2018b}]%
        {grishchenko2018foundations}
\bibfield{author}{\bibinfo{person}{Ilya Grishchenko}, \bibinfo{person}{Matteo
  Maffei}, {et~al\mbox{.}}} \bibinfo{year}{2018}\natexlab{b}.
\newblock \showarticletitle{Foundations and tools for the static analysis of
  ethereum smart contracts}. In \bibinfo{booktitle}{\emph{International
  Conference on Computer Aided Verification}}. Springer,
  \bibinfo{pages}{51--78}.
\newblock


\bibitem[\protect\citeauthoryear{Grishchenko, Maffei,
  et~al\mbox{.}}{Grishchenko et~al\mbox{.}}{2018c}]%
        {grishchenko2018semantic}
\bibfield{author}{\bibinfo{person}{Ilya Grishchenko}, \bibinfo{person}{Matteo
  Maffei}, {et~al\mbox{.}}} \bibinfo{year}{2018}\natexlab{c}.
\newblock \showarticletitle{A semantic framework for the security analysis of
  ethereum smart contracts}. In \bibinfo{booktitle}{\emph{International
  Conference on Principles of Security and Trust}}. Springer,
  \bibinfo{pages}{243--269}.
\newblock


\bibitem[\protect\citeauthoryear{Groce, Feist, et~al\mbox{.}}{Groce
  et~al\mbox{.}}{2020}]%
        {groce2020actual}
\bibfield{author}{\bibinfo{person}{Alex Groce}, \bibinfo{person}{Josselin
  Feist}, {et~al\mbox{.}}} \bibinfo{year}{2020}\natexlab{}.
\newblock \showarticletitle{What are the Actual Flaws in Important Smart
  Contracts (and How Can We Find Them)?}. In
  \bibinfo{booktitle}{\emph{International Conference on Financial Cryptography
  and Data Security}}. Springer, \bibinfo{pages}{634--653}.
\newblock


\bibitem[\protect\citeauthoryear{Grossman, Abraham, et~al\mbox{.}}{Grossman
  et~al\mbox{.}}{2017}]%
        {grossman2017online}
\bibfield{author}{\bibinfo{person}{Shelly Grossman}, \bibinfo{person}{Ittai
  Abraham}, {et~al\mbox{.}}} \bibinfo{year}{2017}\natexlab{}.
\newblock \showarticletitle{Online detection of effectively callback free
  objects with applications to smart contracts}.
\newblock \bibinfo{journal}{\emph{Proceedings of the ACM on Programming
  Languages}} \bibinfo{volume}{2}, \bibinfo{number}{POPL}
  (\bibinfo{year}{2017}), \bibinfo{pages}{1--28}.
\newblock


\bibitem[\protect\citeauthoryear{Gupta, Kumar, et~al\mbox{.}}{Gupta
  et~al\mbox{.}}{2020}]%
        {gupta2020insecurity}
\bibfield{author}{\bibinfo{person}{Bishwas~C Gupta}, \bibinfo{person}{Nitesh
  Kumar}, {et~al\mbox{.}}} \bibinfo{year}{2020}\natexlab{}.
\newblock \showarticletitle{An Insecurity Study of Ethereum Smart Contracts}.
  In \bibinfo{booktitle}{\emph{International Conference on Security, Privacy,
  and Applied Cryptography Engineering}}. Springer, \bibinfo{pages}{188--207}.
\newblock


\bibitem[\protect\citeauthoryear{Gupta, Patel, et~al\mbox{.}}{Gupta
  et~al\mbox{.}}{2021}]%
        {GUPTA2021107583}
\bibfield{author}{\bibinfo{person}{Rajesh Gupta},
  \bibinfo{person}{Mohil~Maheshkumar Patel}, {et~al\mbox{.}}}
  \bibinfo{year}{2021}\natexlab{}.
\newblock \showarticletitle{Deep learning-based malicious smart contract
  detection scheme for internet of things environment}.
\newblock \bibinfo{journal}{\emph{Computers \& Electrical Engineering}}
  (\bibinfo{year}{2021}), \bibinfo{pages}{107583}.
\newblock
\showISSN{0045-7906}


\bibitem[\protect\citeauthoryear{Hajdu and Jovanovi{\'c}}{Hajdu and
  Jovanovi{\'c}}{2019}]%
        {hajdu2019solc}
\bibfield{author}{\bibinfo{person}{{\'A}kos Hajdu} {and} \bibinfo{person}{Dejan
  Jovanovi{\'c}}.} \bibinfo{year}{2019}\natexlab{}.
\newblock \showarticletitle{solc-verify: A modular verifier for solidity smart
  contracts}. In \bibinfo{booktitle}{\emph{Working Conference on Verified
  Software: Theories, Tools, and Experiments}}. Springer,
  \bibinfo{pages}{161--179}.
\newblock


\bibitem[\protect\citeauthoryear{Hajdu, Jovanovi{\'c}, et~al\mbox{.}}{Hajdu
  et~al\mbox{.}}{2020}]%
        {hajdu2020formal}
\bibfield{author}{\bibinfo{person}{{\'A}kos Hajdu}, \bibinfo{person}{Dejan
  Jovanovi{\'c}}, {et~al\mbox{.}}} \bibinfo{year}{2020}\natexlab{}.
\newblock \showarticletitle{Formal specification and verification of solidity
  contracts with events (short paper)}. In \bibinfo{booktitle}{\emph{2nd
  Workshop on Formal Methods for Blockchains (FMBC 2020)}}. Schloss
  Dagstuhl-Leibniz-Zentrum f{\"u}r Informatik.
\newblock


\bibitem[\protect\citeauthoryear{Hao, Ren, et~al\mbox{.}}{Hao
  et~al\mbox{.}}{2020}]%
        {hao2020scscan}
\bibfield{author}{\bibinfo{person}{Xiaohan Hao}, \bibinfo{person}{Wei Ren},
  {et~al\mbox{.}}} \bibinfo{year}{2020}\natexlab{}.
\newblock \showarticletitle{SCScan: A SVM-based Scanning System for
  Vulnerabilities in Blockchain Smart Contracts}. In
  \bibinfo{booktitle}{\emph{2020 IEEE 19th International Conference on Trust,
  Security and Privacy in Computing and Communications (TrustCom)}}. IEEE,
  \bibinfo{pages}{1598--1605}.
\newblock


\bibitem[\protect\citeauthoryear{{He}, {Deng}, et~al\mbox{.}}{{He}
  et~al\mbox{.}}{2020}]%
        {He9143290}
\bibfield{author}{\bibinfo{person}{D. {He}}, \bibinfo{person}{Z. {Deng}},
  {et~al\mbox{.}}} \bibinfo{year}{2020}\natexlab{}.
\newblock \showarticletitle{Smart Contract Vulnerability Analysis and Security
  Audit}.
\newblock \bibinfo{journal}{\emph{IEEE Network}} \bibinfo{volume}{34},
  \bibinfo{number}{5} (\bibinfo{year}{2020}), \bibinfo{pages}{276--282}.
\newblock


\bibitem[\protect\citeauthoryear{He, Balunovi\'{c}, et~al\mbox{.}}{He
  et~al\mbox{.}}{2019}]%
        {ILF2019}
\bibfield{author}{\bibinfo{person}{Jingxuan He}, \bibinfo{person}{Mislav
  Balunovi\'{c}}, {et~al\mbox{.}}} \bibinfo{year}{2019}\natexlab{}.
\newblock \showarticletitle{Learning to Fuzz from Symbolic Execution with
  Application to Smart Contracts}. In \bibinfo{booktitle}{\emph{Proceedings of
  the 2019 ACM SIGSAC Conference on Computer and Communications Security}}.
  \bibinfo{pages}{531–548}.
\newblock
\showISBNx{9781450367479}


\bibitem[\protect\citeauthoryear{He, Zhang, et~al\mbox{.}}{He
  et~al\mbox{.}}{2021}]%
        {he2021eosafe}
\bibfield{author}{\bibinfo{person}{Ningyu He}, \bibinfo{person}{Ruiyi Zhang},
  {et~al\mbox{.}}} \bibinfo{year}{2021}\natexlab{}.
\newblock \showarticletitle{$\{$EOSAFE$\}$: Security Analysis of $\{$EOSIO$\}$
  Smart Contracts}. In \bibinfo{booktitle}{\emph{30th $\{$USENIX$\}$ Security
  Symposium ($\{$USENIX$\}$ Security 21)}}.
\newblock


\bibitem[\protect\citeauthoryear{He, Liao, et~al\mbox{.}}{He
  et~al\mbox{.}}{2022}]%
        {he2022tokencat}
\bibfield{author}{\bibinfo{person}{Zheyuan He}, \bibinfo{person}{Zhou Liao},
  {et~al\mbox{.}}} \bibinfo{year}{2022}\natexlab{}.
\newblock \showarticletitle{TokenCat: Detect Flaw of Authentication on ERC20
  Tokens}. In \bibinfo{booktitle}{\emph{ICC 2022-IEEE International Conference
  on Communications}}. IEEE, \bibinfo{pages}{4999--5004}.
\newblock


\bibitem[\protect\citeauthoryear{Huang, Han, et~al\mbox{.}}{Huang
  et~al\mbox{.}}{2021a}]%
        {huang2021hunting}
\bibfield{author}{\bibinfo{person}{Jianjun Huang}, \bibinfo{person}{Songming
  Han}, {et~al\mbox{.}}} \bibinfo{year}{2021}\natexlab{a}.
\newblock \showarticletitle{Hunting Vulnerable Smart Contracts via Graph
  Embedding Based Bytecode Matching}.
\newblock \bibinfo{journal}{\emph{IEEE Transactions on Information Forensics
  and Security}}  \bibinfo{volume}{16} (\bibinfo{year}{2021}),
  \bibinfo{pages}{2144--2156}.
\newblock


\bibitem[\protect\citeauthoryear{Huang, Jiang, et~al\mbox{.}}{Huang
  et~al\mbox{.}}{2021b}]%
        {huang2021precise}
\bibfield{author}{\bibinfo{person}{Jianjun Huang}, \bibinfo{person}{Jiasheng
  Jiang}, {et~al\mbox{.}}} \bibinfo{year}{2021}\natexlab{b}.
\newblock \showarticletitle{Precise Dynamic Symbolic Execution for Nonuniform
  Data Access in Smart Contracts}.
\newblock \bibinfo{journal}{\emph{IEEE Trans. Comput.}} (\bibinfo{year}{2021}).
\newblock


\bibitem[\protect\citeauthoryear{Huang, Jiang, et~al\mbox{.}}{Huang
  et~al\mbox{.}}{2020}]%
        {huang2020eosfuzzer}
\bibfield{author}{\bibinfo{person}{Yuhe Huang}, \bibinfo{person}{Bo Jiang},
  {et~al\mbox{.}}} \bibinfo{year}{2020}\natexlab{}.
\newblock \bibinfo{title}{EOSFuzzer: Fuzzing EOSIO Smart Contracts for
  Vulnerability Detection}.
\newblock
\newblock
\showeprint[arxiv]{cs.SE/2007.14903}


\bibitem[\protect\citeauthoryear{Jain, Kaneko, et~al\mbox{.}}{Jain
  et~al\mbox{.}}{2022}]%
        {SKLEE2022}
\bibfield{author}{\bibinfo{person}{Namrata Jain}, \bibinfo{person}{Kosuke
  Kaneko}, {et~al\mbox{.}}} \bibinfo{year}{2022}\natexlab{}.
\newblock \showarticletitle{SKLEE: A Dynamic Symbolic Analysis Tool
  for Ethereum Smart Contracts (Tool Paper)}. In
  \bibinfo{booktitle}{\emph{Software Engineering and Formal Methods}}.
  \bibinfo{publisher}{Springer International Publishing},
  \bibinfo{address}{Cham}, \bibinfo{pages}{244--250}.
\newblock
\showISBNx{978-3-031-17108-6}


\bibitem[\protect\citeauthoryear{Jeffrey}{Jeffrey}{1988}]%
        {jeffrey1988principles_datalog}
\bibfield{author}{\bibinfo{person}{D~Ullman Jeffrey}.}
  \bibinfo{year}{1988}\natexlab{}.
\newblock \bibinfo{title}{Principles of Database and Knowledge\_Base systems,
  Volume I}.
\newblock
\newblock


\bibitem[\protect\citeauthoryear{Ji, Liang, et~al\mbox{.}}{Ji
  et~al\mbox{.}}{2021b}]%
        {ji2021security}
\bibfield{author}{\bibinfo{person}{Mingtao Ji}, \bibinfo{person}{GuangJun
  Liang}, {et~al\mbox{.}}} \bibinfo{year}{2021}\natexlab{b}.
\newblock \showarticletitle{Security Analysis of Blockchain Smart Contract:
  Taking Reentrancy Vulnerability as an Example}. In
  \bibinfo{booktitle}{\emph{International Conference on Artificial Intelligence
  and Security}}. Springer, \bibinfo{pages}{492--501}.
\newblock


\bibitem[\protect\citeauthoryear{Ji, He, et~al\mbox{.}}{Ji
  et~al\mbox{.}}{2020}]%
        {ji2020deposafe}
\bibfield{author}{\bibinfo{person}{Ru Ji}, \bibinfo{person}{Ningyu He},
  {et~al\mbox{.}}} \bibinfo{year}{2020}\natexlab{}.
\newblock \showarticletitle{Deposafe: Demystifying the fake deposit
  vulnerability in ethereum smart contracts}. In \bibinfo{booktitle}{\emph{2020
  25th International Conference on Engineering of Complex Computer Systems
  (ICECCS)}}. IEEE, \bibinfo{pages}{125--134}.
\newblock


\bibitem[\protect\citeauthoryear{Ji, Kim, et~al\mbox{.}}{Ji
  et~al\mbox{.}}{2021a}]%
        {ji2021evaluating}
\bibfield{author}{\bibinfo{person}{Suhwan Ji}, \bibinfo{person}{Dohyung Kim},
  {et~al\mbox{.}}} \bibinfo{year}{2021}\natexlab{a}.
\newblock \showarticletitle{Evaluating Countermeasures for Verifying the
  Integrity of Ethereum Smart Contract Applications}.
\newblock \bibinfo{journal}{\emph{IEEE Access}}  \bibinfo{volume}{9}
  (\bibinfo{year}{2021}), \bibinfo{pages}{90029--90042}.
\newblock


\bibitem[\protect\citeauthoryear{Jiang, Liu, et~al\mbox{.}}{Jiang
  et~al\mbox{.}}{2018}]%
        {jiang2018contractfuzzer}
\bibfield{author}{\bibinfo{person}{Bo Jiang}, \bibinfo{person}{Ye Liu},
  {et~al\mbox{.}}} \bibinfo{year}{2018}\natexlab{}.
\newblock \showarticletitle{Contractfuzzer: Fuzzing smart contracts for
  vulnerability detection}. In \bibinfo{booktitle}{\emph{2018 33rd IEEE/ACM
  International Conference on Automated Software Engineering (ASE)}}. IEEE,
  \bibinfo{pages}{259--269}.
\newblock


\bibitem[\protect\citeauthoryear{Jiao, Kan, et~al\mbox{.}}{Jiao
  et~al\mbox{.}}{2020}]%
        {semantics2020Jiao}
\bibfield{author}{\bibinfo{person}{Jiao Jiao}, \bibinfo{person}{Shuanglong
  Kan}, {et~al\mbox{.}}} \bibinfo{year}{2020}\natexlab{}.
\newblock \showarticletitle{Semantic Understanding of Smart Contracts:
  Executable Operational Semantics of Solidity}. In
  \bibinfo{booktitle}{\emph{2020 IEEE Symposium on Security and Privacy (SP)}}.
  \bibinfo{pages}{1695--1712}.
\newblock


\bibitem[\protect\citeauthoryear{Kalra, Goel, et~al\mbox{.}}{Kalra
  et~al\mbox{.}}{2018}]%
        {kalra2018zeus}
\bibfield{author}{\bibinfo{person}{Sukrit Kalra}, \bibinfo{person}{Seep Goel},
  {et~al\mbox{.}}} \bibinfo{year}{2018}\natexlab{}.
\newblock \showarticletitle{ZEUS: Analyzing Safety of Smart Contracts}. In
  \bibinfo{booktitle}{\emph{NDSS}}.
\newblock


\bibitem[\protect\citeauthoryear{Kasampalis, Guth, et~al\mbox{.}}{Kasampalis
  et~al\mbox{.}}{2018}]%
        {kasampalis2018iele}
\bibfield{author}{\bibinfo{person}{Theodoros Kasampalis},
  \bibinfo{person}{Dwight Guth}, {et~al\mbox{.}}}
  \bibinfo{year}{2018}\natexlab{}.
\newblock \bibinfo{booktitle}{\emph{Iele: An intermediate-level blockchain
  language designed and implemented using formal semantics}}.
\newblock \bibinfo{type}{{T}echnical {R}eport}.
\newblock


\bibitem[\protect\citeauthoryear{Khan and Namin}{Khan and Namin}{2020}]%
        {khan2020survey}
\bibfield{author}{\bibinfo{person}{Zulfiqar~Ali Khan} {and}
  \bibinfo{person}{Akbar~Siami Namin}.} \bibinfo{year}{2020}\natexlab{}.
\newblock \showarticletitle{A Survey on Vulnerabilities of Ethereum Smart
  Contracts}.
\newblock  (\bibinfo{year}{2020}).
\newblock
\showeprint[arXiv]{2012.14481}


\bibitem[\protect\citeauthoryear{Khor, Masama, et~al\mbox{.}}{Khor
  et~al\mbox{.}}{2020}]%
        {Khor2020improved}
\bibfield{author}{\bibinfo{person}{JingHuey Khor},
  \bibinfo{person}{Mansur~Aliyu Masama}, {et~al\mbox{.}}}
  \bibinfo{year}{2020}\natexlab{}.
\newblock \showarticletitle{An Improved Gas Efficient Library for Securing IoT
  Smart Contracts Against Arithmetic Vulnerabilities}. In
  \bibinfo{booktitle}{\emph{Proceedings of the 2020 9th International
  Conference on Software and Computer Applications}}.
  \bibinfo{pages}{326--330}.
\newblock


\bibitem[\protect\citeauthoryear{{Kim} and {Ryu}}{{Kim} and {Ryu}}{2020}]%
        {kim9230065}
\bibfield{author}{\bibinfo{person}{S. {Kim}} {and} \bibinfo{person}{S. {Ryu}}.}
  \bibinfo{year}{2020}\natexlab{}.
\newblock \showarticletitle{Analysis of Blockchain Smart Contracts: Techniques
  and Insights}. In \bibinfo{booktitle}{\emph{2020 IEEE Secure Development
  (SecDev)}}. \bibinfo{pages}{65--73}.
\newblock


\bibitem[\protect\citeauthoryear{King}{King}{1976}]%
        {king1976symbolic}
\bibfield{author}{\bibinfo{person}{James~C King}.}
  \bibinfo{year}{1976}\natexlab{}.
\newblock \showarticletitle{Symbolic execution and program testing}.
\newblock \bibinfo{journal}{\emph{Commun. ACM}} \bibinfo{volume}{19},
  \bibinfo{number}{7} (\bibinfo{year}{1976}), \bibinfo{pages}{385--394}.
\newblock


\bibitem[\protect\citeauthoryear{Kozen}{Kozen}{1977}]%
        {Kozen1977}
\bibfield{author}{\bibinfo{person}{Dexter~C. Kozen}.}
  \bibinfo{year}{1977}\natexlab{}.
\newblock \bibinfo{booktitle}{\emph{Rice's Theorem}}.
\newblock \bibinfo{publisher}{Springer Berlin Heidelberg},
  \bibinfo{address}{Berlin, Heidelberg}, \bibinfo{pages}{245--248}.
\newblock
\showISBNx{978-3-642-85706-5}


\bibitem[\protect\citeauthoryear{Krupa, Ries, et~al\mbox{.}}{Krupa
  et~al\mbox{.}}{2021}]%
        {krupa2021security}
\bibfield{author}{\bibinfo{person}{Tomas Krupa}, \bibinfo{person}{Michal Ries},
  {et~al\mbox{.}}} \bibinfo{year}{2021}\natexlab{}.
\newblock \showarticletitle{Security Issues of Smart Contracts in Ethereum
  Platforms}. In \bibinfo{booktitle}{\emph{2021 28th Conference of Open
  Innovations Association (FRUCT)}}. IEEE, \bibinfo{pages}{208--214}.
\newblock


\bibitem[\protect\citeauthoryear{Kushwaha, Joshi, et~al\mbox{.}}{Kushwaha
  et~al\mbox{.}}{2022}]%
        {9667515}
\bibfield{author}{\bibinfo{person}{Satpal~Singh Kushwaha},
  \bibinfo{person}{Sandeep Joshi}, {et~al\mbox{.}}}
  \bibinfo{year}{2022}\natexlab{}.
\newblock \showarticletitle{Systematic Review of Security Vulnerabilities in
  Ethereum Blockchain Smart Contract}.
\newblock \bibinfo{journal}{\emph{IEEE Access}}  \bibinfo{volume}{10}
  (\bibinfo{year}{2022}), \bibinfo{pages}{6605--6621}.
\newblock


\bibitem[\protect\citeauthoryear{Lai and Luo}{Lai and Luo}{2020}]%
        {lai2020static}
\bibfield{author}{\bibinfo{person}{Enmei Lai} {and} \bibinfo{person}{Wenjun
  Luo}.} \bibinfo{year}{2020}\natexlab{}.
\newblock \showarticletitle{Static analysis of integer overflow of smart
  contracts in ethereum}. In \bibinfo{booktitle}{\emph{Proceedings of the 2020
  4th International Conference on Cryptography, Security and Privacy}}.
  \bibinfo{pages}{110--115}.
\newblock


\bibitem[\protect\citeauthoryear{Lamela~Seijas, Nemish,
  et~al\mbox{.}}{Lamela~Seijas et~al\mbox{.}}{2020}]%
        {Seijas2020Marlowe}
\bibfield{author}{\bibinfo{person}{Pablo Lamela~Seijas},
  \bibinfo{person}{Alexander Nemish}, {et~al\mbox{.}}}
  \bibinfo{year}{2020}\natexlab{}.
\newblock \showarticletitle{Marlowe: Implementing and Analysing Financial
  Contracts on Blockchain}. In \bibinfo{booktitle}{\emph{Financial Cryptography
  and Data Security}}. \bibinfo{publisher}{Springer International Publishing},
  \bibinfo{address}{Cham}, \bibinfo{pages}{496--511}.
\newblock
\showISBNx{978-3-030-54455-3}


\bibitem[\protect\citeauthoryear{Lamela~Seijas and Thompson}{Lamela~Seijas and
  Thompson}{2018}]%
        {lamela2018marlowe}
\bibfield{author}{\bibinfo{person}{Pablo Lamela~Seijas} {and}
  \bibinfo{person}{Simon Thompson}.} \bibinfo{year}{2018}\natexlab{}.
\newblock \showarticletitle{Marlowe: Financial contracts on blockchain}. In
  \bibinfo{booktitle}{\emph{International Symposium on Leveraging Applications
  of Formal Methods}}. Springer, \bibinfo{pages}{356--375}.
\newblock


\bibitem[\protect\citeauthoryear{Lamport}{Lamport}{1977}]%
        {Lamport77}
\bibfield{author}{\bibinfo{person}{L. Lamport}.}
  \bibinfo{year}{1977}\natexlab{}.
\newblock \showarticletitle{Proving the Correctness of Multiprocess Programs}.
\newblock \bibinfo{journal}{\emph{IEEE Transactions on Software Engineering}}
  \bibinfo{volume}{SE-3}, \bibinfo{number}{2} (\bibinfo{year}{1977}),
  \bibinfo{pages}{125--143}.
\newblock


\bibitem[\protect\citeauthoryear{Lamport}{Lamport}{1983a}]%
        {lamport1983specifying}
\bibfield{author}{\bibinfo{person}{Leslie Lamport}.}
  \bibinfo{year}{1983}\natexlab{a}.
\newblock \showarticletitle{Specifying concurrent program modules}.
\newblock \bibinfo{journal}{\emph{ACM Transactions on Programming Languages and
  Systems}} \bibinfo{volume}{5}, \bibinfo{number}{2} (\bibinfo{year}{1983}),
  \bibinfo{pages}{190--222}.
\newblock


\bibitem[\protect\citeauthoryear{Lamport}{Lamport}{1983b}]%
        {lamport1983good}
\bibfield{author}{\bibinfo{person}{Leslie Lamport}.}
  \bibinfo{year}{1983}\natexlab{b}.
\newblock \showarticletitle{What good is temporal logic?}. In
  \bibinfo{booktitle}{\emph{IFIP congress}}, Vol.~\bibinfo{volume}{83}.
  \bibinfo{pages}{657--668}.
\newblock


\bibitem[\protect\citeauthoryear{Lerch, Hermann, et~al\mbox{.}}{Lerch
  et~al\mbox{.}}{2014}]%
        {lerch2014flowtwist}
\bibfield{author}{\bibinfo{person}{Johannes Lerch}, \bibinfo{person}{Ben
  Hermann}, {et~al\mbox{.}}} \bibinfo{year}{2014}\natexlab{}.
\newblock \showarticletitle{FlowTwist: efficient context-sensitive inside-out
  taint analysis for large codebases}. In \bibinfo{booktitle}{\emph{Proceedings
  of the 22nd ACM SIGSOFT International Symposium on Foundations of Software
  Engineering}}. \bibinfo{pages}{98--108}.
\newblock


\bibitem[\protect\citeauthoryear{Li and Long}{Li and Long}{2018}]%
        {SOLAR2018LiAo}
\bibfield{author}{\bibinfo{person}{Ao Li} {and} \bibinfo{person}{Fan Long}.}
  \bibinfo{year}{2018}\natexlab{}.
\newblock \bibinfo{title}{Detecting Standard Violation Errors in Smart
  Contracts}.
\newblock
\newblock
\showeprint[arxiv]{1812.07702}


\bibitem[\protect\citeauthoryear{Li, He, et~al\mbox{.}}{Li
  et~al\mbox{.}}{2022a}]%
        {li2022eosioanalyzer}
\bibfield{author}{\bibinfo{person}{Wenyuan Li}, \bibinfo{person}{Jiahao He},
  {et~al\mbox{.}}} \bibinfo{year}{2022}\natexlab{a}.
\newblock \showarticletitle{EOSIOAnalyzer: An Effective Static Analysis
  Vulnerability Detection Framework for EOSIO Smart Contracts}. In
  \bibinfo{booktitle}{\emph{2022 IEEE 46th Annual Computers, Software, and
  Applications Conference (COMPSAC)}}. IEEE, \bibinfo{pages}{746--756}.
\newblock


\bibitem[\protect\citeauthoryear{Li, Wang, et~al\mbox{.}}{Li
  et~al\mbox{.}}{2022c}]%
        {Li2022}
\bibfield{author}{\bibinfo{person}{Wenyin Li}, \bibinfo{person}{Meng Wang},
  {et~al\mbox{.}}} \bibinfo{year}{2022}\natexlab{c}.
\newblock \showarticletitle{Grey-box Fuzzing Based on Execution Feedback for
  EOSIO Smart Contracts}. In \bibinfo{booktitle}{\emph{2022 29th Asia-Pacific
  Software Engineering Conference (APSEC)}}. \bibinfo{pages}{1--10}.
\newblock
\urldef\tempurl%
\url{https://doi.org/10.1109/APSEC57359.2022.00012}
\showDOI{\tempurl}


\bibitem[\protect\citeauthoryear{Li, Chen, et~al\mbox{.}}{Li
  et~al\mbox{.}}{2021}]%
        {Li2021Clue}
\bibfield{author}{\bibinfo{person}{Xiaoqi Li}, \bibinfo{person}{Ting Chen},
  {et~al\mbox{.}}} \bibinfo{year}{2021}\natexlab{}.
\newblock \bibinfo{booktitle}{\emph{CLUE: Towards Discovering Locked
  Cryptocurrencies in Ethereum}}.
\newblock \bibinfo{publisher}{Association for Computing Machinery},
  \bibinfo{address}{New York, NY, USA}, \bibinfo{pages}{1584–1587}.
\newblock
\showISBNx{9781450381048}


\bibitem[\protect\citeauthoryear{Li, Liu, et~al\mbox{.}}{Li
  et~al\mbox{.}}{2020b}]%
        {li2020protect}
\bibfield{author}{\bibinfo{person}{Yue Li}, \bibinfo{person}{Han Liu},
  {et~al\mbox{.}}} \bibinfo{year}{2020}\natexlab{b}.
\newblock \showarticletitle{Protect Your Smart Contract Against Unfair
  Payment}. In \bibinfo{booktitle}{\emph{2020 International Symposium on
  Reliable Distributed Systems (SRDS)}}. IEEE, \bibinfo{pages}{61--70}.
\newblock


\bibitem[\protect\citeauthoryear{Li, Guo, et~al\mbox{.}}{Li
  et~al\mbox{.}}{2020a}]%
        {li2020research}
\bibfield{author}{\bibinfo{person}{Ziyuan Li}, \bibinfo{person}{Wangshu Guo},
  {et~al\mbox{.}}} \bibinfo{year}{2020}\natexlab{a}.
\newblock \showarticletitle{Research on Blockchain Smart Contracts
  Vulnerability and A Code Audit Tool based on Matching Rules}. In
  \bibinfo{booktitle}{\emph{Proceedings of the 2020 International Conference on
  Cyberspace Innovation of Advanced Technologies}}. \bibinfo{pages}{484--489}.
\newblock


\bibitem[\protect\citeauthoryear{Li, Lu, et~al\mbox{.}}{Li
  et~al\mbox{.}}{2022b}]%
        {li2022smartfast}
\bibfield{author}{\bibinfo{person}{Zhaoxuan Li}, \bibinfo{person}{Siqi Lu},
  {et~al\mbox{.}}} \bibinfo{year}{2022}\natexlab{b}.
\newblock \showarticletitle{SmartFast: an accurate and robust formal analysis
  tool for Ethereum smart contracts}.
\newblock \bibinfo{journal}{\emph{Empirical Software Engineering}}
  \bibinfo{volume}{27}, \bibinfo{number}{7} (\bibinfo{year}{2022}),
  \bibinfo{pages}{1--52}.
\newblock


\bibitem[\protect\citeauthoryear{Liao, Tsai, et~al\mbox{.}}{Liao
  et~al\mbox{.}}{2019}]%
        {Soliaudit2019}
\bibfield{author}{\bibinfo{person}{Jian-Wei Liao}, \bibinfo{person}{Tsung-Ta
  Tsai}, {et~al\mbox{.}}} \bibinfo{year}{2019}\natexlab{}.
\newblock \showarticletitle{SoliAudit: Smart Contract Vulnerability Assessment
  Based on Machine Learning and Fuzz Testing}. In
  \bibinfo{booktitle}{\emph{2019 Sixth International Conference on Internet of
  Things: Systems, Management and Security (IOTSMS)}}.
  \bibinfo{pages}{458--465}.
\newblock


\bibitem[\protect\citeauthoryear{Liu, Chen, et~al\mbox{.}}{Liu
  et~al\mbox{.}}{2022a}]%
        {liu2022learning}
\bibfield{author}{\bibinfo{person}{Junrui Liu}, \bibinfo{person}{Yanju Chen},
  {et~al\mbox{.}}} \bibinfo{year}{2022}\natexlab{a}.
\newblock \showarticletitle{Learning Contract Invariants Using Reinforcement
  Learning}. In \bibinfo{booktitle}{\emph{37th IEEE/ACM International
  Conference on Automated Software Engineering (ASE 2022)}}.
\newblock


\bibitem[\protect\citeauthoryear{Liu and Liu}{Liu and Liu}{2019}]%
        {liu2019survey}
\bibfield{author}{\bibinfo{person}{Jing Liu} {and} \bibinfo{person}{Zhentian
  Liu}.} \bibinfo{year}{2019}\natexlab{}.
\newblock \showarticletitle{A survey on security verification of blockchain
  smart contracts}.
\newblock \bibinfo{journal}{\emph{IEEE Access}}  \bibinfo{volume}{7}
  (\bibinfo{year}{2019}), \bibinfo{pages}{77894--77904}.
\newblock


\bibitem[\protect\citeauthoryear{Liu and Li}{Liu and Li}{2022}]%
        {liu2022invcon}
\bibfield{author}{\bibinfo{person}{Ye Liu} {and} \bibinfo{person}{Yi Li}.}
  \bibinfo{year}{2022}\natexlab{}.
\newblock \showarticletitle{InvCon: A Dynamic Invariant Detector for Ethereum
  Smart Contracts}. In \bibinfo{booktitle}{\emph{37th IEEE/ACM International
  Conference on Automated Software Engineering}}. \bibinfo{pages}{1--4}.
\newblock


\bibitem[\protect\citeauthoryear{Liu, Xu, et~al\mbox{.}}{Liu
  et~al\mbox{.}}{2022b}]%
        {Liu2022}
\bibfield{author}{\bibinfo{person}{Yiping Liu}, \bibinfo{person}{Jie Xu},
  {et~al\mbox{.}}} \bibinfo{year}{2022}\natexlab{b}.
\newblock \showarticletitle{Smart Contract Vulnerability Detection Based
  on Symbolic Execution Technology}. In \bibinfo{booktitle}{\emph{Cyber
  Security}}, \bibfield{editor}{\bibinfo{person}{Wei Lu},
  \bibinfo{person}{Yuqing Zhang}, {et~al\mbox{.}}} (Eds.).
  \bibinfo{publisher}{Springer Nature Singapore}, \bibinfo{address}{Singapore},
  \bibinfo{pages}{193--207}.
\newblock
\showISBNx{978-981-16-9229-1}


\bibitem[\protect\citeauthoryear{Liu, Qian, et~al\mbox{.}}{Liu
  et~al\mbox{.}}{2021}]%
        {liu2021combining}
\bibfield{author}{\bibinfo{person}{Zhenguang Liu}, \bibinfo{person}{Peng Qian},
  {et~al\mbox{.}}} \bibinfo{year}{2021}\natexlab{}.
\newblock \showarticletitle{Combining graph neural networks with expert
  knowledge for smart contract vulnerability detection}.
\newblock \bibinfo{journal}{\emph{IEEE Transactions on Knowledge and Data
  Engineering}} (\bibinfo{year}{2021}).
\newblock


\bibitem[\protect\citeauthoryear{L{\'o}pez~Vivar, Castedo,
  et~al\mbox{.}}{L{\'o}pez~Vivar et~al\mbox{.}}{2020}]%
        {lopez2020smart}
\bibfield{author}{\bibinfo{person}{Antonio L{\'o}pez~Vivar},
  \bibinfo{person}{Alberto~Tur{\'e}gano Castedo}, {et~al\mbox{.}}}
  \bibinfo{year}{2020}\natexlab{}.
\newblock \showarticletitle{Smart Contracts: A Review of Security Threats
  Alongside an Analysis of Existing Solutions}.
\newblock \bibinfo{journal}{\emph{Entropy}} \bibinfo{volume}{22},
  \bibinfo{number}{2} (\bibinfo{year}{2020}), \bibinfo{pages}{203}.
\newblock


\bibitem[\protect\citeauthoryear{Lorenz~Breidenbach and
  Sirer}{Lorenz~Breidenbach and Sirer}{2017}]%
        {ParityWalletBug}
\bibfield{author}{\bibinfo{person}{Ari~Juels Lorenz~Breidenbach, Phil~Daian}
  {and} \bibinfo{person}{Emin~Gün Sirer}.} \bibinfo{year}{2017}\natexlab{}.
\newblock \bibinfo{title}{An In-Depth Look at the Parity Multisig Bug}.
\newblock
\newblock
\urldef\tempurl%
\url{https://hackingdistributed.com/2017/07/22/deep-dive-parity-bug/}
\showURL{%
Retrieved 22-11-30 from \tempurl}


\bibitem[\protect\citeauthoryear{Lu, Wang, et~al\mbox{.}}{Lu
  et~al\mbox{.}}{2021}]%
        {lu2021neucheck}
\bibfield{author}{\bibinfo{person}{Ning Lu}, \bibinfo{person}{Bin Wang},
  {et~al\mbox{.}}} \bibinfo{year}{2021}\natexlab{}.
\newblock \showarticletitle{NeuCheck: A more practical Ethereum smart contract
  security analysis tool}.
\newblock \bibinfo{journal}{\emph{Software: Practice and Experience}}
  \bibinfo{volume}{51}, \bibinfo{number}{10} (\bibinfo{year}{2021}),
  \bibinfo{pages}{2065--2084}.
\newblock


\bibitem[\protect\citeauthoryear{Luu, Chu, et~al\mbox{.}}{Luu
  et~al\mbox{.}}{2016}]%
        {Oyente2016}
\bibfield{author}{\bibinfo{person}{Loi Luu}, \bibinfo{person}{Duc-Hiep Chu},
  {et~al\mbox{.}}} \bibinfo{year}{2016}\natexlab{}.
\newblock \showarticletitle{Making Smart Contracts Smarter}. In
  \bibinfo{booktitle}{\emph{Proceedings of the 2016 ACM SIGSAC Conference on
  Computer and Communications Security}}. \bibinfo{publisher}{Association for
  Computing Machinery}, \bibinfo{address}{New York, NY, USA},
  \bibinfo{pages}{254–269}.
\newblock
\showISBNx{9781450341394}


\bibitem[\protect\citeauthoryear{Ma, Xu, et~al\mbox{.}}{Ma
  et~al\mbox{.}}{2021}]%
        {ma2021pluto}
\bibfield{author}{\bibinfo{person}{Fuchen Ma}, \bibinfo{person}{Zhenyang Xu},
  {et~al\mbox{.}}} \bibinfo{year}{2021}\natexlab{}.
\newblock \showarticletitle{Pluto: Exposing Vulnerabilities in Inter-Contract
  Scenarios}.
\newblock \bibinfo{journal}{\emph{IEEE Transactions on Software Engineering}}
  (\bibinfo{year}{2021}).
\newblock


\bibitem[\protect\citeauthoryear{Marescotti, Otoni, et~al\mbox{.}}{Marescotti
  et~al\mbox{.}}{2020}]%
        {marescotti2020accurate}
\bibfield{author}{\bibinfo{person}{Matteo Marescotti}, \bibinfo{person}{Rodrigo
  Otoni}, {et~al\mbox{.}}} \bibinfo{year}{2020}\natexlab{}.
\newblock \showarticletitle{Accurate smart contract verification through direct
  modelling}. In \bibinfo{booktitle}{\emph{International Symposium on
  Leveraging Applications of Formal Methods}}. Springer,
  \bibinfo{pages}{178--194}.
\newblock


\bibitem[\protect\citeauthoryear{Maurer and Lewis}{Maurer and Lewis}{1975}]%
        {maurer1975hash}
\bibfield{author}{\bibinfo{person}{Ward~Douglas Maurer} {and}
  \bibinfo{person}{Ted~G Lewis}.} \bibinfo{year}{1975}\natexlab{}.
\newblock \showarticletitle{Hash table methods}.
\newblock \bibinfo{journal}{\emph{ACM Computing Surveys (CSUR)}}
  \bibinfo{volume}{7}, \bibinfo{number}{1} (\bibinfo{year}{1975}),
  \bibinfo{pages}{5--19}.
\newblock


\bibitem[\protect\citeauthoryear{Mei, Ashraf, et~al\mbox{.}}{Mei
  et~al\mbox{.}}{2019}]%
        {8859497}
\bibfield{author}{\bibinfo{person}{Xiupei Mei}, \bibinfo{person}{Imran Ashraf},
  {et~al\mbox{.}}} \bibinfo{year}{2019}\natexlab{}.
\newblock \showarticletitle{A Fuzz Testing Service for Assuring Smart
  Contracts}. In \bibinfo{booktitle}{\emph{2019 IEEE 19th International
  Conference on Software Quality, Reliability and Security Companion (QRS-C)}}.
  \bibinfo{pages}{544--545}.
\newblock


\bibitem[\protect\citeauthoryear{Mense and Flatscher}{Mense and
  Flatscher}{2018}]%
        {menseVulnerabilities}
\bibfield{author}{\bibinfo{person}{Alexander Mense} {and}
  \bibinfo{person}{Markus Flatscher}.} \bibinfo{year}{2018}\natexlab{}.
\newblock \showarticletitle{Security Vulnerabilities in Ethereum Smart
  Contracts}. \bibinfo{publisher}{Association for Computing Machinery},
  \bibinfo{address}{New York, NY, USA}, \bibinfo{pages}{375–380}.
\newblock
\showISBNx{9781450364799}


\bibitem[\protect\citeauthoryear{Mi, Wang, et~al\mbox{.}}{Mi
  et~al\mbox{.}}{2021}]%
        {mi2021vscl}
\bibfield{author}{\bibinfo{person}{Feng Mi}, \bibinfo{person}{Zhuoyi Wang},
  {et~al\mbox{.}}} \bibinfo{year}{2021}\natexlab{}.
\newblock \showarticletitle{VSCL: Automating Vulnerability Detection in Smart
  Contracts with Deep Learning}. In \bibinfo{booktitle}{\emph{2021 IEEE
  International Conference on Blockchain and Cryptocurrency (ICBC)}}. IEEE,
  \bibinfo{pages}{1--9}.
\newblock


\bibitem[\protect\citeauthoryear{Michelson.org}{Michelson.org}{nd}]%
        {goodmanimichelson}
\bibfield{author}{\bibinfo{person}{Michelson.org}.}
  \bibinfo{year}{n.d.}\natexlab{}.
\newblock \showarticletitle{Michelson: the language of Tezos Smart Contracts}.
\newblock  (\bibinfo{year}{n.d.}).
\newblock
\urldef\tempurl%
\url{https://www.michelson.org/}
\showURL{%
Retrieved December 08, 2022 from \tempurl}


\bibitem[\protect\citeauthoryear{Milo, Nielsen, et~al\mbox{.}}{Milo
  et~al\mbox{.}}{2022}]%
        {milo2022finding}
\bibfield{author}{\bibinfo{person}{Mikkel Milo}, \bibinfo{person}{Eske~Hoy
  Nielsen}, {et~al\mbox{.}}} \bibinfo{year}{2022}\natexlab{}.
\newblock \showarticletitle{Finding smart contract vulnerabilities with
  ConCert's property-based testing framework}.
\newblock  (\bibinfo{year}{2022}).
\newblock
\showeprint[arxiv]{2208.00758}


\bibitem[\protect\citeauthoryear{Momeni, Wang, et~al\mbox{.}}{Momeni
  et~al\mbox{.}}{2019}]%
        {MLmodel2019}
\bibfield{author}{\bibinfo{person}{Pouyan Momeni}, \bibinfo{person}{Yu Wang},
  {et~al\mbox{.}}} \bibinfo{year}{2019}\natexlab{}.
\newblock \showarticletitle{Machine Learning Model for Smart Contracts Security
  Analysis}. In \bibinfo{booktitle}{\emph{2019 17th International Conference on
  Privacy, Security and Trust (PST)}}. \bibinfo{pages}{1--6}.
\newblock


\bibitem[\protect\citeauthoryear{Mossberg, Manzano, et~al\mbox{.}}{Mossberg
  et~al\mbox{.}}{2019}]%
        {mossberg2019manticore}
\bibfield{author}{\bibinfo{person}{Mark Mossberg}, \bibinfo{person}{Felipe
  Manzano}, {et~al\mbox{.}}} \bibinfo{year}{2019}\natexlab{}.
\newblock \showarticletitle{Manticore: A user-friendly symbolic execution
  framework for binaries and smart contracts}. In
  \bibinfo{booktitle}{\emph{2019 34th IEEE/ACM International Conference on
  Automated Software Engineering (ASE)}}. IEEE, \bibinfo{pages}{1186--1189}.
\newblock


\bibitem[\protect\citeauthoryear{Mudgal, Li, et~al\mbox{.}}{Mudgal
  et~al\mbox{.}}{2018}]%
        {mudgal2018deep}
\bibfield{author}{\bibinfo{person}{Sidharth Mudgal}, \bibinfo{person}{Han Li},
  {et~al\mbox{.}}} \bibinfo{year}{2018}\natexlab{}.
\newblock \showarticletitle{Deep learning for entity matching: A design space
  exploration}. In \bibinfo{booktitle}{\emph{Proceedings of the 2018
  International Conference on Management of Data}}. \bibinfo{pages}{19--34}.
\newblock


\bibitem[\protect\citeauthoryear{Nakamoto}{Nakamoto}{2008}]%
        {nakamoto2008bitcoin}
\bibfield{author}{\bibinfo{person}{Satoshi Nakamoto}.}
  \bibinfo{year}{2008}\natexlab{}.
\newblock \showarticletitle{Bitcoin: A peer-to-peer electronic cash system}.
\newblock \bibinfo{journal}{\emph{Decentralized Business Review}}
  (\bibinfo{year}{2008}), \bibinfo{pages}{21260}.
\newblock


\bibitem[\protect\citeauthoryear{Nam and Kil}{Nam and Kil}{2022}]%
        {ATL2022}
\bibfield{author}{\bibinfo{person}{Wonhong Nam} {and}
  \bibinfo{person}{Hyunyoung Kil}.} \bibinfo{year}{2022}\natexlab{}.
\newblock \showarticletitle{Formal Verification of Blockchain Smart Contracts
  via ATL Model Checking}.
\newblock \bibinfo{journal}{\emph{IEEE Access}}  \bibinfo{volume}{10}
  (\bibinfo{year}{2022}), \bibinfo{pages}{8151--8162}.
\newblock


\bibitem[\protect\citeauthoryear{Narayana and Sathiyamurthy}{Narayana and
  Sathiyamurthy}{2021}]%
        {narayana2021automation}
\bibfield{author}{\bibinfo{person}{K~Lakshmi Narayana} {and} \bibinfo{person}{K
  Sathiyamurthy}.} \bibinfo{year}{2021}\natexlab{}.
\newblock \showarticletitle{Automation and smart materials in detecting smart
  contracts vulnerabilities in Blockchain using deep learning}.
\newblock \bibinfo{journal}{\emph{Materials Today: Proceedings}}
  (\bibinfo{year}{2021}).
\newblock


\bibitem[\protect\citeauthoryear{Nassirzadeh, Sun, et~al\mbox{.}}{Nassirzadeh
  et~al\mbox{.}}{2021}]%
        {nassirzadeh2021gas}
\bibfield{author}{\bibinfo{person}{Behkish Nassirzadeh},
  \bibinfo{person}{Huaiying Sun}, {et~al\mbox{.}}}
  \bibinfo{year}{2021}\natexlab{}.
\newblock \showarticletitle{Gas Gauge: A Security Analysis Tool for Smart
  Contract Out-of-Gas Vulnerabilities}.
\newblock  (\bibinfo{year}{2021}).
\newblock
\showeprint[arxiv]{2112.14771}


\bibitem[\protect\citeauthoryear{Nelaturu, Mavridoul, et~al\mbox{.}}{Nelaturu
  et~al\mbox{.}}{2020}]%
        {nelaturu2020verified}
\bibfield{author}{\bibinfo{person}{Keerthi Nelaturu},
  \bibinfo{person}{Anastasia Mavridoul}, {et~al\mbox{.}}}
  \bibinfo{year}{2020}\natexlab{}.
\newblock \showarticletitle{Verified development and deployment of multiple
  interacting smart contracts with VeriSolid}. In
  \bibinfo{booktitle}{\emph{2020 IEEE International Conference on Blockchain
  and Cryptocurrency (ICBC)}}. IEEE, \bibinfo{pages}{1--9}.
\newblock


\bibitem[\protect\citeauthoryear{Nguyen, Nguyen, et~al\mbox{.}}{Nguyen
  et~al\mbox{.}}{2022}]%
        {MANDO2022}
\bibfield{author}{\bibinfo{person}{Hoang~H. Nguyen}, \bibinfo{person}{Nhat-Minh
  Nguyen}, {et~al\mbox{.}}} \bibinfo{year}{2022}\natexlab{}.
\newblock \showarticletitle{MANDO: Multi-Level Heterogeneous Graph Embeddings
  for Fine-Grained Detection of Smart Contract Vulnerabilities}. In
  \bibinfo{booktitle}{\emph{2022 IEEE 9th International Conference on Data
  Science and Advanced Analytics (DSAA)}}. \bibinfo{pages}{1--10}.
\newblock
\urldef\tempurl%
\url{https://doi.org/10.1109/DSAA54385.2022.10032337}
\showDOI{\tempurl}


\bibitem[\protect\citeauthoryear{Nguyen, Pham, et~al\mbox{.}}{Nguyen
  et~al\mbox{.}}{2020}]%
        {nguyen2020sfuzz}
\bibfield{author}{\bibinfo{person}{Tai~D Nguyen}, \bibinfo{person}{Long~H
  Pham}, {et~al\mbox{.}}} \bibinfo{year}{2020}\natexlab{}.
\newblock \showarticletitle{sfuzz: An efficient adaptive fuzzer for solidity
  smart contracts}. In \bibinfo{booktitle}{\emph{Proceedings of the ACM/IEEE
  42nd International Conference on Software Engineering}}.
  \bibinfo{pages}{778--788}.
\newblock


\bibitem[\protect\citeauthoryear{Nha and Thuan}{Nha and Thuan}{2022}]%
        {Nha2022}
\bibfield{author}{\bibinfo{person}{Bui~Tong Nha} {and}
  \bibinfo{person}{Nguyen~Dinh Thuan}.} \bibinfo{year}{2022}\natexlab{}.
\newblock \showarticletitle{Methodology Interaction by Machine Learning Model
  to Detect Vulnerability in Smart Contract of Blockchain}. In
  \bibinfo{booktitle}{\emph{2022 RIVF International Conference on Computing and
  Communication Technologies (RIVF)}}. \bibinfo{pages}{112--117}.
\newblock
\urldef\tempurl%
\url{https://doi.org/10.1109/RIVF55975.2022.10013832}
\showDOI{\tempurl}


\bibitem[\protect\citeauthoryear{Nikolic}{Nikolic}{2018}]%
        {MAIAN}
\bibfield{author}{\bibinfo{person}{Ivica Nikolic}.}
  \bibinfo{year}{2018}\natexlab{}.
\newblock \bibinfo{title}{MAIAN}.
\newblock
  \bibinfo{howpublished}{\url{https://github.com/ivicanikolicsg/MAIAN}}.
\newblock


\bibitem[\protect\citeauthoryear{Nikoli{\'c}, Kolluri,
  et~al\mbox{.}}{Nikoli{\'c} et~al\mbox{.}}{2018}]%
        {nikolic2018finding}
\bibfield{author}{\bibinfo{person}{Ivica Nikoli{\'c}}, \bibinfo{person}{Aashish
  Kolluri}, {et~al\mbox{.}}} \bibinfo{year}{2018}\natexlab{}.
\newblock \showarticletitle{Finding the greedy, prodigal, and suicidal
  contracts at scale}. In \bibinfo{booktitle}{\emph{Proceedings of the 34th
  annual computer security applications conference}}.
  \bibinfo{pages}{653--663}.
\newblock


\bibitem[\protect\citeauthoryear{Nipkow, Paulson, et~al\mbox{.}}{Nipkow
  et~al\mbox{.}}{2002}]%
        {nipkow2002isabelle}
\bibfield{author}{\bibinfo{person}{Tobias Nipkow}, \bibinfo{person}{Lawrence~C
  Paulson}, {et~al\mbox{.}}} \bibinfo{year}{2002}\natexlab{}.
\newblock \bibinfo{booktitle}{\emph{Isabelle/HOL: a proof assistant for
  higher-order logic}}. Vol.~\bibinfo{volume}{2283}.
\newblock \bibinfo{publisher}{Springer Science \& Business Media}.
\newblock


\bibitem[\protect\citeauthoryear{Nishida, Saito, et~al\mbox{.}}{Nishida
  et~al\mbox{.}}{2022}]%
        {nishida2022helmholtz}
\bibfield{author}{\bibinfo{person}{Yuki Nishida}, \bibinfo{person}{Hiromasa
  Saito}, {et~al\mbox{.}}} \bibinfo{year}{2022}\natexlab{}.
\newblock \showarticletitle{Helmholtz: A Verifier for Tezos Smart Contracts
  Based on Refinement Types}.
\newblock \bibinfo{journal}{\emph{New Generation Computing}}
  (\bibinfo{year}{2022}), \bibinfo{pages}{1--34}.
\newblock


\bibitem[\protect\citeauthoryear{Noor~Aidee, Johar, et~al\mbox{.}}{Noor~Aidee
  et~al\mbox{.}}{2021}]%
        {aidee2021vulnerability}
\bibfield{author}{\bibinfo{person}{Nurul~Aida Noor~Aidee},
  \bibinfo{person}{Md~Gapar~Md Johar}, {et~al\mbox{.}}}
  \bibinfo{year}{2021}\natexlab{}.
\newblock \showarticletitle{Vulnerability Assessment on Ethereum Based Smart
  Contract Applications}. In \bibinfo{booktitle}{\emph{2021 IEEE International
  Conference on Automatic Control Intelligent Systems (I2CACIS)}}.
  \bibinfo{pages}{13--18}.
\newblock


\bibitem[\protect\citeauthoryear{of~Bits~Blog}{of~Bits~Blog}{2019}]%
        {gridLock}
\bibfield{author}{\bibinfo{person}{Trail of Bits~Blog}.}
  \bibinfo{year}{2019}\natexlab{}.
\newblock \bibinfo{title}{Avoiding Smart Contract “Gridlock” with Slither}.
\newblock
\newblock
\urldef\tempurl%
\url{https://blog.trailofbits.com/2019/07/03/avoiding-smart-contract-gridlock-with-slither/}
\showURL{%
Retrieved December 08, 2022 from \tempurl}


\bibitem[\protect\citeauthoryear{of~the Ether~Throne}{of~the
  Ether~Throne}{2016}]%
        {KOTH}
\bibfield{author}{\bibinfo{person}{King of~the Ether~Throne}.}
  \bibinfo{year}{2016}\natexlab{}.
\newblock \bibinfo{title}{King of the Ether Throne: Post mortem investigation}.
\newblock
\newblock
\urldef\tempurl%
\url{https://www.kingoftheether.com/postmortem.html}
\showURL{%
Retrieved 2022-11-30 from \tempurl}


\bibitem[\protect\citeauthoryear{Park}{Park}{2019}]%
        {vyperIssue}
\bibfield{author}{\bibinfo{person}{Daejun Park}.}
  \bibinfo{year}{2019}\natexlab{}.
\newblock \bibinfo{title}{Vyper Issue 1761: Potentially insufficient gas
  stipend for precompiled contract calls}.
\newblock
\newblock
\urldef\tempurl%
\url{https://github.com/vyperlang/vyper/issues/1761}
\showURL{%
Retrieved March 21, 2022 from \tempurl}


\bibitem[\protect\citeauthoryear{Park, Zhang, et~al\mbox{.}}{Park
  et~al\mbox{.}}{2020}]%
        {park2020end}
\bibfield{author}{\bibinfo{person}{Daejun Park}, \bibinfo{person}{Yi Zhang},
  {et~al\mbox{.}}} \bibinfo{year}{2020}\natexlab{}.
\newblock \showarticletitle{End-to-end formal verification of ethereum 2.0
  deposit smart contract}. In \bibinfo{booktitle}{\emph{International
  Conference on Computer Aided Verification}}. Springer,
  \bibinfo{pages}{151--164}.
\newblock


\bibitem[\protect\citeauthoryear{Park, Zhang, et~al\mbox{.}}{Park
  et~al\mbox{.}}{2018}]%
        {Daejun2018FormalVerification}
\bibfield{author}{\bibinfo{person}{Daejun Park}, \bibinfo{person}{Yi Zhang},
  {et~al\mbox{.}}} \bibinfo{year}{2018}\natexlab{}.
\newblock \showarticletitle{A Formal Verification Tool for Ethereum VM
  Bytecode}. In \bibinfo{booktitle}{\emph{Proceedings of the 2018 26th ACM
  Joint Meeting on European Software Engineering Conference and Symposium on
  the Foundations of Software Engineering}}. \bibinfo{publisher}{Association
  for Computing Machinery}, \bibinfo{address}{New York, NY, USA},
  \bibinfo{pages}{912–915}.
\newblock
\showISBNx{9781450355735}


\bibitem[\protect\citeauthoryear{Parvez}{Parvez}{2016}]%
        {parvez2016combining}
\bibfield{author}{\bibinfo{person}{Muhammad~Riyad Parvez}.}
  \bibinfo{year}{2016}\natexlab{}.
\newblock \emph{\bibinfo{title}{Combining static analysis and targeted symbolic
  execution for scalable bug-finding in application binaries}}.
\newblock \bibinfo{thesistype}{Master's\ thesis}. \bibinfo{school}{University
  of Waterloo}.
\newblock


\bibitem[\protect\citeauthoryear{Patrignani and Blackshear}{Patrignani and
  Blackshear}{2021}]%
        {patrignani2021robust}
\bibfield{author}{\bibinfo{person}{Marco Patrignani} {and} \bibinfo{person}{Sam
  Blackshear}.} \bibinfo{year}{2021}\natexlab{}.
\newblock \showarticletitle{Robust Safety for Move}.
\newblock  (\bibinfo{year}{2021}).
\newblock
\showeprint[arxiv]{2110.05043}


\bibitem[\protect\citeauthoryear{Peng, Akca, et~al\mbox{.}}{Peng
  et~al\mbox{.}}{2019}]%
        {peng2019sif}
\bibfield{author}{\bibinfo{person}{Chao Peng}, \bibinfo{person}{Sefa Akca},
  {et~al\mbox{.}}} \bibinfo{year}{2019}\natexlab{}.
\newblock \showarticletitle{SIF: A framework for solidity contract
  instrumentation and analysis}. In \bibinfo{booktitle}{\emph{2019 26th
  Asia-Pacific Software Engineering Conference (APSEC)}}. IEEE,
  \bibinfo{pages}{466--473}.
\newblock


\bibitem[\protect\citeauthoryear{Perez and Livshits}{Perez and
  Livshits}{2021}]%
        {perez2021smart}
\bibfield{author}{\bibinfo{person}{Daniel Perez} {and}
  \bibinfo{person}{Benjamin Livshits}.} \bibinfo{year}{2021}\natexlab{}.
\newblock \showarticletitle{Smart contract vulnerabilities: Vulnerable does not
  imply exploited}. In \bibinfo{booktitle}{\emph{30th USENIX Security Symposium
  (USENIX Security 21)}}. \bibinfo{pages}{1325--1341}.
\newblock


\bibitem[\protect\citeauthoryear{Permenev, Dimitrov, et~al\mbox{.}}{Permenev
  et~al\mbox{.}}{2020}]%
        {permenev2020verx}
\bibfield{author}{\bibinfo{person}{Anton Permenev}, \bibinfo{person}{Dimitar
  Dimitrov}, {et~al\mbox{.}}} \bibinfo{year}{2020}\natexlab{}.
\newblock \showarticletitle{Verx: Safety verification of smart contracts}. In
  \bibinfo{booktitle}{\emph{2020 IEEE Symposium on Security and Privacy (SP)}}.
  IEEE, \bibinfo{pages}{1661--1677}.
\newblock


\bibitem[\protect\citeauthoryear{Piantadosi, Rosa, et~al\mbox{.}}{Piantadosi
  et~al\mbox{.}}{2022}]%
        {piantadosi2022detecting}
\bibfield{author}{\bibinfo{person}{Valentina Piantadosi},
  \bibinfo{person}{Giovanni Rosa}, {et~al\mbox{.}}}
  \bibinfo{year}{2022}\natexlab{}.
\newblock \showarticletitle{Detecting functional and security-related issues in
  smart contracts: A systematic literature review}.
\newblock \bibinfo{journal}{\emph{Software: Practice and Experience}}
  (\bibinfo{year}{2022}).
\newblock


\bibitem[\protect\citeauthoryear{Praitheeshan, Pan, et~al\mbox{.}}{Praitheeshan
  et~al\mbox{.}}{2020}]%
        {praitheeshan2020security}
\bibfield{author}{\bibinfo{person}{Purathani Praitheeshan},
  \bibinfo{person}{Lei Pan}, {et~al\mbox{.}}} \bibinfo{year}{2020}\natexlab{}.
\newblock \showarticletitle{Security evaluation of smart contract-based
  on-chain ethereum wallets}. In \bibinfo{booktitle}{\emph{International
  Conference on Network and System Security}}. Springer,
  \bibinfo{pages}{22--41}.
\newblock


\bibitem[\protect\citeauthoryear{Praitheeshan, Pan, et~al\mbox{.}}{Praitheeshan
  et~al\mbox{.}}{2019}]%
        {praitheeshan2019security}
\bibfield{author}{\bibinfo{person}{Purathani Praitheeshan},
  \bibinfo{person}{Lei Pan}, {et~al\mbox{.}}} \bibinfo{year}{2019}\natexlab{}.
\newblock \showarticletitle{Security analysis methods on Ethereum smart
  contract vulnerabilities: a survey}.
\newblock  (\bibinfo{year}{2019}).
\newblock
\showeprint[arxiv]{1908.08605}


\bibitem[\protect\citeauthoryear{Praitheeshan, Pan, et~al\mbox{.}}{Praitheeshan
  et~al\mbox{.}}{2021}]%
        {praitheeshan2021solguard}
\bibfield{author}{\bibinfo{person}{Purathani Praitheeshan},
  \bibinfo{person}{Lei Pan}, {et~al\mbox{.}}} \bibinfo{year}{2021}\natexlab{}.
\newblock \showarticletitle{SolGuard: Preventing external call issues in smart
  contract-based multi-agent robotic systems}.
\newblock \bibinfo{journal}{\emph{Information Sciences}}  \bibinfo{volume}{579}
  (\bibinfo{year}{2021}), \bibinfo{pages}{150--166}.
\newblock


\bibitem[\protect\citeauthoryear{Qian, Liu, et~al\mbox{.}}{Qian
  et~al\mbox{.}}{2020}]%
        {qian2020towards}
\bibfield{author}{\bibinfo{person}{Peng Qian}, \bibinfo{person}{Zhenguang Liu},
  {et~al\mbox{.}}} \bibinfo{year}{2020}\natexlab{}.
\newblock \showarticletitle{Towards automated reentrancy detection for smart
  contracts based on sequential models}.
\newblock \bibinfo{journal}{\emph{IEEE Access}}  \bibinfo{volume}{8}
  (\bibinfo{year}{2020}), \bibinfo{pages}{19685--19695}.
\newblock


\bibitem[\protect\citeauthoryear{Rahimian and Clark}{Rahimian and
  Clark}{2021}]%
        {rahimian2021tokenhook}
\bibfield{author}{\bibinfo{person}{Reza Rahimian} {and} \bibinfo{person}{Jeremy
  Clark}.} \bibinfo{year}{2021}\natexlab{}.
\newblock \bibinfo{title}{TokenHook: Secure ERC-20 smart contract}.
\newblock
\newblock
\showeprint[arxiv]{cs.CR/2107.02997}


\bibitem[\protect\citeauthoryear{Rameder, Di~Angelo, et~al\mbox{.}}{Rameder
  et~al\mbox{.}}{2022}]%
        {rameder2022review}
\bibfield{author}{\bibinfo{person}{Heidelinde Rameder}, \bibinfo{person}{Monika
  Di~Angelo}, {et~al\mbox{.}}} \bibinfo{year}{2022}\natexlab{}.
\newblock \showarticletitle{Review of automated vulnerability analysis of smart
  contracts on Ethereum}.
\newblock \bibinfo{journal}{\emph{Front. Blockchain}}  \bibinfo{volume}{5}
  (\bibinfo{year}{2022}).
\newblock


\bibitem[\protect\citeauthoryear{Reis, Crocker, et~al\mbox{.}}{Reis
  et~al\mbox{.}}{2020}]%
        {reis2020tezla}
\bibfield{author}{\bibinfo{person}{Jo{\~a}o~Santos Reis}, \bibinfo{person}{Paul
  Crocker}, {et~al\mbox{.}}} \bibinfo{year}{2020}\natexlab{}.
\newblock \showarticletitle{Tezla, an intermediate representation for static
  analysis of Michelson smart contracts}.
\newblock  (\bibinfo{year}{2020}).
\newblock
\showeprint[arxiv]{2005.11839}


\bibitem[\protect\citeauthoryear{Ribeiro, Ad{\~a}o, et~al\mbox{.}}{Ribeiro
  et~al\mbox{.}}{2020}]%
        {ribeiro2020formal}
\bibfield{author}{\bibinfo{person}{Maria Ribeiro}, \bibinfo{person}{Pedro
  Ad{\~a}o}, {et~al\mbox{.}}} \bibinfo{year}{2020}\natexlab{}.
\newblock \showarticletitle{Formal Verification of Ethereum Smart Contracts
  Using Isabelle/HOL}.
\newblock In \bibinfo{booktitle}{\emph{Logic, Language, and Security}}.
  \bibinfo{publisher}{Springer}, \bibinfo{pages}{71--97}.
\newblock


\bibitem[\protect\citeauthoryear{Saad, Spaulding, et~al\mbox{.}}{Saad
  et~al\mbox{.}}{2019}]%
        {saad2019exploring}
\bibfield{author}{\bibinfo{person}{Muhammad Saad}, \bibinfo{person}{Jeffrey
  Spaulding}, {et~al\mbox{.}}} \bibinfo{year}{2019}\natexlab{}.
\newblock \showarticletitle{Exploring the Attack Surface of Blockchain: {A}
  Systematic Overview}.
\newblock  (\bibinfo{year}{2019}).
\newblock
\showeprint[arXiv]{1904.03487}


\bibitem[\protect\citeauthoryear{Samreen and Alalfi}{Samreen and
  Alalfi}{2020}]%
        {samreen2020reentrancy}
\bibfield{author}{\bibinfo{person}{Noama~Fatima Samreen} {and}
  \bibinfo{person}{Manar~H Alalfi}.} \bibinfo{year}{2020}\natexlab{}.
\newblock \showarticletitle{Reentrancy vulnerability identification in Ethereum
  smart contracts}. In \bibinfo{booktitle}{\emph{2020 IEEE International
  Workshop on Blockchain Oriented Software Engineering (IWBOSE)}}. IEEE,
  \bibinfo{pages}{22--29}.
\newblock


\bibitem[\protect\citeauthoryear{Samreen and Alalfi}{Samreen and
  Alalfi}{2021}]%
        {samreen2021smartscan}
\bibfield{author}{\bibinfo{person}{Noama~Fatima Samreen} {and}
  \bibinfo{person}{Manar~H Alalfi}.} \bibinfo{year}{2021}\natexlab{}.
\newblock \showarticletitle{SmartScan: An approach to detect Denial of Service
  Vulnerability in Ethereum Smart Contracts}.
\newblock  (\bibinfo{year}{2021}).
\newblock
\showeprint[arxiv]{2105.02852}


\bibitem[\protect\citeauthoryear{Sayeed, Marco-Gisbert, et~al\mbox{.}}{Sayeed
  et~al\mbox{.}}{2020}]%
        {sayeed2020Attacks}
\bibfield{author}{\bibinfo{person}{Sarwar Sayeed}, \bibinfo{person}{Hector
  Marco-Gisbert}, {et~al\mbox{.}}} \bibinfo{year}{2020}\natexlab{}.
\newblock \showarticletitle{Smart Contract: Attacks and Protections}.
\newblock \bibinfo{journal}{\emph{IEEE Access}}  \bibinfo{volume}{8}
  (\bibinfo{year}{2020}), \bibinfo{pages}{24416--24427}.
\newblock


\bibitem[\protect\citeauthoryear{Schiffl, Grundmann, et~al\mbox{.}}{Schiffl
  et~al\mbox{.}}{2021}]%
        {schiffl2021towards}
\bibfield{author}{\bibinfo{person}{Jonas Schiffl}, \bibinfo{person}{Matthias
  Grundmann}, {et~al\mbox{.}}} \bibinfo{year}{2021}\natexlab{}.
\newblock \showarticletitle{Towards Correct Smart Contracts: A Case Study on
  Formal Verification of Access Control}. In
  \bibinfo{booktitle}{\emph{Proceedings of the 26th ACM Symposium on Access
  Control Models and Technologies}}. \bibinfo{pages}{125--130}.
\newblock


\bibitem[\protect\citeauthoryear{Schneidewind, Grishchenko,
  et~al\mbox{.}}{Schneidewind et~al\mbox{.}}{2020a}]%
        {schneidewind2020ethor}
\bibfield{author}{\bibinfo{person}{Clara Schneidewind}, \bibinfo{person}{Ilya
  Grishchenko}, {et~al\mbox{.}}} \bibinfo{year}{2020}\natexlab{a}.
\newblock \showarticletitle{ethor: Practical and provably sound static analysis
  of ethereum smart contracts}. In \bibinfo{booktitle}{\emph{Proceedings of the
  2020 ACM SIGSAC Conference on Computer and Communications Security}}.
  \bibinfo{pages}{621--640}.
\newblock


\bibitem[\protect\citeauthoryear{Schneidewind, Scherer,
  et~al\mbox{.}}{Schneidewind et~al\mbox{.}}{2020b}]%
        {schneidewind2020good}
\bibfield{author}{\bibinfo{person}{Clara Schneidewind}, \bibinfo{person}{Markus
  Scherer}, {et~al\mbox{.}}} \bibinfo{year}{2020}\natexlab{b}.
\newblock \showarticletitle{The Good, the Bad and the Ugly: Pitfalls and Best
  Practices in Automated Sound Static Analysis of Ethereum Smart Contracts}. In
  \bibinfo{booktitle}{\emph{International Symposium on Leveraging Applications
  of Formal Methods}}. Springer, \bibinfo{pages}{212--231}.
\newblock


\bibitem[\protect\citeauthoryear{Seijas, Smith, et~al\mbox{.}}{Seijas
  et~al\mbox{.}}{2020}]%
        {seijas2020efficient}
\bibfield{author}{\bibinfo{person}{Pablo~Lamela Seijas}, \bibinfo{person}{David
  Smith}, {et~al\mbox{.}}} \bibinfo{year}{2020}\natexlab{}.
\newblock \showarticletitle{Efficient Static Analysis of Marlowe Contracts}. In
  \bibinfo{booktitle}{\emph{International Symposium on Leveraging Applications
  of Formal Methods}}. Springer, \bibinfo{pages}{161--177}.
\newblock


\bibitem[\protect\citeauthoryear{Sen}{Sen}{2007}]%
        {concolicTesting}
\bibfield{author}{\bibinfo{person}{Koushik Sen}.}
  \bibinfo{year}{2007}\natexlab{}.
\newblock \showarticletitle{Concolic Testing}. \bibinfo{publisher}{Association
  for Computing Machinery}, \bibinfo{address}{New York, NY, USA},
  \bibinfo{pages}{571–572}.
\newblock
\showISBNx{9781595938824}


\bibitem[\protect\citeauthoryear{Sen, Hajra, et~al\mbox{.}}{Sen
  et~al\mbox{.}}{2020}]%
        {sen2020supervised}
\bibfield{author}{\bibinfo{person}{Pratap~Chandra Sen},
  \bibinfo{person}{Mahimarnab Hajra}, {et~al\mbox{.}}}
  \bibinfo{year}{2020}\natexlab{}.
\newblock \showarticletitle{Supervised classification algorithms in machine
  learning: A survey and review}.
\newblock In \bibinfo{booktitle}{\emph{Emerging technology in modelling and
  graphics}}. \bibinfo{publisher}{Springer}, \bibinfo{pages}{99--111}.
\newblock


\bibitem[\protect\citeauthoryear{Sergey, Kumar, et~al\mbox{.}}{Sergey
  et~al\mbox{.}}{2018a}]%
        {Scilla2018}
\bibfield{author}{\bibinfo{person}{Ilya Sergey}, \bibinfo{person}{Amrit Kumar},
  {et~al\mbox{.}}} \bibinfo{year}{2018}\natexlab{a}.
\newblock \bibinfo{title}{Scilla: a Smart Contract Intermediate-Level
  LAnguage}.
\newblock
\newblock
\showeprint[arxiv]{1801.00687}


\bibitem[\protect\citeauthoryear{Sergey, Kumar, et~al\mbox{.}}{Sergey
  et~al\mbox{.}}{2018b}]%
        {sergey2018SCProperties}
\bibfield{author}{\bibinfo{person}{Ilya Sergey}, \bibinfo{person}{Amrit Kumar},
  {et~al\mbox{.}}} \bibinfo{year}{2018}\natexlab{b}.
\newblock \showarticletitle{Temporal Properties of Smart Contracts}. In
  \bibinfo{booktitle}{\emph{Leveraging Applications of Formal Methods,
  Verification and Validation. Industrial Practice}}.
  \bibinfo{publisher}{Springer International Publishing},
  \bibinfo{address}{Cham}, \bibinfo{pages}{323--338}.
\newblock
\showISBNx{978-3-030-03427-6}


\bibitem[\protect\citeauthoryear{Sergey, Kumar, et~al\mbox{.}}{Sergey
  et~al\mbox{.}}{2018c}]%
        {sergey2018temporal}
\bibfield{author}{\bibinfo{person}{Ilya Sergey}, \bibinfo{person}{Amrit Kumar},
  {et~al\mbox{.}}} \bibinfo{year}{2018}\natexlab{c}.
\newblock \showarticletitle{Temporal properties of smart contracts}. In
  \bibinfo{booktitle}{\emph{International Symposium on Leveraging Applications
  of Formal Methods}}. Springer, \bibinfo{pages}{323--338}.
\newblock


\bibitem[\protect\citeauthoryear{Sergey, Nagaraj, et~al\mbox{.}}{Sergey
  et~al\mbox{.}}{2019}]%
        {Scilla2019}
\bibfield{author}{\bibinfo{person}{Ilya Sergey}, \bibinfo{person}{Vaivaswatha
  Nagaraj}, {et~al\mbox{.}}} \bibinfo{year}{2019}\natexlab{}.
\newblock \showarticletitle{Safer Smart Contract Programming with Scilla}.
\newblock \bibinfo{journal}{\emph{Proc. ACM Program. Lang.}}
  \bibinfo{volume}{3}, \bibinfo{number}{OOPSLA}, Article
  \bibinfo{articleno}{185} (\bibinfo{date}{oct} \bibinfo{year}{2019}),
  \bibinfo{numpages}{30}~pages.
\newblock


\bibitem[\protect\citeauthoryear{Shakya, Mukherjee, et~al\mbox{.}}{Shakya
  et~al\mbox{.}}{2022}]%
        {SmartMixModel2022}
\bibfield{author}{\bibinfo{person}{Supriya Shakya}, \bibinfo{person}{Arnab
  Mukherjee}, {et~al\mbox{.}}} \bibinfo{year}{2022}\natexlab{}.
\newblock \showarticletitle{SmartMixModel: Machine Learning-based Vulnerability
  Detection of Solidity Smart Contracts}. In \bibinfo{booktitle}{\emph{2022
  IEEE International Conference on Blockchain (Blockchain)}}.
  \bibinfo{pages}{37--44}.
\newblock


\bibitem[\protect\citeauthoryear{Siegel}{Siegel}{2022}]%
        {DAOAttack}
\bibfield{author}{\bibinfo{person}{David Siegel}.}
  \bibinfo{year}{2022}\natexlab{}.
\newblock \bibinfo{title}{CoinDesk: Understanding The DAO Attack}.
\newblock
\newblock
\urldef\tempurl%
\url{https://www.coindesk.com/learn/2016/06/25/understanding-the-dao-attack/}
\showURL{%
Retrieved 2022-11-30 from \tempurl}


\bibitem[\protect\citeauthoryear{Singh, Parizi, et~al\mbox{.}}{Singh
  et~al\mbox{.}}{2020}]%
        {singh2020blockchain}
\bibfield{author}{\bibinfo{person}{Amritraj Singh}, \bibinfo{person}{Reza~M
  Parizi}, {et~al\mbox{.}}} \bibinfo{year}{2020}\natexlab{}.
\newblock \showarticletitle{Blockchain smart contracts formalization:
  Approaches and challenges to address vulnerabilities}.
\newblock \bibinfo{journal}{\emph{Computers and Security}}
  \bibinfo{volume}{88} (\bibinfo{year}{2020}), \bibinfo{pages}{101654}.
\newblock


\bibitem[\protect\citeauthoryear{Smaragdakis, Grech, et~al\mbox{.}}{Smaragdakis
  et~al\mbox{.}}{2021}]%
        {smaragdakis2021symbolic}
\bibfield{author}{\bibinfo{person}{Yannis Smaragdakis},
  \bibinfo{person}{Neville Grech}, {et~al\mbox{.}}}
  \bibinfo{year}{2021}\natexlab{}.
\newblock \showarticletitle{Symbolic value-flow static analysis: deep, precise,
  complete modeling of Ethereum smart contracts}.
\newblock \bibinfo{journal}{\emph{Proceedings of the ACM on Programming
  Languages}} \bibinfo{volume}{5}, \bibinfo{number}{OOPSLA}
  (\bibinfo{year}{2021}), \bibinfo{pages}{1--30}.
\newblock


\bibitem[\protect\citeauthoryear{So, Lee, et~al\mbox{.}}{So
  et~al\mbox{.}}{2020}]%
        {so2020verismart}
\bibfield{author}{\bibinfo{person}{Sunbeom So}, \bibinfo{person}{Myungho Lee},
  {et~al\mbox{.}}} \bibinfo{year}{2020}\natexlab{}.
\newblock \showarticletitle{VeriSmart: A highly precise safety verifier for
  Ethereum smart contracts}. In \bibinfo{booktitle}{\emph{2020 IEEE Symposium
  on Security and Privacy (SP)}}. IEEE, \bibinfo{pages}{1678--1694}.
\newblock


\bibitem[\protect\citeauthoryear{Song, Matulevicius, et~al\mbox{.}}{Song
  et~al\mbox{.}}{2022}]%
        {Son2022}
\bibfield{author}{\bibinfo{person}{Kunjian Song}, \bibinfo{person}{Nedas
  Matulevicius}, {et~al\mbox{.}}} \bibinfo{year}{2022}\natexlab{}.
\newblock \showarticletitle{ESBMC-Solidity: An SMT-Based Model Checker for
  Solidity Smart Contracts}. In \bibinfo{booktitle}{\emph{Proceedings of the
  ACM/IEEE 44th International Conference on Software Engineering: Companion
  Proceedings}} \emph{(\bibinfo{series}{ICSE '22})}.
  \bibinfo{publisher}{Association for Computing Machinery},
  \bibinfo{address}{New York, NY, USA}, \bibinfo{pages}{65–69}.
\newblock
\showISBNx{9781450392235}
\urldef\tempurl%
\url{https://doi.org/10.1145/3510454.3516855}
\showDOI{\tempurl}


\bibitem[\protect\citeauthoryear{Staderini, Palli, et~al\mbox{.}}{Staderini
  et~al\mbox{.}}{2020}]%
        {staderini2020classification}
\bibfield{author}{\bibinfo{person}{Mirko Staderini}, \bibinfo{person}{Caterina
  Palli}, {et~al\mbox{.}}} \bibinfo{year}{2020}\natexlab{}.
\newblock \showarticletitle{Classification of Ethereum Vulnerabilities and
  their Propagations}. In \bibinfo{booktitle}{\emph{2020 Second International
  Conference on Blockchain Computing and Applications (BCCA)}}. IEEE,
  \bibinfo{pages}{44--51}.
\newblock


\bibitem[\protect\citeauthoryear{Staff}{Staff}{2021}]%
        {cryptopedia}
\bibfield{author}{\bibinfo{person}{Cryptopedia Staff}.}
  \bibinfo{year}{2021}\natexlab{}.
\newblock \bibinfo{title}{Digital Assets: Cryptocurrencies vs. Tokens}.
\newblock
\newblock
\urldef\tempurl%
\url{https://www.gemini.com/cryptopedia/cryptocurrencies-vs-tokens-difference}
\showURL{%
Retrieved April 26, 2022 from \tempurl}


\bibitem[\protect\citeauthoryear{Steffen, Bichsel, et~al\mbox{.}}{Steffen
  et~al\mbox{.}}{2022}]%
        {steffen2022zeestar}
\bibfield{author}{\bibinfo{person}{Samuel Steffen}, \bibinfo{person}{Benjamin
  Bichsel}, {et~al\mbox{.}}} \bibinfo{year}{2022}\natexlab{}.
\newblock \showarticletitle{ZeeStar: Private Smart Contracts by Homomorphic
  Encryption and Zero-knowledge Proofs}. In \bibinfo{booktitle}{\emph{2022 IEEE
  Symposium on Security and Privacy (SP)}}. IEEE Computer Society,
  \bibinfo{pages}{1543--1543}.
\newblock


\bibitem[\protect\citeauthoryear{Steffen, Bichsel, et~al\mbox{.}}{Steffen
  et~al\mbox{.}}{2019}]%
        {zkay1.0}
\bibfield{author}{\bibinfo{person}{Samuel Steffen}, \bibinfo{person}{Benjamin
  Bichsel}, {et~al\mbox{.}}} \bibinfo{year}{2019}\natexlab{}.
\newblock \showarticletitle{Zkay: Specifying and Enforcing Data Privacy in
  Smart Contracts}. In \bibinfo{booktitle}{\emph{Proceedings of the 2019 ACM
  SIGSAC Conference on Computer and Communications Security}}.
  \bibinfo{publisher}{Association for Computing Machinery},
  \bibinfo{address}{New York, NY, USA}, \bibinfo{pages}{1759–1776}.
\newblock
\showISBNx{9781450367479}


\bibitem[\protect\citeauthoryear{Stephens, Ferles, et~al\mbox{.}}{Stephens
  et~al\mbox{.}}{2021}]%
        {stephens2021smartpulse}
\bibfield{author}{\bibinfo{person}{Jon Stephens}, \bibinfo{person}{Kostas
  Ferles}, {et~al\mbox{.}}} \bibinfo{year}{2021}\natexlab{}.
\newblock \showarticletitle{SmartPulse: Automated Checking of Temporal
  Properties in Smart Contracts}. In \bibinfo{booktitle}{\emph{IEEE S\&P}}.
\newblock


\bibitem[\protect\citeauthoryear{Sun and Yu}{Sun and Yu}{2020}]%
        {sun2020formal}
\bibfield{author}{\bibinfo{person}{Tianyu Sun} {and} \bibinfo{person}{Wensheng
  Yu}.} \bibinfo{year}{2020}\natexlab{}.
\newblock \showarticletitle{A formal verification framework for security issues
  of blockchain smart contracts}.
\newblock \bibinfo{journal}{\emph{Electronics}} \bibinfo{volume}{9},
  \bibinfo{number}{2} (\bibinfo{year}{2020}), \bibinfo{pages}{255}.
\newblock


\bibitem[\protect\citeauthoryear{S{\"u}r{\"u}c{\"u}, Yeprem,
  et~al\mbox{.}}{S{\"u}r{\"u}c{\"u} et~al\mbox{.}}{2022}]%
        {Onur2022}
\bibfield{author}{\bibinfo{person}{Onur S{\"u}r{\"u}c{\"u}},
  \bibinfo{person}{Uygar Yeprem}, {et~al\mbox{.}}}
  \bibinfo{year}{2022}\natexlab{}.
\newblock \showarticletitle{{A survey on ethereum smart contract vulnerability
  detection using machine learning}}. In \bibinfo{booktitle}{\emph{Disruptive
  Technologies in Information Sciences VI}},
  \bibfield{editor}{\bibinfo{person}{Misty Blowers},
  \bibinfo{person}{Russell~D. Hall}, {et~al\mbox{.}}} (Eds.),
  Vol.~\bibinfo{volume}{12117}. International Society for Optics and Photonics,
  \bibinfo{publisher}{SPIE}, \bibinfo{pages}{121170C}.
\newblock
\urldef\tempurl%
\url{https://doi.org/10.1117/12.2618899}
\showDOI{\tempurl}


\bibitem[\protect\citeauthoryear{Sutton, Greene, et~al\mbox{.}}{Sutton
  et~al\mbox{.}}{2007}]%
        {sutton2007fuzzing}
\bibfield{author}{\bibinfo{person}{Michael Sutton}, \bibinfo{person}{Adam
  Greene}, {et~al\mbox{.}}} \bibinfo{year}{2007}\natexlab{}.
\newblock \bibinfo{booktitle}{\emph{Fuzzing: brute force vulnerability
  discovery}}.
\newblock \bibinfo{publisher}{Pearson Education}.
\newblock


\bibitem[\protect\citeauthoryear{Tang, Zhou, et~al\mbox{.}}{Tang
  et~al\mbox{.}}{2021}]%
        {tang2021vulnerabilities}
\bibfield{author}{\bibinfo{person}{Xiangyan Tang}, \bibinfo{person}{Ke Zhou},
  {et~al\mbox{.}}} \bibinfo{year}{2021}\natexlab{}.
\newblock \showarticletitle{The Vulnerabilities in Smart Contracts: A Survey}.
  In \bibinfo{booktitle}{\emph{International Conference on Artificial
  Intelligence and Security}}. Springer, \bibinfo{pages}{177--190}.
\newblock


\bibitem[\protect\citeauthoryear{Tantikul and Ngamsuriyaroj}{Tantikul and
  Ngamsuriyaroj}{2020}]%
        {tantikul2020exploring}
\bibfield{author}{\bibinfo{person}{Phitchayaphong Tantikul} {and}
  \bibinfo{person}{Sudsanguan Ngamsuriyaroj}.} \bibinfo{year}{2020}\natexlab{}.
\newblock \showarticletitle{Exploring Vulnerabilities in Solidity Smart
  Contract.}. In \bibinfo{booktitle}{\emph{ICISSP}}. \bibinfo{pages}{317--324}.
\newblock


\bibitem[\protect\citeauthoryear{Technologies}{Technologies}{2017}]%
        {ParityWallet}
\bibfield{author}{\bibinfo{person}{Parity Technologies}.}
  \bibinfo{year}{2017}\natexlab{}.
\newblock \bibinfo{title}{Parity Technologies. Security Alert - Parity Wallet}.
\newblock
\newblock
\urldef\tempurl%
\url{https://www.parity.io/security-alert/}
\showURL{%
Retrieved January 1, 2021 from \tempurl}


\bibitem[\protect\citeauthoryear{Tikhomirov, Voskresenskaya,
  et~al\mbox{.}}{Tikhomirov et~al\mbox{.}}{2018}]%
        {tikhomirov2018smartcheck}
\bibfield{author}{\bibinfo{person}{Sergei Tikhomirov},
  \bibinfo{person}{Ekaterina Voskresenskaya}, {et~al\mbox{.}}}
  \bibinfo{year}{2018}\natexlab{}.
\newblock \showarticletitle{Smartcheck: Static analysis of ethereum smart
  contracts}. In \bibinfo{booktitle}{\emph{Proceedings of the 1st International
  Workshop on Emerging Trends in Software Engineering for Blockchain}}.
  \bibinfo{pages}{9--16}.
\newblock


\bibitem[\protect\citeauthoryear{Tolmach, Li, et~al\mbox{.}}{Tolmach
  et~al\mbox{.}}{2021}]%
        {tolmach2020survey}
\bibfield{author}{\bibinfo{person}{Palina Tolmach}, \bibinfo{person}{Yi Li},
  {et~al\mbox{.}}} \bibinfo{year}{2021}\natexlab{}.
\newblock \showarticletitle{A survey of smart contract formal specification and
  verification}.
\newblock \bibinfo{journal}{\emph{ACM Computing Surveys (CSUR)}}
  \bibinfo{volume}{54}, \bibinfo{number}{7} (\bibinfo{year}{2021}),
  \bibinfo{pages}{1--38}.
\newblock


\bibitem[\protect\citeauthoryear{Torres, Iannillo, et~al\mbox{.}}{Torres
  et~al\mbox{.}}{2021}]%
        {torres2021confuzzius}
\bibfield{author}{\bibinfo{person}{Christof~Ferreira Torres},
  \bibinfo{person}{Antonio~Ken Iannillo}, {et~al\mbox{.}}}
  \bibinfo{year}{2021}\natexlab{}.
\newblock \bibinfo{title}{ConFuzzius: A Data Dependency-Aware Hybrid Fuzzer for
  Smart Contracts}.
\newblock
\newblock
\showeprint[arxiv]{cs.CR/2005.12156}


\bibitem[\protect\citeauthoryear{Torres, Sch\"{u}tte, et~al\mbox{.}}{Torres
  et~al\mbox{.}}{2018}]%
        {Osiris2018}
\bibfield{author}{\bibinfo{person}{Christof~Ferreira Torres},
  \bibinfo{person}{Julian Sch\"{u}tte}, {et~al\mbox{.}}}
  \bibinfo{year}{2018}\natexlab{}.
\newblock \showarticletitle{Osiris: Hunting for Integer Bugs in Ethereum Smart
  Contracts}. \bibinfo{publisher}{Association for Computing Machinery},
  \bibinfo{address}{New York, NY, USA}, \bibinfo{pages}{664–676}.
\newblock
\showISBNx{9781450365697}


\bibitem[\protect\citeauthoryear{Tripp, Pistoia, et~al\mbox{.}}{Tripp
  et~al\mbox{.}}{2009}]%
        {tripp2009taj}
\bibfield{author}{\bibinfo{person}{Omer Tripp}, \bibinfo{person}{Marco
  Pistoia}, {et~al\mbox{.}}} \bibinfo{year}{2009}\natexlab{}.
\newblock \showarticletitle{TAJ: effective taint analysis of web applications}.
\newblock \bibinfo{journal}{\emph{ACM Sigplan Notices}} \bibinfo{volume}{44},
  \bibinfo{number}{6} (\bibinfo{year}{2009}), \bibinfo{pages}{87--97}.
\newblock


\bibitem[\protect\citeauthoryear{Tsankov, Dan, et~al\mbox{.}}{Tsankov
  et~al\mbox{.}}{2018}]%
        {tsankov2018securify}
\bibfield{author}{\bibinfo{person}{Petar Tsankov}, \bibinfo{person}{Andrei
  Dan}, {et~al\mbox{.}}} \bibinfo{year}{2018}\natexlab{}.
\newblock \showarticletitle{Securify: Practical security analysis of smart
  contracts}. In \bibinfo{booktitle}{\emph{Proceedings of the 2018 ACM SIGSAC
  Conference on Computer and Communications Security}}.
  \bibinfo{pages}{67--82}.
\newblock


\bibitem[\protect\citeauthoryear{Vacca, Di~Sorbo, et~al\mbox{.}}{Vacca
  et~al\mbox{.}}{2020}]%
        {vacca2020systematic}
\bibfield{author}{\bibinfo{person}{Anna Vacca}, \bibinfo{person}{Andrea
  Di~Sorbo}, {et~al\mbox{.}}} \bibinfo{year}{2020}\natexlab{}.
\newblock \showarticletitle{A systematic literature review of blockchain and
  smart contract development: Techniques, tools, and open challenges}.
\newblock \bibinfo{journal}{\emph{Journal of Systems and Software}}
  (\bibinfo{year}{2020}), \bibinfo{pages}{110891}.
\newblock


\bibitem[\protect\citeauthoryear{Van~Sprundel}{Van~Sprundel}{2005}]%
        {van2005fuzzing}
\bibfield{author}{\bibinfo{person}{Ilja Van~Sprundel}.}
  \bibinfo{year}{2005}\natexlab{}.
\newblock \showarticletitle{Fuzzing: Breaking software in an automated
  fashion}.
\newblock \bibinfo{journal}{\emph{Decmember 8th}} (\bibinfo{year}{2005}).
\newblock


\bibitem[\protect\citeauthoryear{Vogelsteller and Buterin}{Vogelsteller and
  Buterin}{2015}]%
        {ERC20Token}
\bibfield{author}{\bibinfo{person}{Fabian Vogelsteller} {and}
  \bibinfo{person}{Vitalik Buterin}.} \bibinfo{year}{2015}\natexlab{}.
\newblock \bibinfo{title}{ERC-20: Token Standard}.
\newblock
\newblock
\urldef\tempurl%
\url{https://eips.ethereum.org/EIPS/eip-20}
\showURL{%
Retrieved March 9, 2023 from \tempurl}


\bibitem[\protect\citeauthoryear{Wang, Li, et~al\mbox{.}}{Wang
  et~al\mbox{.}}{2019b}]%
        {wang2019vultron}
\bibfield{author}{\bibinfo{person}{Haijun Wang}, \bibinfo{person}{Yi Li},
  {et~al\mbox{.}}} \bibinfo{year}{2019}\natexlab{b}.
\newblock \showarticletitle{Vultron: catching vulnerable smart contracts once
  and for all}. In \bibinfo{booktitle}{\emph{2019 IEEE/ACM 41st International
  Conference on Software Engineering: New Ideas and Emerging Results
  (ICSE-NIER)}}. IEEE, \bibinfo{pages}{1--4}.
\newblock


\bibitem[\protect\citeauthoryear{Wang, Liu, et~al\mbox{.}}{Wang
  et~al\mbox{.}}{2020a}]%
        {wang2020oracle}
\bibfield{author}{\bibinfo{person}{Haijun Wang}, \bibinfo{person}{Ye Liu},
  {et~al\mbox{.}}} \bibinfo{year}{2020}\natexlab{a}.
\newblock \showarticletitle{Oracle-supported dynamic exploit generation for
  smart contracts}.
\newblock \bibinfo{journal}{\emph{IEEE Transactions on Dependable and Secure
  Computing}} (\bibinfo{year}{2020}).
\newblock


\bibitem[\protect\citeauthoryear{Wang, Zhang, et~al\mbox{.}}{Wang
  et~al\mbox{.}}{2019c}]%
        {NPChecker2019}
\bibfield{author}{\bibinfo{person}{Shuai Wang}, \bibinfo{person}{Chengyu
  Zhang}, {et~al\mbox{.}}} \bibinfo{year}{2019}\natexlab{c}.
\newblock \showarticletitle{Detecting Nondeterministic Payment Bugs in Ethereum
  Smart Contracts}.
\newblock \bibinfo{journal}{\emph{Proc. ACM Program. Lang.}}
  \bibinfo{volume}{3}, \bibinfo{number}{OOPSLA}, Article
  \bibinfo{articleno}{189} (\bibinfo{date}{oct} \bibinfo{year}{2019}),
  \bibinfo{numpages}{29}~pages.
\newblock


\bibitem[\protect\citeauthoryear{Wang, Song, et~al\mbox{.}}{Wang
  et~al\mbox{.}}{2020b}]%
        {wang2020contractward}
\bibfield{author}{\bibinfo{person}{Wei Wang}, \bibinfo{person}{Jingjing Song},
  {et~al\mbox{.}}} \bibinfo{year}{2020}\natexlab{b}.
\newblock \showarticletitle{Contractward: Automated vulnerability detection
  models for ethereum smart contracts}.
\newblock \bibinfo{journal}{\emph{IEEE Transactions on Network Science and
  Engineering}} (\bibinfo{year}{2020}).
\newblock


\bibitem[\protect\citeauthoryear{Wang, He, et~al\mbox{.}}{Wang
  et~al\mbox{.}}{2021a}]%
        {wang2021security}
\bibfield{author}{\bibinfo{person}{Yajing Wang}, \bibinfo{person}{Jingsha He},
  {et~al\mbox{.}}} \bibinfo{year}{2021}\natexlab{a}.
\newblock \showarticletitle{Security enhancement technologies for smart
  contracts in the blockchain: A survey}.
\newblock \bibinfo{journal}{\emph{Transactions on Emerging Telecommunications
  Technologies}} (\bibinfo{year}{2021}), \bibinfo{pages}{e4341}.
\newblock


\bibitem[\protect\citeauthoryear{Wang, Lahiri, et~al\mbox{.}}{Wang
  et~al\mbox{.}}{2019a}]%
        {wang2019formal}
\bibfield{author}{\bibinfo{person}{Yuepeng Wang}, \bibinfo{person}{Shuvendu~K
  Lahiri}, {et~al\mbox{.}}} \bibinfo{year}{2019}\natexlab{a}.
\newblock \showarticletitle{Formal verification of workflow policies for smart
  contracts in azure blockchain}. In \bibinfo{booktitle}{\emph{Working
  Conference on Verified Software: Theories, Tools, and Experiments}}.
  Springer, \bibinfo{pages}{87--106}.
\newblock


\bibitem[\protect\citeauthoryear{Wang, Wen, et~al\mbox{.}}{Wang
  et~al\mbox{.}}{2021b}]%
        {wang2021mar}
\bibfield{author}{\bibinfo{person}{Zexu Wang}, \bibinfo{person}{Bin Wen},
  {et~al\mbox{.}}} \bibinfo{year}{2021}\natexlab{b}.
\newblock \showarticletitle{MAR: A Dynamic Symbol Execution Detection Method
  for Smart Contract Reentry Vulnerability}. In
  \bibinfo{booktitle}{\emph{International Conference on Blockchain and
  Trustworthy Systems}}. Springer, \bibinfo{pages}{418--429}.
\newblock


\bibitem[\protect\citeauthoryear{Wang, Zheng, et~al\mbox{.}}{Wang
  et~al\mbox{.}}{2022}]%
        {wang2022gvd}
\bibfield{author}{\bibinfo{person}{Ziling Wang}, \bibinfo{person}{Qinyuan
  Zheng}, {et~al\mbox{.}}} \bibinfo{year}{2022}\natexlab{}.
\newblock \showarticletitle{GVD-net: Graph embedding-based Machine Learning
  Model for Smart Contract Vulnerability Detection}. In
  \bibinfo{booktitle}{\emph{2022 International Conference on Algorithms, Data
  Mining, and Information Technology (ADMIT)}}. IEEE, \bibinfo{pages}{99--103}.
\newblock


\bibitem[\protect\citeauthoryear{Wei, Wang, et~al\mbox{.}}{Wei
  et~al\mbox{.}}{2020}]%
        {wei2020smart}
\bibfield{author}{\bibinfo{person}{Zaoyu Wei}, \bibinfo{person}{Jiaqi Wang},
  {et~al\mbox{.}}} \bibinfo{year}{2020}\natexlab{}.
\newblock \showarticletitle{Smart Contract Fuzzing Based on Taint Analysis and
  Genetic Algorithms}.
\newblock \bibinfo{journal}{\emph{Journal of Quantum Computing}}
  \bibinfo{volume}{2}, \bibinfo{number}{1} (\bibinfo{year}{2020}),
  \bibinfo{pages}{11}.
\newblock


\bibitem[\protect\citeauthoryear{Weiss and Sch{\"u}tte}{Weiss and
  Sch{\"u}tte}{2019}]%
        {Annotary2019}
\bibfield{author}{\bibinfo{person}{Konrad Weiss} {and} \bibinfo{person}{Julian
  Sch{\"u}tte}.} \bibinfo{year}{2019}\natexlab{}.
\newblock \showarticletitle{Annotary: A concolic execution system for
  developing secure smart contracts}. In \bibinfo{booktitle}{\emph{European
  Symposium on Research in Computer Security}}. Springer,
  \bibinfo{pages}{747--766}.
\newblock


\bibitem[\protect\citeauthoryear{Wesley, Christakis, et~al\mbox{.}}{Wesley
  et~al\mbox{.}}{2021}]%
        {wesley2021compositional}
\bibfield{author}{\bibinfo{person}{Scott Wesley}, \bibinfo{person}{Maria
  Christakis}, {et~al\mbox{.}}} \bibinfo{year}{2021}\natexlab{}.
\newblock \showarticletitle{Compositional Verification of Smart Contracts
  Through Communication Abstraction}. In
  \bibinfo{booktitle}{\emph{International Static Analysis Symposium}}.
  Springer, \bibinfo{pages}{429--452}.
\newblock


\bibitem[\protect\citeauthoryear{Wood et~al\mbox{.}}{Wood
  et~al\mbox{.}}{2014}]%
        {wood2014ethereum}
\bibfield{author}{\bibinfo{person}{Gavin Wood} {et~al\mbox{.}}}
  \bibinfo{year}{2014}\natexlab{}.
\newblock \showarticletitle{Ethereum: A secure decentralised generalised
  transaction ledger}.
\newblock \bibinfo{journal}{\emph{Ethereum project yellow paper}}
  \bibinfo{volume}{151}, \bibinfo{number}{2014} (\bibinfo{year}{2014}),
  \bibinfo{pages}{1--32}.
\newblock


\bibitem[\protect\citeauthoryear{Workshop}{Workshop}{2004}]%
        {TheLivenessManifestor}
\bibfield{author}{\bibinfo{person}{Beyond Safety~International Workshop}.}
  \bibinfo{year}{2004}\natexlab{}.
\newblock \bibinfo{title}{Liveness Manifestor}.
\newblock
\newblock
\urldef\tempurl%
\url{https://cs.nyu.edu/acsys/beyond-safety/liveness.htm}
\showURL{%
Retrieved December 13, 2022 from \tempurl}


\bibitem[\protect\citeauthoryear{W\"{u}stholz and Christakis}{W\"{u}stholz and
  Christakis}{2020}]%
        {wustholz2020harvey}
\bibfield{author}{\bibinfo{person}{Valentin W\"{u}stholz} {and}
  \bibinfo{person}{Maria Christakis}.} \bibinfo{year}{2020}\natexlab{}.
\newblock \bibinfo{booktitle}{\emph{Harvey: A Greybox Fuzzer for Smart
  Contracts}}.
\newblock \bibinfo{publisher}{Association for Computing Machinery},
  \bibinfo{address}{New York, NY, USA}, \bibinfo{pages}{1398–1409}.
\newblock
\showISBNx{9781450370431}


\bibitem[\protect\citeauthoryear{Xi, Chen, et~al\mbox{.}}{Xi
  et~al\mbox{.}}{2003}]%
        {xi2003guarded}
\bibfield{author}{\bibinfo{person}{Hongwei Xi}, \bibinfo{person}{Chiyan Chen},
  {et~al\mbox{.}}} \bibinfo{year}{2003}\natexlab{}.
\newblock \showarticletitle{Guarded recursive datatype constructors}. In
  \bibinfo{booktitle}{\emph{Proceedings of the 30th ACM SIGPLAN-SIGACT
  symposium on Principles of programming languages}}.
  \bibinfo{pages}{224--235}.
\newblock


\bibitem[\protect\citeauthoryear{Xing, Chen, et~al\mbox{.}}{Xing
  et~al\mbox{.}}{2020}]%
        {xing2020new}
\bibfield{author}{\bibinfo{person}{Cipai Xing}, \bibinfo{person}{Zhuorong
  Chen}, {et~al\mbox{.}}} \bibinfo{year}{2020}\natexlab{}.
\newblock \showarticletitle{A new scheme of vulnerability analysis in smart
  contract with machine learning}.
\newblock \bibinfo{journal}{\emph{Wireless Networks}} (\bibinfo{year}{2020}),
  \bibinfo{pages}{1--10}.
\newblock


\bibitem[\protect\citeauthoryear{Xu, Liu, et~al\mbox{.}}{Xu
  et~al\mbox{.}}{2022}]%
        {HAMBiLSTM2022}
\bibfield{author}{\bibinfo{person}{Guangxia Xu}, \bibinfo{person}{Lei Liu},
  {et~al\mbox{.}}} \bibinfo{year}{2022}\natexlab{}.
\newblock \showarticletitle{Reentrancy Vulnerability Detection of Smart
  Contract Based on Bidirectional Sequential Neural Network with Hierarchical
  Attention Mechanism}. In \bibinfo{booktitle}{\emph{2022 International
  Conference on Blockchain Technology and Information Security}}. IEEE,
  \bibinfo{pages}{56--59}.
\newblock


\bibitem[\protect\citeauthoryear{Xu, Hu, et~al\mbox{.}}{Xu
  et~al\mbox{.}}{2021}]%
        {xu2021novel}
\bibfield{author}{\bibinfo{person}{Yingjie Xu}, \bibinfo{person}{Gengran Hu},
  {et~al\mbox{.}}} \bibinfo{year}{2021}\natexlab{}.
\newblock \showarticletitle{A Novel Machine Learning-Based Analysis Model for
  Smart Contract Vulnerability}.
\newblock \bibinfo{journal}{\emph{Security and Communication Networks}}
  \bibinfo{volume}{2021} (\bibinfo{year}{2021}).
\newblock


\bibitem[\protect\citeauthoryear{Xue, Ma, et~al\mbox{.}}{Xue
  et~al\mbox{.}}{2020}]%
        {xue2020cross}
\bibfield{author}{\bibinfo{person}{Yinxing Xue}, \bibinfo{person}{Mingliang
  Ma}, {et~al\mbox{.}}} \bibinfo{year}{2020}\natexlab{}.
\newblock \showarticletitle{Cross-contract static analysis for detecting
  practical reentrancy vulnerabilities in smart contracts}. In
  \bibinfo{booktitle}{\emph{2020 35th IEEE/ACM International Conference on
  Automated Software Engineering}}. IEEE, \bibinfo{pages}{1029--1040}.
\newblock


\bibitem[\protect\citeauthoryear{Yamashita, Nomura, et~al\mbox{.}}{Yamashita
  et~al\mbox{.}}{2019}]%
        {yamashitaPotentialRisks}
\bibfield{author}{\bibinfo{person}{Kazuhiro Yamashita},
  \bibinfo{person}{Yoshihide Nomura}, {et~al\mbox{.}}}
  \bibinfo{year}{2019}\natexlab{}.
\newblock \showarticletitle{Potential Risks of Hyperledger Fabric Smart
  Contracts}. In \bibinfo{booktitle}{\emph{2019 IEEE International Workshop on
  Blockchain Oriented Software Engineering (IWBOSE)}}. \bibinfo{pages}{1--10}.
\newblock


\bibitem[\protect\citeauthoryear{Yan, Wang, et~al\mbox{.}}{Yan
  et~al\mbox{.}}{2022}]%
        {YanXingrun2022}
\bibfield{author}{\bibinfo{person}{Xingrun Yan}, \bibinfo{person}{Shuo Wang},
  {et~al\mbox{.}}} \bibinfo{year}{2022}\natexlab{}.
\newblock \showarticletitle{A Semantic Analysis-Based Method for Smart Contract
  Vulnerability}. In \bibinfo{booktitle}{\emph{2022 IEEE 8th Intl Conference on
  Big Data Security on Cloud (BigDataSecurity), IEEE Intl Conference on High
  Performance and Smart Computing, (HPSC) and IEEE Intl Conference on
  Intelligent Data and Security (IDS)}}. \bibinfo{pages}{23--28}.
\newblock
\urldef\tempurl%
\url{https://doi.org/10.1109/BigDataSecurityHPSCIDS54978.2022.00015}
\showDOI{\tempurl}


\bibitem[\protect\citeauthoryear{Yang, Zhang, et~al\mbox{.}}{Yang
  et~al\mbox{.}}{2022}]%
        {Yang2022}
\bibfield{author}{\bibinfo{person}{Huiwen Yang}, \bibinfo{person}{Jiaming
  Zhang}, {et~al\mbox{.}}} \bibinfo{year}{2022}\natexlab{}.
\newblock \showarticletitle{Smart Contract Vulnerability Detection based on
  Abstract Syntax Tree}. In \bibinfo{booktitle}{\emph{2022 8th International
  Symposium on System Security, Safety, and Reliability (ISSSR)}}.
  \bibinfo{pages}{169--170}.
\newblock
\urldef\tempurl%
\url{https://doi.org/10.1109/ISSSR56778.2022.00032}
\showDOI{\tempurl}


\bibitem[\protect\citeauthoryear{Yang and Lei}{Yang and Lei}{2019}]%
        {FSPVME2019}
\bibfield{author}{\bibinfo{person}{Zheng Yang} {and} \bibinfo{person}{Hang
  Lei}.} \bibinfo{year}{2019}\natexlab{}.
\newblock \showarticletitle{A General Formal Memory Framework for Smart
  Contracts Verification based on Higher-Order Logic Theorem Proving}.
\newblock \bibinfo{journal}{\emph{International Journal of Performability
  Engineering}} \bibinfo{volume}{15}, \bibinfo{number}{11}, Article
  \bibinfo{articleno}{2998} (\bibinfo{year}{2019}),
  \bibinfo{numpages}{9}~pages.
\newblock


\bibitem[\protect\citeauthoryear{Yang and Lei}{Yang and Lei}{2020}]%
        {yang2020lolisa}
\bibfield{author}{\bibinfo{person}{Zheng Yang} {and} \bibinfo{person}{Hang
  Lei}.} \bibinfo{year}{2020}\natexlab{}.
\newblock \showarticletitle{Lolisa: Formal syntax and semantics for a subset of
  the solidity programming language in Mathematical Tool Coq}.
\newblock \bibinfo{journal}{\emph{Mathematical Problems in Engineering}}
  \bibinfo{volume}{2020} (\bibinfo{year}{2020}).
\newblock


\bibitem[\protect\citeauthoryear{Yang, Lei, et~al\mbox{.}}{Yang
  et~al\mbox{.}}{2020a}]%
        {yang2020hybrid}
\bibfield{author}{\bibinfo{person}{Zheng Yang}, \bibinfo{person}{Hang Lei},
  {et~al\mbox{.}}} \bibinfo{year}{2020}\natexlab{a}.
\newblock \showarticletitle{A hybrid formal verification system in coq for
  ensuring the reliability and security of ethereum-based service smart
  contracts}.
\newblock \bibinfo{journal}{\emph{IEEE Access}}  \bibinfo{volume}{8}
  (\bibinfo{year}{2020}), \bibinfo{pages}{21411--21436}.
\newblock


\bibitem[\protect\citeauthoryear{Yang, Liu, et~al\mbox{.}}{Yang
  et~al\mbox{.}}{2020b}]%
        {yang2020seraph}
\bibfield{author}{\bibinfo{person}{Zhiqiang Yang}, \bibinfo{person}{Han Liu},
  {et~al\mbox{.}}} \bibinfo{year}{2020}\natexlab{b}.
\newblock \showarticletitle{Seraph: enabling cross-platform security analysis
  for evm and wasm smart contracts}. In \bibinfo{booktitle}{\emph{2020 IEEE/ACM
  42nd International Conference on Software Engineering: Companion
  Proceedings}}. IEEE, \bibinfo{pages}{21--24}.
\newblock


\bibitem[\protect\citeauthoryear{Yao, Li, et~al\mbox{.}}{Yao
  et~al\mbox{.}}{2022}]%
        {Yao2022}
\bibfield{author}{\bibinfo{person}{Yao Yao}, \bibinfo{person}{Hui Li},
  {et~al\mbox{.}}} \bibinfo{year}{2022}\natexlab{}.
\newblock \showarticletitle{An Improved Vulnerability Detection System of Smart
  Contracts Based on Symbolic Execution}. In \bibinfo{booktitle}{\emph{2022
  IEEE International Conference on Big Data (Big Data)}}.
  \bibinfo{pages}{3225--3234}.
\newblock
\urldef\tempurl%
\url{https://doi.org/10.1109/BigData55660.2022.10020730}
\showDOI{\tempurl}


\bibitem[\protect\citeauthoryear{Yuan and Xie}{Yuan and Xie}{2022}]%
        {SVChecker2022}
\bibfield{author}{\bibinfo{person}{Ye Yuan} {and} \bibinfo{person}{TongYi
  Xie}.} \bibinfo{year}{2022}\natexlab{}.
\newblock \showarticletitle{SVChecker: a deep learning-based system for smart
  contract vulnerability detection}. In \bibinfo{booktitle}{\emph{International
  Conference on Computer Application and Information Security (ICCAIS 2021)}},
  Vol.~\bibinfo{volume}{12260}. SPIE, \bibinfo{pages}{226--231}.
\newblock


\bibitem[\protect\citeauthoryear{Zeng, He, et~al\mbox{.}}{Zeng
  et~al\mbox{.}}{2022}]%
        {EtherGIS2020}
\bibfield{author}{\bibinfo{person}{Qingren Zeng}, \bibinfo{person}{Jiahao He},
  {et~al\mbox{.}}} \bibinfo{year}{2022}\natexlab{}.
\newblock \showarticletitle{EtherGIS: A Vulnerability Detection Framework for
  Ethereum Smart Contracts Based on Graph Learning Features}. In
  \bibinfo{booktitle}{\emph{2022 IEEE 46th Annual Computers, Software, and
  Applications Conference}}. IEEE, \bibinfo{pages}{1742--1749}.
\newblock


\bibitem[\protect\citeauthoryear{Zhang, Wang, et~al\mbox{.}}{Zhang
  et~al\mbox{.}}{2022c}]%
        {Zhang2022}
\bibfield{author}{\bibinfo{person}{Lejun Zhang}, \bibinfo{person}{Jinlong
  Wang}, {et~al\mbox{.}}} \bibinfo{year}{2022}\natexlab{c}.
\newblock \showarticletitle{A Novel Smart Contract Vulnerability Detection
  Method Based on Information Graph and Ensemble Learning}.
\newblock \bibinfo{journal}{\emph{Sensors}} \bibinfo{volume}{22},
  \bibinfo{number}{9} (\bibinfo{year}{2022}).
\newblock
\showISSN{1424-8220}
\urldef\tempurl%
\url{https://doi.org/10.3390/s22093581}
\showDOI{\tempurl}


\bibitem[\protect\citeauthoryear{Zhang, Wang, et~al\mbox{.}}{Zhang
  et~al\mbox{.}}{2020b}]%
        {zhang2020ethploit}
\bibfield{author}{\bibinfo{person}{Qingzhao Zhang}, \bibinfo{person}{Yizhuo
  Wang}, {et~al\mbox{.}}} \bibinfo{year}{2020}\natexlab{b}.
\newblock \showarticletitle{Ethploit: From fuzzing to efficient exploit
  generation against smart contracts}. In \bibinfo{booktitle}{\emph{2020 IEEE
  27th International Conference on Software Analysis, Evolution and
  Reengineering (SANER)}}. IEEE, \bibinfo{pages}{116--126}.
\newblock


\bibitem[\protect\citeauthoryear{Zhang, Banescu, et~al\mbox{.}}{Zhang
  et~al\mbox{.}}{2019}]%
        {MPro2019}
\bibfield{author}{\bibinfo{person}{William Zhang}, \bibinfo{person}{Sebastian
  Banescu}, {et~al\mbox{.}}} \bibinfo{year}{2019}\natexlab{}.
\newblock \showarticletitle{MPro: Combining Static and Symbolic Analysis for
  Scalable Testing of Smart Contract}. In \bibinfo{booktitle}{\emph{2019 IEEE
  30th International Symposium on Software Reliability Engineering (ISSRE)}}.
  \bibinfo{pages}{456--462}.
\newblock


\bibitem[\protect\citeauthoryear{Zhang, Li, et~al\mbox{.}}{Zhang
  et~al\mbox{.}}{2022b}]%
        {zhang2022smart}
\bibfield{author}{\bibinfo{person}{Xuesen Zhang}, \bibinfo{person}{Jianhua Li},
  {et~al\mbox{.}}} \bibinfo{year}{2022}\natexlab{b}.
\newblock \showarticletitle{Smart Contract Vulnerability Detection Method based
  on Bi-LSTM Neural Network}. In \bibinfo{booktitle}{\emph{2022 IEEE
  International Conference on Advances in Electrical Engineering and Computer
  Applications (AEECA)}}. IEEE, \bibinfo{pages}{38--41}.
\newblock


\bibitem[\protect\citeauthoryear{Zhang, Ma, et~al\mbox{.}}{Zhang
  et~al\mbox{.}}{2020a}]%
        {zhang2020smartshield}
\bibfield{author}{\bibinfo{person}{Yuyao Zhang}, \bibinfo{person}{Siqi Ma},
  {et~al\mbox{.}}} \bibinfo{year}{2020}\natexlab{a}.
\newblock \showarticletitle{Smartshield: Automatic smart contract protection
  made easy}. In \bibinfo{booktitle}{\emph{2020 IEEE 27th International
  Conference on Software Analysis, Evolution and Reengineering (SANER)}}. IEEE,
  \bibinfo{pages}{23--34}.
\newblock


\bibitem[\protect\citeauthoryear{Zhang, Lei, et~al\mbox{.}}{Zhang
  et~al\mbox{.}}{2022a}]%
        {zhang2022reentrancy}
\bibfield{author}{\bibinfo{person}{Zhuo Zhang}, \bibinfo{person}{Yan Lei},
  {et~al\mbox{.}}} \bibinfo{year}{2022}\natexlab{a}.
\newblock \showarticletitle{Reentrancy Vulnerability Detection and
  Localization: A Deep Learning Based Two-phase Approach}. In
  \bibinfo{booktitle}{\emph{37th IEEE/ACM International Conference on Automated
  Software Engineering}}. \bibinfo{pages}{1--13}.
\newblock


\bibitem[\protect\citeauthoryear{Zhdarkin and Anureev}{Zhdarkin and
  Anureev}{2021}]%
        {zhdarkin2021development}
\bibfield{author}{\bibinfo{person}{Evgenii Zhdarkin} {and}
  \bibinfo{person}{Igor Anureev}.} \bibinfo{year}{2021}\natexlab{}.
\newblock \showarticletitle{Development and Verification of Smart-Contracts for
  the ScientificCoin Platform}. In \bibinfo{booktitle}{\emph{2021 IEEE 22nd
  International Conference of Young Professionals in Electron Devices and
  Materials (EDM)}}. IEEE, \bibinfo{pages}{528--532}.
\newblock


\bibitem[\protect\citeauthoryear{Zhong, Cheang, et~al\mbox{.}}{Zhong
  et~al\mbox{.}}{2020}]%
        {zhong2020move}
\bibfield{author}{\bibinfo{person}{Jingyi~Emma Zhong}, \bibinfo{person}{Kevin
  Cheang}, {et~al\mbox{.}}} \bibinfo{year}{2020}\natexlab{}.
\newblock \showarticletitle{The move prover}. In
  \bibinfo{booktitle}{\emph{International Conference on Computer Aided
  Verification}}. Springer, \bibinfo{pages}{137--150}.
\newblock


\bibitem[\protect\citeauthoryear{Zhou, Hua, et~al\mbox{.}}{Zhou
  et~al\mbox{.}}{2018}]%
        {zhou2018security}
\bibfield{author}{\bibinfo{person}{Ence Zhou}, \bibinfo{person}{Song Hua},
  {et~al\mbox{.}}} \bibinfo{year}{2018}\natexlab{}.
\newblock \showarticletitle{Security assurance for smart contract}. In
  \bibinfo{booktitle}{\emph{2018 9th IFIP International Conference on New
  Technologies, Mobility and Security (NTMS)}}. IEEE, \bibinfo{pages}{1--5}.
\newblock


\bibitem[\protect\citeauthoryear{Zhou, Zheng, et~al\mbox{.}}{Zhou
  et~al\mbox{.}}{2022}]%
        {Zhou2022}
\bibfield{author}{\bibinfo{person}{Qihao Zhou}, \bibinfo{person}{Kan Zheng},
  {et~al\mbox{.}}} \bibinfo{year}{2022}\natexlab{}.
\newblock \showarticletitle{Vulnerability Analysis of Smart Contract for
  Blockchain-Based IoT Applications: A Machine Learning Approach}.
\newblock \bibinfo{journal}{\emph{IEEE Internet of Things Journal}}
  \bibinfo{volume}{9}, \bibinfo{number}{24} (\bibinfo{year}{2022}),
  \bibinfo{pages}{24695--24707}.
\newblock
\urldef\tempurl%
\url{https://doi.org/10.1109/JIOT.2022.3196269}
\showDOI{\tempurl}


\bibitem[\protect\citeauthoryear{Zhou, Liu, et~al\mbox{.}}{Zhou
  et~al\mbox{.}}{2021}]%
        {teng2021SmartGift}
\bibfield{author}{\bibinfo{person}{Teng Zhou}, \bibinfo{person}{Kui Liu},
  {et~al\mbox{.}}} \bibinfo{year}{2021}\natexlab{}.
\newblock \showarticletitle{SmartGift: Learning to Generate Practical Inputs
  for Testing Smart Contracts}. In \bibinfo{booktitle}{\emph{2021 IEEE
  International Conference on Software Maintenance and Evolution (ICSME)}}.
  \bibinfo{pages}{23--34}.
\newblock


\bibitem[\protect\citeauthoryear{Zhu, Zheng, et~al\mbox{.}}{Zhu
  et~al\mbox{.}}{2018}]%
        {zhu2018research}
\bibfield{author}{\bibinfo{person}{Liehuang Zhu}, \bibinfo{person}{Baokun
  Zheng}, {et~al\mbox{.}}} \bibinfo{year}{2018}\natexlab{}.
\newblock \showarticletitle{Research on the Security of Blockchain Data: {A}
  Survey}.
\newblock  (\bibinfo{year}{2018}).
\newblock
\showeprint[arXiv]{1812.02009}


\bibitem[\protect\citeauthoryear{Zhuang, Liu, et~al\mbox{.}}{Zhuang
  et~al\mbox{.}}{2020}]%
        {zhuang2020smart}
\bibfield{author}{\bibinfo{person}{Yuan Zhuang}, \bibinfo{person}{Zhenguang
  Liu}, {et~al\mbox{.}}} \bibinfo{year}{2020}\natexlab{}.
\newblock \showarticletitle{Smart Contract Vulnerability Detection using Graph
  Neural Network.}. In \bibinfo{booktitle}{\emph{IJCAI}}.
  \bibinfo{pages}{3283--3290}.
\newblock


\bibitem[\protect\citeauthoryear{Zipfel}{Zipfel}{2020}]%
        {hashCollisionIncident}
\bibfield{author}{\bibinfo{person}{Kaden Zipfel}.}
  \bibinfo{year}{2020}\natexlab{}.
\newblock \bibinfo{title}{Hash collisions with multiple variable lengths
  arguments}.
\newblock
\newblock
\urldef\tempurl%
\url{https://medium.com/swlh/new-smart-contract-weakness-hash-collisions-with-multiple-variable-length-arguments-dc7b9c84e493}
\showURL{%
Retrieved December 08, 2022 from \tempurl}


\end{thebibliography}
